\documentclass[twocolumn]{aastex62}
\pdfoutput=1
\usepackage{gensymb,amsmath}
\usepackage{epstopdf}
\usepackage{url}
\usepackage[]{natbib}

\shorttitle{Resolved Properties of J0901}
\shortauthors{Sharon et al.}
\accepted{to ApJ 5/17/2019}
\begin{document}
\title{Resolved Molecular Gas and Star Formation Properties of the Strongly Lensed $z=2.26$ Galaxy SDSS\,J0901+1814\footnote{Based on observations carried out with the IRAM Plateau de Bure Interferometer. IRAM is supported by INSU/CNRS (France), MPG (Germany) and IGN (Spain).}}

\correspondingauthor{Chelsea E. Sharon}
\email{chelsea.sharon@yale-nus.edu.sg}

\author{Chelsea E. Sharon}
\affil{Yale-NUS College, Singapore, 138527}
\affil{Department of Physics \& Astronomy, McMaster University, Hamilton, ON, L8S-4M1, Canada}
\author{Amitpal S. Tagore}
\affil{Jodrell Bank Centre for Astrophysics, The University of Manchester, Manchester, M13 9PL, UK}
\author{Andrew J. Baker}
\affil{Department of Physics and Astronomy, Rutgers, the State University of New Jersey, Piscataway, NJ, 08854-8019, USA}
\author{Jesus Rivera}
\affil{Department of Physics and Astronomy, Rutgers, the State University of New Jersey, Piscataway, NJ, 08854-8019, USA}
\author{Charles R. Keeton}
\affil{Department of Physics and Astronomy, Rutgers, the State University of New Jersey, Piscataway, NJ, 08854-8019, USA}
\author{Dieter Lutz}
\affil{Max-Planck-Institut f\"{u}r extraterrestrische Physik (MPE), Giessenbachstr. 1, 85748 Garching, Germany}
\author{Reinhard Genzel}
\affil{Max-Planck-Institut f\"{u}r extraterrestrische Physik (MPE), Giessenbachstr. 1, 85748 Garching, Germany}
\author{David J. Wilner}
\affil{Harvard-Smithsonian Center for Astrophysics, Cambridge, MA, 02138, USA}
\author{Erin K. S. Hicks}
\affil{Department of Physics and Astronomy, University of Alaska, Anchorage, AK, 99508, USA}
\author{Sahar S. Allam}
\affil{Center for Particle Astrophysics, Fermi National Accelerator Laboratory, P.O. Box 500, Batavia, IL, 60510, USA}
\author{Douglas L. Tucker}
\affil{Center for Particle Astrophysics, Fermi National Accelerator Laboratory, P.O. Box 500, Batavia, IL, 60510, USA}

\begin{abstract} 

We present $\sim1^{\prime\prime}$ resolution ($\sim2\,{\rm kpc}$ in the source plane) observations of the \mbox{CO(1--0)}, \mbox{CO(3--2)}, ${\rm H\alpha}$, and [N\,{\sc ii}] lines in the strongly-lensed $z=2.26$ star-forming galaxy SDSS\,J0901+1814. We use these observations to constrain the lensing potential of a foreground group of galaxies, and our source-plane reconstructions indicate that SDSS\,J0901+1814 is a nearly face-on ($i\approx30\degree$) massive disk with $r_{1/2}\gtrsim4\,{\rm kpc}$ for its molecular gas. Using our new magnification factors ($\mu_{tot}\approx30$), we find that SDSS\,J0901+1814 has a star formation rate (SFR) of $268^{+63}_{-61}\,{M_\sun}\,{\rm yr^{-1}}$, $M_{\rm gas}=(1.6^{+0.3}_{-0.2})\times10^{11}(\alpha_{\rm CO}/4.6)\,{M_\sun}$, and $M_\star=(9.5^{+3.8}_{-2.8})\times10^{10}\,{M_\sun}$, which places it on the star-forming galaxy ``main sequence." We use our matched high-angular resolution gas and SFR tracers (CO and ${\rm H\alpha}$, respectively) to perform a spatially resolved (pixel-by-pixel) analysis of SDSS\,J0901+1814 in terms of the Schmidt-Kennicutt relation. After correcting for the large fraction of obscured star formation (${\rm SFR_{H\alpha}}/{\rm SFR_{TIR}}=0.054^{+0.015}_{-0.014}$), we find SDSS\,J0901+1814 is offset from ``normal" star-forming galaxies to higher star formation efficiencies independent of assumptions for the CO-to-${\rm H_2}$ conversion factor. Our mean best-fit index for the Schmidt-Kennicutt relation for SDSS\,J0901+1814, evaluated with different CO lines and smoothing levels,  is $\bar n=1.54\pm0.13$; however, the index may be affected by gravitational lensing, and we find $\bar n=1.24\pm0.02$ when analyzing the source-plane reconstructions. While the Schmidt-Kennicutt index largely appears unaffected by which of the two CO transitions we use to trace the molecular gas, the source-plane reconstructions and dynamical modeling suggest that the CO(1--0) emission is more spatially extended than the CO(3--2) emission.
\end{abstract}


\section{Introduction \label{sec:intro}}

Star forming galaxies at high redshift have been selected using a variety of methods, most notably on the basis of rest-ultraviolet (UV) colors (e.\/g.\/, Lyman break galaxies, LBGs; \citealt{steidel1996,giavalisco2002}) and large (observed frame) submillimeter fluxes (e.\/g.\/, submillimeter galaxies, SMGs; \citealt{blain2002,casey2014}). Historically, galaxies selected using these two methods have been described as two separate populations, with UV-bright galaxies characterized as ``normal" galaxies at high redshifts (mostly disks, and falling along a star-forming ``main sequence" (MS) in star formation rate (SFR) vs.~stellar mass; e.\/g\/, \citealt{forster2006,genzel2006,bouche2007,wright2007,noeske2007,elbaz2007,daddi2007,genzel2008,forster2009,wisnioski2015}) whose SFRs are an order of magnitude lower than those of dusty starbursts that are SMGs \citep[e.\/g.\/,][]{rodighiero2011}. However, fits to spectral energy distributions (SEDs; e.\/g.\/, \citealt{wuyts2011b}), the decomposition of many of the brightest SMGs into multiples, and stacking \citep[e.\/g.\/,][]{lindner2012,decarli2014,walter2014} suggest there is substantial overlap in the underlying physical properties of UV- and IR-bright high-$z$ galaxies, at least for higher masses.

Despite the increasing evidence of overlap between these populations, comparing their directly observable properties remains difficult. The substantial dust masses that give SMGs their large far-infrared (FIR) luminosities obscure their UV-emission \citep[e.\/g.\/,][]{smail2002,hodge2012}, including common short-wavelength SFR tracers such as ${\rm H\alpha}$. Similarly, UV-bright galaxies are comparatively dust and gas poor, and therefore frequently require substantial investments of telescope time and/or magnification from gravitational lensing to achieve mere detections of dust and molecular gas (e.\/g.\/, \citealt{baker2001,baker2004,coppin2007,daddi2010a,saintonge2013,dessauges2015}). Only recently have larger samples of high-redshift optical/UV color-selected galaxies been detected in CO (e.\/g.\/, \citealt{tacconi2013,freundlich2017}; see also: \citealt{genzel2015,tacconi2017} and references therein), the canonical tracer of molecular gas, in numbers comparable to those of dusty galaxies (see \citealt{carilli2013b} for a review of gas in high-redshift galaxies). Of the UV or optical color-selected galaxies with CO detections, few have spatially resolved or multi-$J$ CO detections \citep[e.\/g.\/,][]{genzel2013}. With the wide bandwidths and the sensitivities of telescopes like the Atacama Large Millimeter/submillimeter Array, it has been suggested that dust continuum measurements may be a more efficient way to measure the masses of galaxies' interstellar media (ISMs; e.\/g.\/, \citealt{scoville2014,scoville2016}), including their molecular gas components, even for UV-bright/dust-poor systems. However, using different observables to trace the same intrinsic galaxy parameter (e.\/g.\/, infrared vs.~UV-tracers of the total SFR, or dust vs.~CO tracers of the molecular gas) may generate false differences between galaxy populations due to systematic factors like extinction or AGN contamination (\citealt{kennicutt2012}). 

The lack of data at complementary wavelengths also makes resolved multi-wavelength analyses applied to low-redshift galaxies, such as the Schmidt-Kennicutt relation (the correlation between galaxies' SFR and gas mass surface densities; e.\/g.\/, \citealt{schmidt1959,kennicutt1989,kennicutt1998}) significantly less common at high redshift. High-resolution CO observations are critical for evaluating where high-redshift galaxies fall on the true surface density version of the Schmidt-Kennicutt relation, where $\Sigma_{\rm SFR}$ and $\Sigma_{\rm gas}$ can be compared on a pixel-by-pixel basis within individual galaxies (as done for local galaxies; e.\/g.\/, \citealt{kennicutt2007,bigiel2008,wei2010,bigiel2011,leroy2013}). Many high-redshift analyses use star formation and gas properties averaged over the entire galaxy \citep[e.\/g.\/,][]{kennicutt1989,buat1989,kennicutt1998,genzel2010,daddi2010b,tacconi2013} or avoid the additional uncertainties in source size and scaling factors by using the total luminosities of the star formation and gas tracers \citep[e.\/g.\/,][]{young1986,solomon1988,gao2004}. These different methods for determining SFRs and gas masses make it difficult to compare studies that focus on different galaxy populations, leading to significant uncertainties in the power-law index of the Schmidt-Kennicutt relation and the relative placement of different galaxy types in the $\Sigma_{\rm SFR}$--$\Sigma_{\rm gas}$ plane. Accurately characterizing the Schmidt-Kennicutt relation is important, since offsets imply a difference in star formation efficiency (SFE), and the power law index probes the underlying physical processes of star formation (for example, a linear correlation would imply supply-limited star formation, whereas super-linear correlations occur if star formation depends on cloud-cloud collisions or total gas free-fall collapse times; e.\/g.\/, \citealt{larson1992,tan1999,krumholz2005c,ostriker2011}). Systematic differences in the Schmidt-Kennicutt relation between different galaxy populations would imply important differences in their star formation processes.

\citet{kennicutt2012} present a compilation of disk-averaged SFR and gas mass surface densities whose values have been calculated in a uniform manner across different galaxy types (including normal disk galaxies and dusty starburst galaxies selected in the IR), and find a power law index of $n\sim1.4$. However, this result may be an artifact of combining galaxies of different interaction states. For a sample of $z\sim1$--$3$ MS galaxies, \citet{tacconi2013} find an index consistent with unity and only a slight offset between their high-redshift sample and a low-redshift sample with similar masses. However, SMGs and other ultra-/luminous infrared galaxies (U/LIRGs) are further offset above the correlation for star-forming disk galaxies even when similar CO-to-${\rm H_2}$ conversion factors are used for all galaxy populations (see also \citealt{bigiel2008,daddi2010b,genzel2010,genzel2015,tacconi2017}). In analyses of the resolved star formation properties of nearby disks, a near-unity index for the Schmidt-Kennicutt relation is also found in regimes where the molecular gas dominates the total gas mass surface density ($\Sigma_{\rm gas}>9\,M_\sun\,{\rm pc^{-2}}$; e.\/g.\/, \citealt{bigiel2008,bigiel2010,schruba2011}). The surface density version of the Schmidt-Kennicutt relation has been evaluated \emph{within} only eight high-redshift galaxies: SMM\,J14011+0252 at $z=2.56$ \citep{sharon2013}, EGS 13011166 at $z=1.53$ \citep{genzel2013}, HLS0918 at $z=5.24$ \citep{rawle2014}, GN20 at $z=4.05$ \citep{hodge2015}, PLCK G244.8+54.9 at $z=3.00$ \citep{canameras2017b}, AzTEC-1 at $z=4.34$ \citep{tadaki2018}, and the two components of HATLAS J084933 at $z=2.41$ \citep{gomez2018}\footnote{\citet{freundlich2013} and \citet{sharda2017} also examine the Schmidt-Kennicutt relation at $z>1$, but they analyze individually resolved clumps within high-redshift galaxies rather than performing full pixel-by-pixel comparisons.}. These studies find a range of Schmidt-Kennicutt relation indices ($n=1$--$2$). It is particularly worth noting that \citet{genzel2013} find that their measured index depends strongly on which spatially-resolved extinction correction they apply to their ${\rm H\alpha}$ measurements.

Comparisons between the Schmidt-Kennicutt relations for high- and low-redshift galaxies may be affected by the different CO lines observed \citep{narayanan2011}; the molecular gas in local galaxies is probed via the \mbox{CO(1--0)} and/or \mbox{CO(2--1)} lines, while the molecular gas at high redshift has typically been probed via mid-$J$ CO lines (i.\/e.\/, \mbox{CO(3--2)}, \mbox{CO(4--3)}, and \mbox{CO(5--4)}). Different transitions have different excitation temperatures and critical densities and are therefore sensitive to different density regimes in the molecular ISM \citep{krumholz2007,narayanan2008d,narayanan2011}, making the observed index dependent on the physical conditions of the star-forming gas. Using either global luminosities or mean surface densities, substantial differences in Schmidt-Kennicutt indices have been found using molecular gas tracers with different critical densities in local galaxies (all with $n<1.5$; e.\/g.\/, \citealt{gao2004,narayanan2005,gracia-carpio2008,bussmann2008,iono2009,juneau2009,greve2014,kamenetzky2016}), but no significant difference in index has been found between \mbox{CO(1--0)} and \mbox{CO(3--2)} studies of $z>1$ galaxies \citep{tacconi2013,sharon2016}. So far there have been no comparisons between Schmidt-Kennicutt indices for different molecular gas tracers in spatially resolved studies of high-redshift galaxies.

Here we present high-resolution ($\sim1^{\prime\prime}$ observed; $\sim2\,{\rm kpc}$ in the source plane) observations of the molecular gas and star formation tracers in the UV-bright galaxy SDSS\,J0901+1814 (J0901 hereafter). J0901 was discovered by \citet{diehl2009} in a systematic search of the Sloan Digital Sky Survey \citep{york2000} for strongly lensed galaxies (identified as blue arcs near known brightest cluster galaxies or luminous red galaxies). Followup observations at the Astrophysics Research Consortium (ARC) $3.5\,{\rm m}$ telescope at Apache Point Observatory confirmed that J0901 is a $z=2.26$ galaxy \citep{diehl2009,hainline2009} that is multiply imaged (into a pair of bright arcs to the north and south that nearly connect to the east, and a fainter western counter-image) by a $z=0.35$ luminous red galaxy. Single-slit spectroscopy at rest-frame optical wavelengths using Keck\,II/NIRSPEC show large [N\,{\sc ii}] ($\lambda=6583\,{\rm \AA}$)/${\rm H\alpha}$ ratios in the two brightest images \citep{hainline2009}, indicating the presence of an AGN (e.\/g.\/, \citealt{baldwin1981,kauffman2003}) that includes a prominent broad-line component \citep{genzel2014}. However, the strong PAH features detected in {\it Spitzer}/IRS spectra and weak continuum features in the (observed frame) mid-IR suggest that the AGN contribution to the IR luminosity of J0901 is negligible \citep{fadely2010}. Further observations have revealed that J0901 is one of the brightest high-redshift UV-selected galaxies in terms of its dust emission \citep[e.\/g.\/,][]{baker2001,coppin2007}; \citet{saintonge2013} estimate a total IR luminosity (magnification corrected) of $L_{\rm IR}\sim7\times10^{12}(\mu/8)\,L_\odot$ using {\it Herschel}/PACS and SPIRE photometry. The substantial dust content implied by the IR luminosity makes J0901 a natural target for observations of molecular emission lines and other gas-phase coolants; \citet{rhoads2014} observe a double-peaked profile in (spatially unresolved) {\it Herschel}/HIFI observations of the [C\,{\sc ii}] $158\,{\rm \mu m}$ line and infer that J0901 is a rotating disk galaxy. The additional spatial resolution provided by gravitational lensing allows us to resolve the velocity structure of J0901 and verify its structure in this paper, as well as study the variation of gas and star formation conditions with J0901. 

We describe our observations of J0901 and basic measurements in Sections~\ref{sec:obs} and \ref{sec:results}, respectively. In Section~\ref{sec:analysis} we describe our lens model for J0901 (Section~\ref{sec:lensing}); the resulting magnification-corrected gas mass, stellar mass, SFR, and dynamical mass (Section~\ref{sec:masses}); resolved analyses of CO excitation (Section~\ref{sec:COratios}), metallicity (Section~\ref{sec:metallicity}), the Schmidt-Kennicutt relation (Section~\ref{sec:KSlaw}), and the SFR-CO excitation correlation (Section~\ref{sec:sfrco}); and finally, the potential radio continuum emission from the central AGN (Section~\ref{sec:VLAcont}). Our results are summarized in Section~\ref{sec:concl}. We assume the WMAP7+BAO+$H_0$ mean $\Lambda$CDM cosmology throughout this paper, with $\Omega_\Lambda=0.725$ and $H_0=70.2\,{\rm km\,s^{-1}\,Mpc^{-1}}$ \citep{komatsu2011}.

\section{Observations \& Reduction}
\label{sec:obs}

\subsection{IRAM Plateau de Bure Interferometer}
\label{sec:pdbi}

We observed \mbox{CO(3--2)} emission from J0901 using the IRAM Plateau de Bure Interferometer (PdBI; \citealt{guilloteau1992}) in four separate configurations. Three tracks in a five-antenna version of the compact D configuration were obtained in September and October 2008 (project ID S040; PI Baker), with a single pointing centered on the southern image that had been strongly detected in 1.2\,mm continuum photometry ($6.4 \pm 0.6\,{\rm mJy}$) with the Max-Planck Millimeter Bolometer (MAMBO) array \citep{kreysa1998}. The PdBI data confirmed that all three images were detected at high significance in \mbox{CO(3--2)}, motivating the acquisition of four further tracks from 2009 November through 2010 February with all six PdBI antennas in their more extended C (1), B (1), and A (2) configurations (project ID T0AB; PI Baker). All observations targeted a J2000 position of $\alpha {\rm (J2000)} = 09^{\rm h}01^{\rm m}22.^{\rm s}59$ and $\delta {\rm (J2000)} = 18\degree14^{\prime}24.20^{\prime\prime}$, and a redshifted \mbox{CO(3--2)} line frequency of $106.082\,{\rm GHz}$ in the upper sideband. We employed a narrow-band correlator mode with $5\,{\rm MHz}$ channels and a total bandwidth of $1\,{\rm GHz}$, which recorded both horizontal and vertical polarizations. The final combination of seven tracks yielded 52 distinct baselines with lengths ranging from $24\,{\rm m}$ to $760\,{\rm m}$, and a total on-source integration time equivalent to $18.0\,{\rm hr}$ with a six-telescope array.

Phase and amplitude variation were tracked by interleaving observations of J0901 and the bright quasar PG\,0851+202, only $2.4^\circ$ away on the sky. Bandpass calibrators included PG\,0851+202, 3C273, and 0932+392; our overall flux scale was tied to observations of MWC349 and the quasars 3C273 and 0923+392, which are regularly monitored with IRAM facilities, and is accurate to $\sim 10\%$. Calibration and flagging for data quality used the CLIC program within the IRAM GILDAS package \citep{guilloteau2000}. The resulting $uv$ data set was exported to FITS format and imaged with AIPS. We created an initial set of channel maps to explore possible $uv$ weighting schemes, and after comparing these settled on a robustness of 1, which delivered slightly higher resolution than natural weighting without compromising image fidelity or flux recovery. Our final data cube has a synthesized beam of $1.33^{\prime\prime} \times 0.98^{\prime\prime}$ at a position angle of $41.1^\circ$, and a mean rms noise of $0.62\,{\rm mJy\,beam^{-1}}$ per $5\,{\rm MHz} \leftrightarrow 14.1\,{\rm km\,s^{-1}}$ channel. Following confirmation that it contained no continuum emission at the sensitivity/resolution of these observations (as expected), the resulting data cube was cleaned with the IMAGR task in AIPS, corrected for primary beam attenuation, and analyzed further with a custom set of IDL scripts.

\subsection{Karl G. Jansky Very Large Array}
\label{sec:vla}

We observed J0901 at the Karl G.~Jansky Very Large Array (VLA) in three different configurations (project IDs AB1347, AS1057, AS1144; PIs Baker, Sharon); the configurations, maximum baselines, observation dates, numbers of antennas used, and weather conditions are summarized in Table~\ref{tab:j0901vlaobs}. The minimum $uv$-radius of the full dataset is $3.67\,{\rm k\lambda}$. We observed with the WIDAR correlator in the ``OSRO Dual Polarization" mode using the lowest spectral resolution (256 channels $\times~500\,{\rm kHz}$ resolution) and a single intermediate frequency pair (IF pair B/D). The total bandwidth of $128\,{\rm MHz}$ was centered at the observed frequency of \mbox{CO(1--0)} for $z=2.2586$ ($35.363\,{\rm GHz}$). Observations were centered at $\alpha {\rm (J2000)} = 09^{\rm h}01^{\rm m}23.^{\rm s}00$, $\delta {\rm (J2000)} = +18\degree14^{\prime}24.0^{\prime\prime}$, the position of the southernmost and brightest (at optical wavelengths) of the three lensed images \citep{diehl2009}. At the beginning of each track, we observed 3C\,138 as both passband and flux calibrator, adopting $S_\nu = 1.1786\,{\rm Jy}$ using the CASA\footnote{\href{http://casa.nrao.edu}{http://casa.nrao.edu}} \citep{mcmullin2007} package's default ``Perley-Butler 2010" flux standard. Phase and amplitude fluctuations were tracked by alternating between the source and a nearby quasar, J0854+2006, with a cycle time of 6 minutes. A total of 16 hours was spent on source across the various configurations in Table~\ref{tab:j0901vlaobs}.

\floattable
\begin{deluxetable*}{clcl}
\tablewidth{0pt}
\tablecaption{J0901 VLA Observations \label{tab:j0901vlaobs}}
\tablehead{ {Configuration} & {Date} & {$N_{\rm Ant}$} & {Weather} \\
{Max. Baseline} & {} & {} }
\startdata
D & 2010 April 3& 17 &Clear\\
{1.031 km} & 2010 April 15& 20 & Clear\\
{} & 2010 May 4& 19 & Clear\\
{} & 2010 May 8 & 19 & Average sky cover 25\%; mixed clouds\\
{} & 2010 May 15 & 20 & Sky cover 20\%; cumuliform clouds\\
\hline
B & 2011 February 14 & 26 & Sky cover $<30\%$; stratiform clouds\\
{10.306 km} & {} & {}\\
\hline
C & 2012 January 29 & 26 & Clear\\
{3.289 km} & 2012 January 30 & 25 & Clear\\
{} & 2012 January 31 & 25 & Clear\\
{} & 2012 March 26 & 25 & Sky cover 90\%; stratiform clouds\\
{} & 2012 March 30 & 27 & Sky cover 20\%; stratiform clouds\\
\enddata
\end{deluxetable*}

We performed calibration in CASA version 3.3.0, mapping in CASA version 4.1.0, and subsequently used CASA version 4.2.2 for image smoothing and some later analysis steps. A Hogbom cleaning algorithm was used to construct the image model; model components were restricted to an arc-shaped region that encompassed the northern and southern images, and a circle at the position of the western image, for all channels. The final data cube was created to match the channelization of the \mbox{CO(3--2)} data (rest frame spectral resolution of $14.129\,{\rm km\,s^{-1}}$). Since the naturally weighted channel maps synthesized beam ($0.^{\prime\prime}79\times 0.^{\prime\prime}68$ at a position angle of $-70.76^{\degree}$) already provided higher angular resolution than our \mbox{CO(3--2)} data, we chose not to pursue still higher resolution (at the cost of degraded SNR) with alternative weighting schemes. Since the spatial extent of J0901 is a substantial fraction of the VLA antenna primary beam FWHM, we applied a primary beam correction in order to retrieve the correct flux from the source (a $\sim10\%$ correction for the northern image). The average noise for each channel is $0.136\,{\rm mJy\,beam^{-1}}$ (prior to correcting for the primary beam).

\subsection{Submillimeter Array}
\label{sec:sma}

We observed J0901 in continuum emission at the Submillimeter Array using the $345\,{\rm GHz}$ receivers on 2010 May 20 and 2011 March 26 (project ID 2010A-S068, 2010B-S068; PI Baker). The observations were taken with the array in its compact configuration, using seven antennas (maximum baseline $66.5\,{\rm m}$) in 2010 and eight antennas (maximum baseline $75.25\,{\rm m}$) in 2011. We observed with the standard correlator setup that provided a maximum bandwidth of $4\,{\rm GHz}$ per sideband (for a single receiver), with a channel width of $3.25\,{\rm MHz}$. The central frequency of the correlator was tuned to $312\,{\rm GHz}$. During the observations, phase and amplitude variations were tracked with interleaved observations of the quasars 0854+201 and 0840+132. Mars and Titan were observed as flux calibrators, and the quasar 3C279 was used for bandpass calibration.

Data calibration and mapping were carried out in CASA version 4.1.0 after using the \texttt{sma2casa.py} and \texttt{smaImportFix.py} scripts\footnote{\href{http://www.cfa.harvard.edu/sma/casa/}{http://www.cfa.harvard.edu/sma/casa/}} to perform the initial system temperature correction and convert the data format to CASA measurement sets. The naturally weighted continuum map has a total bandwidth of $7.96\,{\rm GHz}$ and total time on source of $9.15\,{\rm hours}$, resulting in an RMS noise of $0.75\,{\rm mJy\,beam^{-1}}$ for a $2.^{\prime\prime}09\times 2.^{\prime\prime}09$ synthesized beam.

\subsection{SINFONI/VLT}
\label{sec:sinfoni}

We obtained integral field observations of H$\alpha$ emission from J0901 using the Spectrograph for Integral Field Observations in the Near Infrared (SINFONI) instrument \citep{eisenhauer2003,bonnet2004} on the Very Large Telescope (VLT) of the European Southern Observatory (ESO; program 087.A-0972, PI Baker). Observations were obtained in seeing-limited mode with $0.25^{\prime\prime}$ pixels, for which the SINFONI field of view is $8^{\prime\prime} \times 8^{\prime\prime}$. Data were taken at three pointings corresponding to the northern (observed 2012 January 7), southern (observed 2012 January 8), and western images (observed 2011 November 21 and December 17), targeted via blind offsets from a reference star; for each pointing, $8 \times 300\,{\rm s}$ exposures alternated between source and offset sky positions, with small dithers between successive exposures to facilitate background subtraction. The total on-source integration time was therefore 1200\,s per pointing (2400\,s per pointing for the fainter western image, which was deliberately visited twice). All data were reduced with standard ESO pipeline routines using the Gasgano interface. Point spread function (PSF) and flux calibration relied on contemporaneous observations of a nearby star with published 2MASS photometry.

After the pipeline calibration, we used noise clipping to identify and mask out cosmic rays and channels affected by sky lines. Since the three images of  J0901 were observed on different nights, the PSFs were slightly different for the three images ($\Delta{\rm FWHM}<0.^{\prime\prime}1$). We smoothed the observations to the largest PSF among the three images (the western image; $0.^{\prime\prime}75\times0.^{\prime\prime}65$ at $11.5^{\degree}$), and we also created versions smoothed to the CO beam size (for multi-line comparisons) if this was larger than the ${\rm H\alpha}$ PSF. The three pointings were then combined into a common cube, with no additional astrometric corrections applied to the blind offset positions. In order to make preliminary maps of the noise, continuum emission, line emission, and detector defects, we performed a linear fit to each pixel (excluding the channels with ${\rm H}\alpha$, [N\,{\sc ii}], or sky lines) and subtracted the fit cube from the data. This process over-subtracts the background (due to edges of skylines and cosmic rays that are not excluded), so we use these preliminary maps to mask out J0901, foreground galaxies, and chip defects, and then calculate the median sky level per channel within the three sub-images. The sky level is then subtracted from the data cube and then the data are re-fit to produce our final continuum-subtracted data cube and continuum map. Chip defects not removed by this process are still somewhat noticeable near the edges of the images (particularly the regions where dither patterns did not overlap), but they dominate the continuum image due to its low noise, so we mask out the outer $1.^{\prime\prime}25$ of the three sub-images for the continuum map. We calculate the standard deviation of each pixel (excluding channels with emission lines) to produce an average noise map. We then perform an additional astrometric correction using the integrated ${\rm H}\alpha$ and \mbox{CO(3--2)} maps and an imaging cross-correlation algorithm provided by Adam Ginsburg\footnote{\texttt{pixshift}: \href{http://casa.colorado.edu/~ginsbura/corrfit.htm}{http://casa.colorado.edu/$\sim$ginsbura/corrfit.htm}} to find and remove a $1.^{\prime\prime}32$ offset between the near-IR and radio data. 

The spectral resolution of SINFONI is $\lambda/\Delta\lambda\sim4000$; the channel widths are $36.75\,{\rm km\,s^{-1}}$ at the frequency of the ${\rm H\alpha}$ line. We apply a $16\,{\rm km\,s^{-1}}$ correction to convert velocities to the same local kinematic standard of rest used in the radio data. Since the three sub-images were observed on different dates, we use the average heliocentric corrections for the observations (which range from $12$--$22\,{\rm km\,s^{-1}}$) when analyzing the aggregate data, but for the analysis of the spectral line profiles in each sub-image, we apply their individual velocity corrections.

\subsection{Hubble Space Telescope}
\label{sec:hubble}

We also use {\it Hubble Space Telescope} ({\it HST}) observations of J0901 to constrain the lens model. J0901 was observe in Cycle 17 (Program ID 11602, PI S. Allam). Imaging was performed with {\it HST}'s Wide Field Camera 3 (WFC3) using filters F475W, F814W, F606W, F160W, and F110W. We processed the data using the standard AstroDrizzle reduction pipeline\footnote{Part of DrizzlePac: \href{http://drizzlepac.stsci.edu}{http://drizzlepac.stsci.edu}}. In order to use these data for lens modeling, we also remove contaminating light from the foreground lens galaxies using GALFIT \citep{peng2010}.

\section{Results}
\label{sec:results}

\begin{figure*}
\plotone{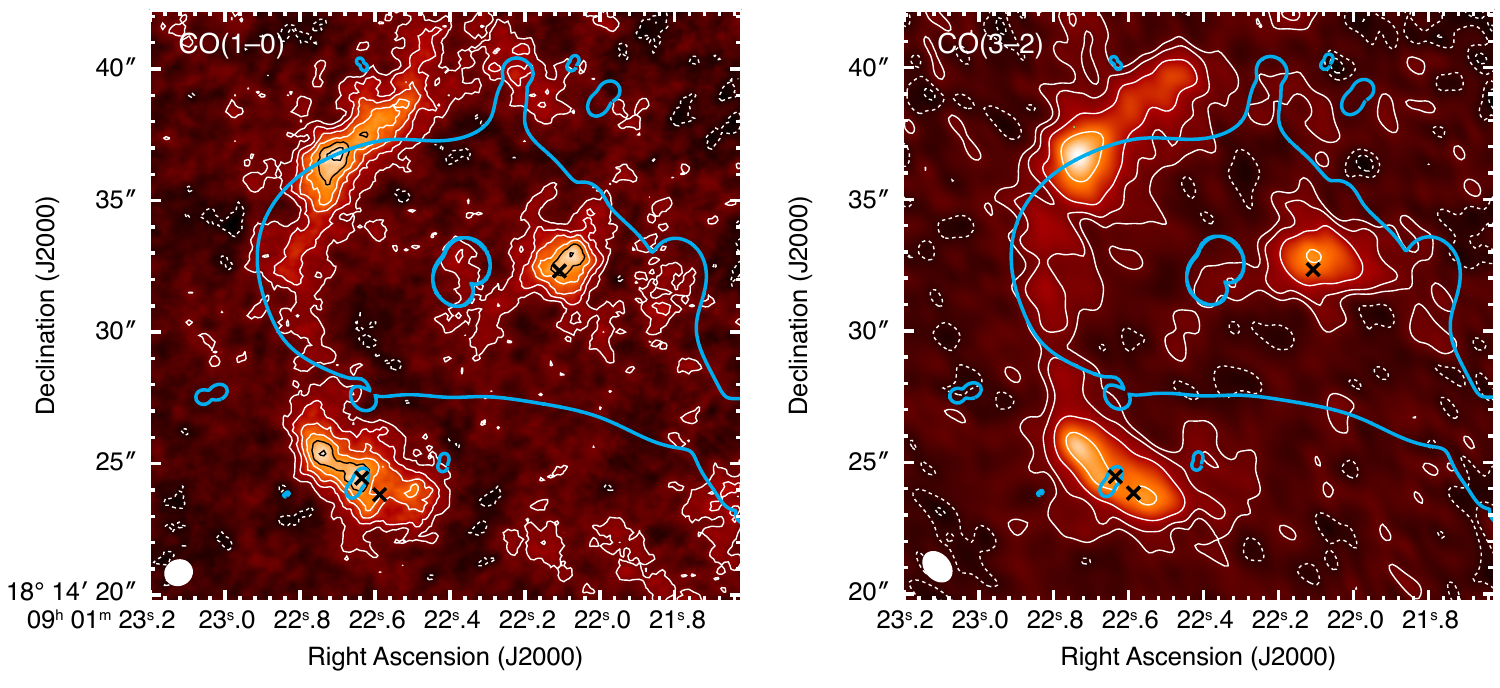}
\caption{Integrated \mbox{CO(1--0)} (left) and \mbox{CO(3--2)} (right) intensity maps (with primary beam corrections applied); the \mbox{CO(1--0)} map is the ``natural" map that has not been $uv$-clipped to match the \mbox{CO(3--2)} map. Contours are multiples of $\pm2\sigma_\text{1--0}$ for the \mbox{CO(1--0)} map and are powers of $2\times\pm\sigma_\text{3--2}$ (i.\/e.\/ $\pm2\sigma$, $\pm4\sigma$, $\pm8\sigma$, etc.) for the \mbox{CO(3--2)} map ($\sigma_\text{1--0}=9.1\,{\rm mJy\,km\,s^{-1}\,beam^{-1}}$; $\sigma_\text{3--2}=41\,{\rm mJy\,km\,s^{-1}\,beam^{-1}}$). Negative contours are dotted and the synthesized beams are shown at lower left. Blue lines indicate the lens model critical curves (Section~\ref{sec:lensing}). Black crosses mark the mean dynamical center determined from the source-plane reconstructions and dynamical modeling (see Section~\ref{sec:masses}). \label{fig:j0901intline}}
\end{figure*}

We successfully detect the three images of J0901 in both the \mbox{CO(1--0)} and \mbox{CO(3--2)} maps (Figure~\ref{fig:j0901intline}). In order to make a fair comparison between the two maps, we also analyze versions of the data cubes (including the VLT data) that have been smoothed to a common beam/PSF (the smallest Gaussian resolution FWHM that all datasets can be smoothed to is $1.^{\prime\prime}34\times1.^{\prime\prime}10$ at a position angle of $41.10^{\degree}$, which is close to the native resolution of the \mbox{CO(3--2)} data); we refer to the two sets of maps as the ``natural" and ``matched" maps below. For the matched \mbox{CO(1--0)} data, in addition to smoothing to the common beam, we also exclude baselines that have $uv$ distances smaller than the minimum for the \mbox{CO(3--2)} data; the $uv$-clipping ensures that flux distributed on large spatial scales that cannot be detected at the PdBI is also excluded from the \mbox{CO(1--0)} maps. The smoothing most strongly changes the surface brightness distribution in the southern image for the \mbox{CO(1--0)} data, increasing the peak surface brightness by $\sim30\%$ and thus exaggerating the asymmetry between the two peaks in brightness (see Figure~\ref{fig:j0901intline}). However, the $uv$-clipping removes only a small fraction of the total \mbox{CO(1--0)} flux ($<10\%$). 

\begin{figure*}
\epsscale{1}
\plotone{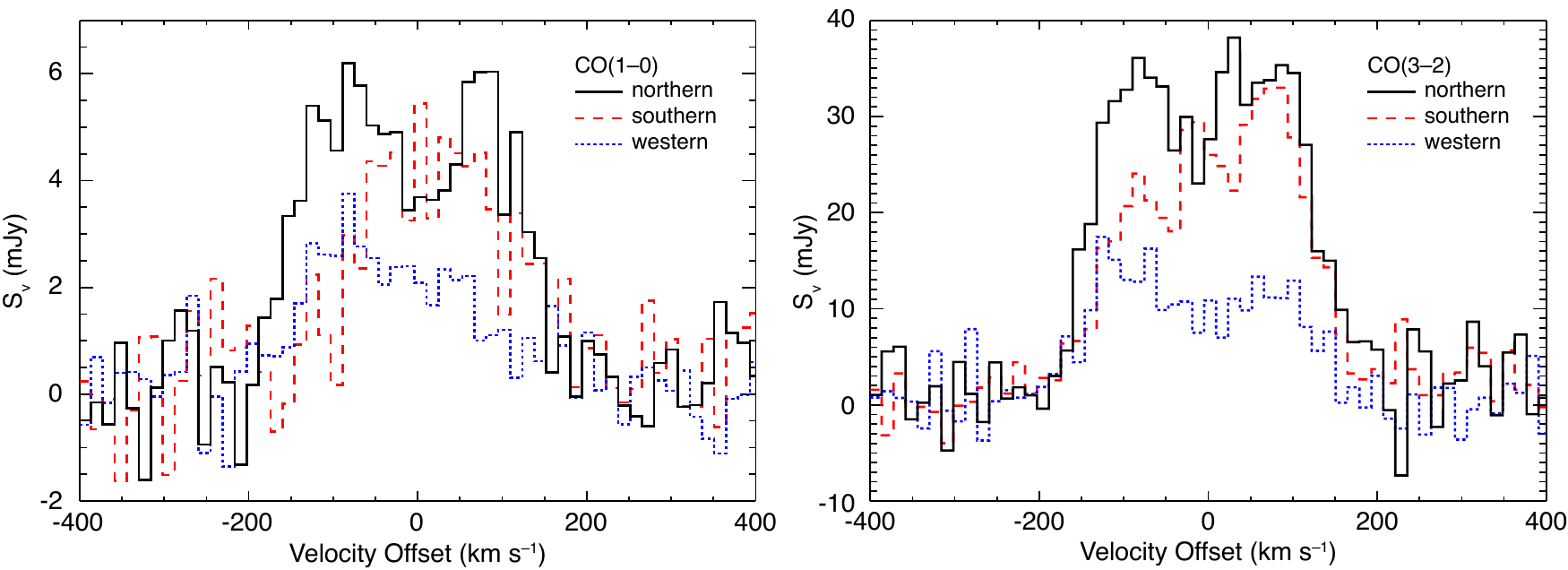}
\caption{VLA \mbox{CO(1--0)} spectra (left) and PdBI \mbox{CO(3--2)} spectra (right) extracted from the ``natural" maps for the northern (black/solid), southern (red/dashed), and western (blue/dotted) images, plotted relative to the $z=2.2586$ ${\rm H\alpha}$ systemic redshift from \citet{hainline2009}. \label{fig:j0901spectra}}
\end{figure*}

\begin{figure*}
\plotone{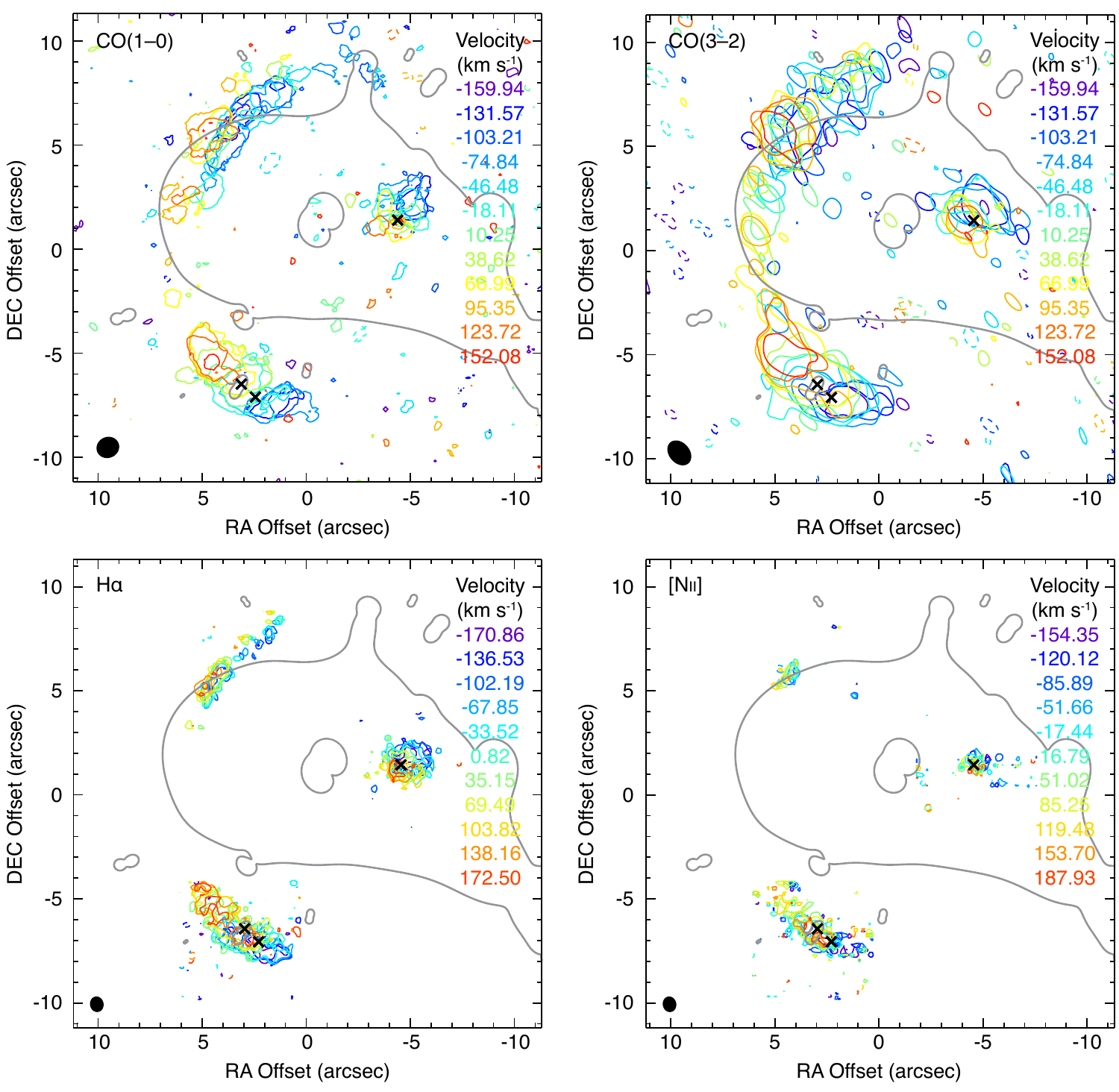}
\caption{Overlaid contours of the natural resolution \mbox{CO(1--0)} (upper left), \mbox{CO(3--2)} (upper right), ${\rm H\alpha}$ (lower left), and [N{\sc ii}] (lower right) channel maps, colorized by their velocities relative to the $z=2.2586$ ${\rm H\alpha}$ systemic redshift from \citet{hainline2009}. The images are centered at $\alpha {\rm (J2000)} = 09^{\rm h}01^{\rm m}22.^{\rm s}42$ and $\delta {\rm (J2000)} = +18\degree14^{\prime}30.^{\prime\prime}9$. For the two CO lines we show only the $\pm3\sigma$ contours ($\sigma_\text{1--0}=0.21\,{\rm mJy\,beam}^{-1}$, $\sigma_\text{3--2}=0.82\,{\rm mJy\,beam}^{-1}$) where the channels have been rebinned by a factor of two to $28.37\,{\rm km\,s^{-1}}$. For the VLT/SINFONI data we show only the $\pm2\sigma$ contours ($\sigma_{\rm VLT}=2.2\times10^{-16}\,{\rm erg\,s^{-1}\,cm^{-2}\,\mu m^{-1}}$) and have not done any additional channel binning. Negative contours are dotted. For clarity we do not use the primary beam-corrected data for the two CO lines and we mask out the outer $1.^{\prime\prime}25$ (10 pixels from the dither pattern) for the VLT data. Synthesized beams and PSFs are shown at lower left. Gray lines indicate the lens model critical curves (Section~\ref{sec:lensing}). Black crosses mark the mean dynamical center determined from the source-plane reconstructions and lens modeling (see Section~\ref{sec:masses}). \label{fig:j0901renzos}}
\end{figure*}

The measured line fluxes are summarized in Table~\ref{tab:j0901emission} and are extracted over identical image areas for the three maps; the uncertainties include an assumed $10\%$ flux calibration error. For the spectra in Figure~\ref{fig:j0901spectra}, we use the natural maps \footnote{Due to the velocity structure of J0901 and the small synthesized beam of the natural \mbox{CO(1--0)} map, we extract the spectra over slightly smaller regions; the larger regions used in the rest of the analysis include enough signal-free pixels in the individual channel maps to significantly increase the noise for the integrated spectra.}. We find that the spectra of the \mbox{CO(1--0)} and \mbox{CO(3--2)} lines have a consistent ${\rm FWZI}\approx350\,{\rm km\,s^{-1}}$ centered at the ${\rm H\alpha}$-determined systemic redshift from \citet{hainline2009}, but that the shapes of the two CO line profiles differ for the same images. The different relative line profiles for the two CO lines in all three images suggest that differential lensing is occurring; i.\/e.\/, the spatial variation of the magnification factor across J0901 is amplifying the light in regions with different \mbox{CO(3--2)}/\mbox{CO(1--0)} line ratios \citep[e.\/g.\/,][]{blain1999,serjeant2012}. In Figure~\ref{fig:j0901renzos} we show the overlaid channel maps of the natural \mbox{CO(1--0)} and \mbox{CO(3--2)} lines, rebinned by a factor of two; there is a clear velocity gradient across the three images, suggesting J0901 is either disk-like or a merging galaxy.

\begin{figure*}
\plotone{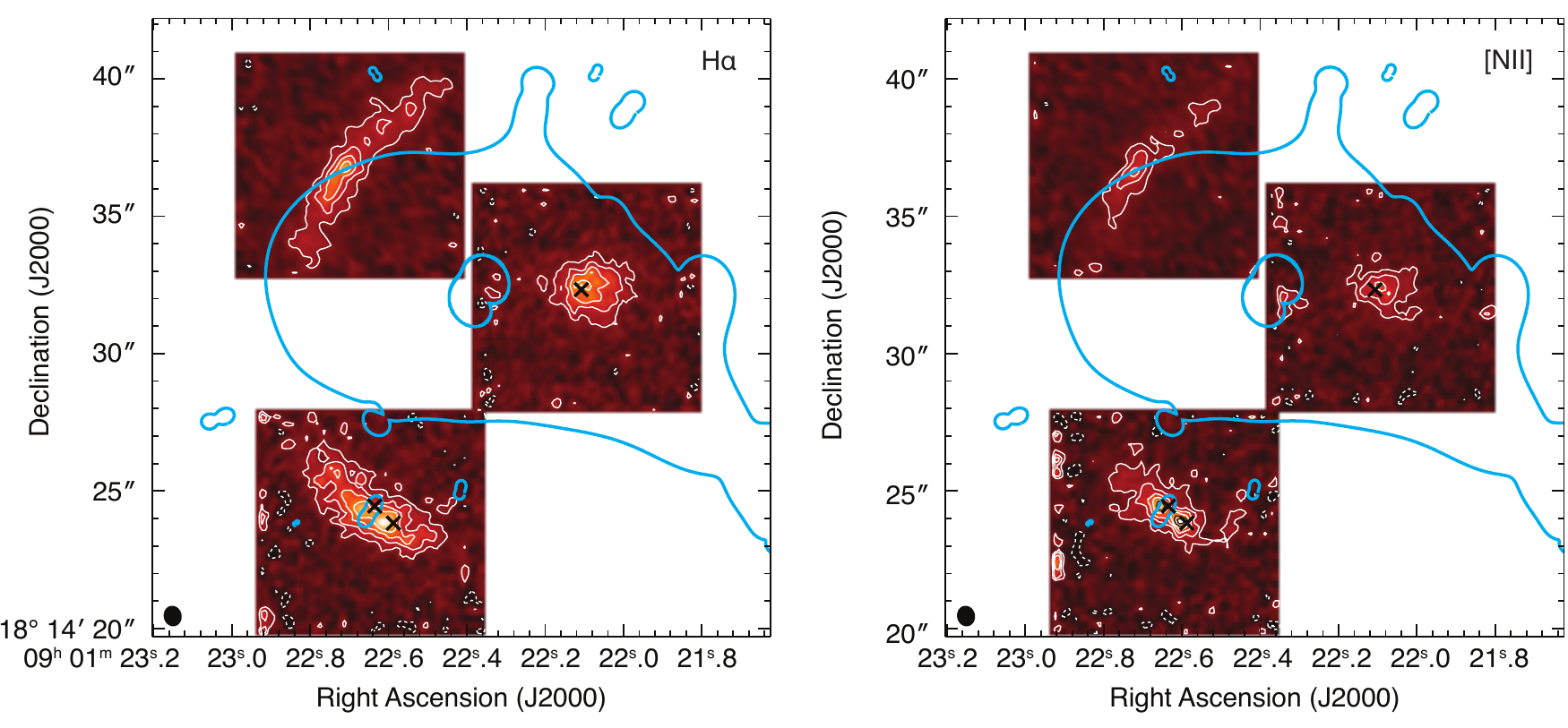}
\caption{Integrated ${\rm H\alpha}$ (left) and [N\,{\sc ii}] (right) intensity maps of J0901. Due to SINFONI's small field of view, the three images of J0901 were observed separately and have been smoothed to the same PSF ($0.^{\prime\prime}75\times0.^{\prime\prime}65$) shown at the bottom left corners. Contours are multiples of $\pm2\bar{\sigma}$ (where $\bar{\sigma}=8.0\times10^{-16}\,{\rm erg\,s^{-1}\,cm^{-2}}$ is the average noise for the three sub-images); negative contours are dashed. Cyan lines indicate the lens model critical curves (Section~\ref{sec:lensing}). Black crosses mark the mean dynamical center determined from the source-plane reconstructions and dynamical modeling (see Section~\ref{sec:masses}). \label{fig:j0901halpha}}
\end{figure*}

We also detect the three images of J0901 in ${\rm H\alpha}$ and [N\,{\sc ii}] using the VLT/SINFONI data (Fig.~\ref{fig:j0901halpha}). The measured line fluxes are given in Table~\ref{tab:j0901emission}; the statistical uncertainties are determined by weighted Gaussian fits to the line shapes. The spectra for the ${\rm H\alpha}$ and [N\,{\sc ii}] lines do not show the double-peaked structure seen in the CO lines. However, the FWHMs derived from fitting Gaussians to the ${\rm H\alpha}$ and [N\,{\sc ii}] line profiles (Table~\ref{tab:j0901fits}; after accounting for instrumental broadening) are consistent with single Gaussian fits to the CO line profiles.

\begin{figure}
\epsscale{1.15}
\plotone{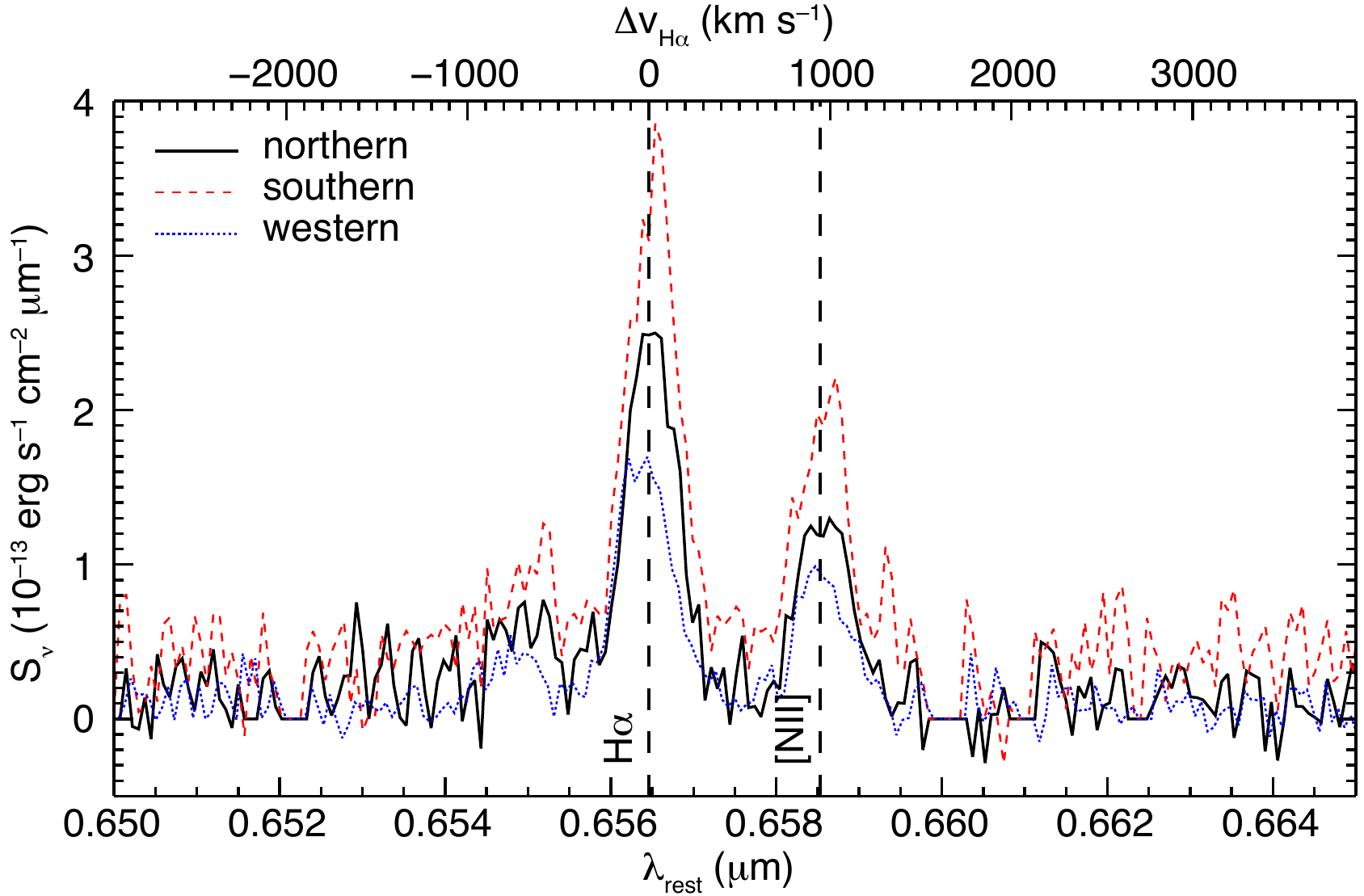}
\caption{VLT spectra showing the H$\alpha$ and [N\,{\sc ii}] lines (as well as continuum emission) for the northern (black/solid), southern (red/dashed), and western (blue/dotted) images, plotted relative to rest wavelength using the $z=2.2586$ ${\rm H\alpha}$ systemic redshift from \citet{hainline2009}. Channels with zero emission correspond to sky-line masks. \label{fig:halphaspectra}}
\end{figure}

\begin{figure*}
\epsscale{1.15}
\plotone{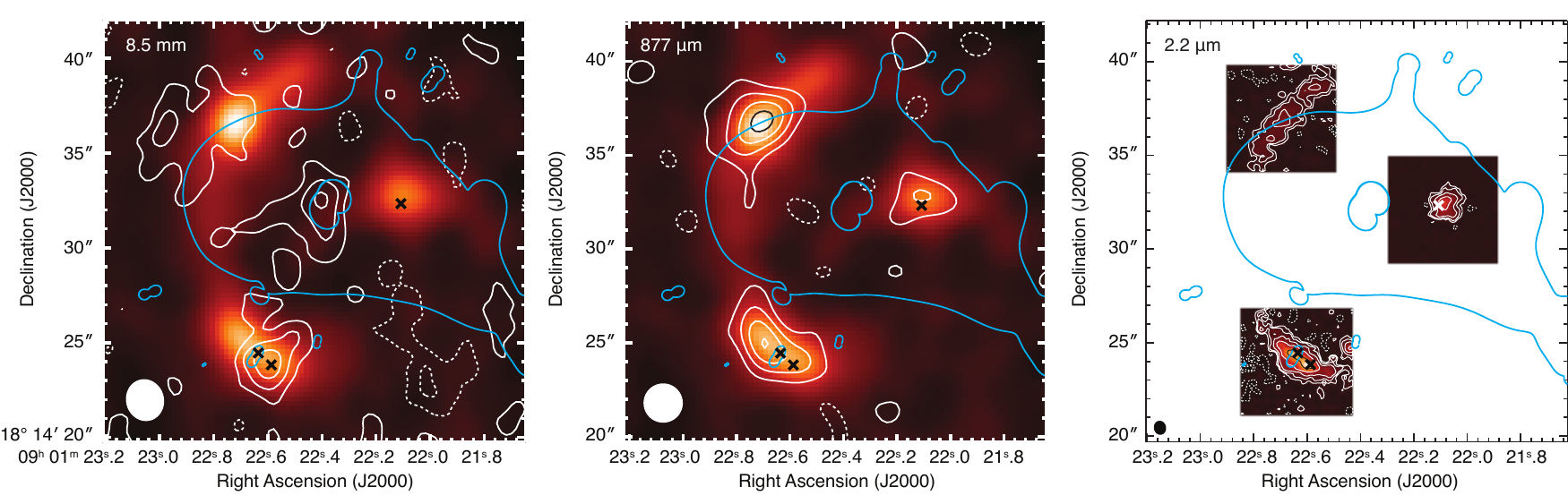}
\caption{ VLA $8.5\,{\rm mm}$ (left/contours; plotted over the CO(3--2) integrated line map smoothed to the same resolution), SMA $877\,{\rm \mu m}$ (center/contours; plotted over the CO(3--2) integrated line map smoothed to the same resolution), and VLT $2.2\,{\rm \mu m}$ continuum maps of J0901 (where the wavelengths listed are in the observed frame). For the VLA data, a $1^{\prime\prime}$ taper was applied, resulting in a $2.^{\prime\prime}22\times 2.^{\prime\prime}00$ resolution map (beam shown at bottom left). Contours are multiples of $\pm1.5\sigma$ ($\sigma=26.6\,{\rm \mu Jy\,beam^{-1}}$). The SMA $2.^{\prime\prime}09\times 2.^{\prime\prime}09$ beam FWHM is shown at lower left. Contours are multiples of $\pm2\sigma$ ($\sigma=0.75\,{\rm mJy\,beam^{-1}}$). Due to SINFONI's small field of view, the three images of J0901 were observed separately and have been smoothed to same PSF ($0.^{\prime\prime}75\times0.^{\prime\prime}65$) shown at the bottom left. Contours are powers of $2\times\pm\bar{\sigma}$ (i.\/e.\/, $\pm2\bar{\sigma}$, $\pm4\bar{\sigma}$, $\pm8\bar{\sigma}$, etc.; where $\bar{\sigma}=5.2\times10^{-18}\,{\rm erg\,s^{-1}\,cm^{-2}\,\mu m^{-1}}$ is the average noise for the three sub-images). An additional $1.^{\prime\prime}25$ was masked around the image edges compared to the integrated line maps for clarity (the edges have significant defects which are only apparent in the high S/N of the continuum map). The other bright continuum sources are members of the lensing cluster. For all maps, negative contours are dashed and crosses mark the mean dynamical center determined from the source-plane reconstructions and lens modeling (see Section~\ref{sec:masses}). Cyan lines indicate the lens model critical curves (Section~\ref{sec:lensing}). \label{fig:j0901cont}}
\end{figure*}

We successfully detect continuum emission from J0901 at the SMA ($295\,{\rm \mu m}$ rest frame), the VLA ($2.6\,{\rm mm}$ rest frame), and the VLT ($0.66\,{\rm \mu m}$ rest frame; Fig.~\ref{fig:j0901cont}). Our continuum flux measurements are given in Table~\ref{tab:j0901emission}. We detect all three images for both the SMA and VLT continuum maps. For the VLA continuum map, we definitely detect rest-$2.6\,{\rm mm}$ continuum emission from the southern image, we marginally detect the northern image, and we do not detect the western image (Fig.~\ref{fig:j0901cont}). For both the VLA and VLT maps, we also detect continuum emission from the lensing group galaxies (corresponding to rest wavelengths of $\sim 3.5\,{\rm mm}$ and $\sim 0.27\,{\rm \mu m}$ at the redshift of the lensing group), although most group members are masked out in the VLT continuum image since they are near the edges of the field of view. For the VLA and SMA data, we compare the distribution of the continuum emission to the \mbox{CO(3--2)} line emission (smoothed to the continuum maps' spatial resolutions; the results are qualitatively similar when comparing to the smoothed \mbox{CO(1--0)} line maps). The rest $295\,{\rm \mu m}$ continuum emission peaks at the same location as the CO emission for the three lensed images. However, for the northern image, the $295\,{\rm \mu m}$ continuum emission is not as spatially extended as the CO. The missing extended emission is either below the sensitivity of our current maps, or the dust distribution does not perfectly trace the molecular gas within J0901 (regardless of any complications caused by lensing). While the SNR for the VLA continuum map is limited, the rest $2.6\,{\rm mm}$ emission in the southern image is offset from the peak in CO emission. As the rest $2.6\,{\rm mm}$ continuum emission would likely trace either star formation or a central AGN, the offset is somewhat peculiar.

\floattable
\begin{deluxetable*}{ccccccc}
\tablewidth{0pt}
\tablecaption{J0901 emission line and continuum measurements (magnification-corrected where indicated) \label{tab:j0901emission}}
\tablehead{ {Line/Map} & {Parameter} & {Units} & {North} & {South} & {West} & {Total}}
\startdata
CO(1--0) & $S_\text{1--0}\Delta v$ & ${\rm Jy\,km\,s^{-1}}$ & $1.41\pm0.16$ & $0.94\pm0.12$ & $0.60\pm0.08$ & $2.95\pm0.32$ \\
natural & $L^\prime_\text{CO(1--0)}$ & $10^{10}\,{\rm K\,km\,s^{-1}\,pc^2}$ & $38.4\pm4.5$ & $25.5\pm3.2$ & $16.3\pm2.2$ & $3.53^{+0.57}_{-0.45}$\tablenotemark{b} \\
& $r_{3,1}$ & & $0.74\pm0.11$ & $0.84\pm0.14$ & $0.62\pm0.11$ & $0.75\pm0.11$\\
\hline
CO(1--0) & $S_\text{1--0}\Delta v$ & ${\rm Jy\,km\,s^{-1}}$ & $1.34\pm0.16$ & $0.87\pm0.11$ & $0.59\pm0.08$ & $2.80\pm0.30$ \\
matched & $L^\prime_\text{CO(1--0)}$ & $10^{10}\,{\rm K\,km\,s^{-1}\,pc^2}$ & $36.4\pm2.1$ & $23.8\pm3.0$ & $16.0\pm2.1$ & $1.62^{+0.35}_{-0.27}$\tablenotemark{b} \\
& $r_{3,1}$ & & $0.78\pm0.12$ & $0.91\pm0.15$ & $0.63\pm0.12$ & $0.79\pm0.12$\\
\hline
CO(3--2) & $S_\text{3--2}\Delta v$ & ${\rm Jy\,km\,s^{-1}}$ & $9.35\pm1.00$ & $7.10\pm0.76$ & $3.36\pm0.41$ & $19.8\pm2.0$ \\
& $L^\prime_\text{CO(3--2)}$ & $10^{10}\,{\rm K\,km\,s^{-1}\,pc^2}$ & $28.3\pm3.0$ & $21.5\pm2.3$ & $10.2\pm1.2$ & $1.99^{+0.32}_{-0.29}$\tablenotemark{b} \\
\hline
H$\alpha$ & $S_{\rm H\alpha}\Delta v$ & ${\rm 10^{-16}\,erg\,s^{-1}\,cm^{-2}}$ & $5.72\pm0.57$ & $7.69\pm0.74$ & $4.29\pm0.36$ & $18.46\pm1.01$ \\
& $L_{\rm H\alpha}$ & ${\rm 10^{42}\,erg\,s^{-1}}$ & $24.8\pm2.5$ & $33.3\pm3.2$ & $18.6\pm1.6$ & $2.70^{+0.39}_{-0.32}$\tablenotemark{b} \\
\hline
[N\,{\sc ii}] & $S_\text{[N\,{\sc ii}]}\Delta v$ & ${\rm 10^{-16}\,erg\,s^{-1}\,cm^{-2}}$ & $2.80\pm0.57$ & $4.00\pm0.74$ & $2.16\pm0.35$ & $9.60\pm1.01$ \\
& $L_\text{[N\,{\sc ii}]}$ & ${\rm 10^{42}\,erg\,s^{-1}}$ & $12.1\pm2.5$ & $17.4\pm3.2$ & $9.4\pm1.5$ & $1.53^{+0.36}_{-0.28}$\tablenotemark{b} \\
\hline
$8.5\,{\rm mm}$ & $S_{8.5\,{\rm mm}}$ & ${\rm mJy}$ & $0.33\pm0.09$ & $0.25\pm0.07$ & $<0.08$\tablenotemark{a} & $0.66\pm0.12$ \\
\hline
$877\,{\rm \mu m}$ & $S_{877\,{\rm \mu m}}$ & ${\rm mJy}$ & $17.7\pm5.3$ & $13.0\pm3.8$ & $4.3\pm2.9$ & $35.0\pm8.8$ \\
\hline
$2.2\,{\rm \mu m}$ & $S_{2.2\,{\rm \mu m}}$ & ${\rm 10^{-14}\,erg\,s^{-1}\,cm^{-2}\,\mu m^{-1}}$ & $1.40\pm0.08$ & $3.10\pm0.10$ & $0.76\pm0.10$ & $5.26\pm0.14$ \\
\enddata
\tablenotetext{a}{$3\sigma$ upper limit assuming a point-like flux distribution.}
\tablenotetext{b}{The total line luminosities are magnification corrected assuming the corresponding magnification factors listed in Table~\ref{tab:mag} (i.\/e.\/ the ``natural" magnification factors calculated using the native resolution data presented in the bulk of this table, or the ``matched" magnification factors for the \mbox{CO(1--0)} data $uv$-clipped to create the matching resolution datasets).}
\tablecomments{The VLT observations include statistical uncertainties only. The integrated line fluxes are from Gaussian fits to the spectra. Since each image was observed on a different night, the spectra were corrected for their different heliocentric velocities before being combined. Therefore, the total integrated line fluxes/luminosities differ slightly from the sum from the individual images.}
\end{deluxetable*}

\floattable
\begin{deluxetable*}{clcccc}
\tablewidth{0pt}
\tablecaption{Gaussian fits to the spectral lines \label{tab:j0901fits}}
\tablehead{ {Line/Map} & {Parameter} & {North} & {South} & {West} & {Total}}
\startdata
CO(1--0) & $S_{\nu,{\rm peak}}$\tablenotemark{a} & $5.92\pm0.55$/$5.54\pm0.61$ & $4.53\pm0.36$ & $2.72\pm0.27$ & $10.1\pm2.0$/$10.5\pm1.2$ \\
natural & FWHM \tablenotemark{b}& $135\pm22$/$110\pm20$ & $210\pm22$ & $226\pm29$ & $134\pm27$/$161\pm35$ \\
{} & $v_{\rm offset}$\tablenotemark{b} & $-81\pm9$/$80\pm9$ & $24\pm9$ & $-38\pm12$ & $-81\pm18$/$64\pm22$ \\
\hline
CO(1--0) & $S_{\nu,{\rm peak}}$\tablenotemark{a} & $5.60\pm0.53$/$5.41\pm0.57$ & $4.24\pm0.35$ & $2.47\pm0.26$ & $9.4\pm1.4$/$10.4\pm0.8$ \\
matched & FWHM\tablenotemark{b} & $124\pm20$/$109\pm18$ & $196\pm21$ & $227\pm30$ & $118\pm22$/$157\pm28$ \\
{} & $v_{\rm offset}$\tablenotemark{b} & $-81\pm8$/$77\pm8$ & $26\pm8$ & $-36\pm12$ & $-83\pm13$/$61\pm15$ \\
\hline
CO(3--2) & $S_{\nu,{\rm peak}}$\tablenotemark{a} & $33.5\pm3.2$/$36.6\pm2.3$ & $29.8\pm1.3$ & $15.0\pm1.6$/$12.3\pm1.6$ & $69.0\pm4.0$/$79.7\pm3.4$ \\
{} & FWHM \tablenotemark{b} & $105\pm14$/$151\pm19$ & $237\pm13$ & $125\pm22$/$124\pm27$ & $122\pm12$/$138\pm12$ \\
{} & $v_{\rm offset}$\tablenotemark{b} & $-92\pm7$/$59\pm8$ & $19\pm5$ & $-95\pm9$/$71\pm12$ & $-86\pm6$/$66\pm6$ \\
\hline
${\rm H\alpha}$ & $S_{\nu,{\rm peak}}$\tablenotemark{c} & $2.41\pm0.15$ & $3.13\pm0.19$ & $1.63\pm0.09$ & $7.12\pm0.25$ \\
{} & FWHM\tablenotemark{b} & $312\pm24$ & $323\pm24$ & $347\pm23$ & $341\pm14$ \\
{} & $v_{\rm offset}$\tablenotemark{b} & $13\pm10$ & $25\pm10$ & $-29\pm9$ & $9\pm6$ \\
\hline
[N{\sc ii}] & $S_{\nu,{\rm peak}}$\tablenotemark{c}& $1.19\pm0.15$ & $1.65\pm0.19$ & $0.87\pm0.09$ & $3.78\pm0.25$ \\
{} & FWHM \tablenotemark{b}& $308\pm49$ & $318\pm45$ & $324\pm41$ & $333\pm27$ \\
{} & $v_{\rm offset}$\tablenotemark{b} & $20\pm20$ & $19\pm19$ & $5\pm17$ & $16\pm11$ \\
\enddata
\tablenotetext{a}{In units of ${\rm mJy}$.}
\tablenotetext{b}{In units of ${\rm km\,s^{-1}}$.}
\tablenotetext{c}{In units of ${\rm 10^{-13}\,erg\,s^{-1}\,cm^{-2}\,\mu m^{-1}}$.}
\tablecomments{Multiple values are listed for double Gaussian fits where those fits preferred two spectral peaks. Centroid velocity offsets are measured relative to the $z=2.2586$ ${\rm H\alpha}$ systemic redshift from \citet{hainline2009}.}
\end{deluxetable*}

\section{Analysis}
\label{sec:analysis}

\subsection{Lens modeling and source-plane reconstruction}
\label{sec:lensing}

\subsubsection{Methods}

J0901 is lensed by a group of galaxies, which needs to be accounted for explicitly in order to reconstruct the galaxy's source-plane structure. Our lens model therefore comprises one component representing the group halo and others representing the group members. The former is described by an elliptical power-law density distribution, whose (spherical) convergence profile is given by

\begin{equation}
\label{eq:plaw}
\kappa(\vec{x}) = \frac{b^{2-\alpha}}{2|\vec{x}|^{2-\alpha}},
\end{equation}

\noindent
where $b$ is the Einstein radius. The group members within two Einstein radii are represented by singular isothermal ellipsoids (SIEs) given by equation~(\ref{eq:plaw}) with $\alpha=1$. In this case, $b$ not only represents the Einstein radius, but is also related to the velocity dispersion $\sigma_v$ by $b\propto\sigma_v^2$.\footnote{This relation does not strictly hold for elliptical mass distributions, but the corrections are negligible for small ellipticities \citep[e.\/g.\/,][]{chae2003, huterer2005}. The proportionality constant depends on the ellipticity.}

While the position and ellipticity of the group halo are allowed to vary, the group members' positions and ellipticities are fixed to the observed values. Additionally, a log-normal prior about the nominal Faber-Jackson relation \citep{faber1976} is placed on their velocity dispersions. For any two galaxies $G_1$ and $G_2$, equation~(\ref{eq:plaw}) and the Faber-Jackson relation give $b_2/b_1 \propto \sigma_{v,2}^2/\sigma_{v,1}^2 \propto \sqrt{L_2}/\sqrt{L_1}$, where $L_i$ is the observed luminosity of $G_i$. Using mass as a proxy for luminosity, we set priors, noting that \citet{gallazzi2006} find that the scatter in the logarithmic mass-velocity dispersion relation is $\approx 0.07$ for early-type galaxies selected from the Sloan Digital Sky Survey \citep{abazajian2004}. We also note that the presence of a galaxy at the location of the southern image represents a unique challenge given its close proximity. Due to its small halo mass, fits with a SIE model are challenging since deflections due to that potential never reach zero. Since this is a smaller galaxy in a dense environment, its mass profile may be tidally truncated, and we therefore adopt a truncated, elliptical pseudo-Jaffe profile \citep{keeton2001b} to represent this component. The spherical convergence profile for this model is given by

\begin{equation}
\label{eq:jaffe}
\kappa(\vec{x}) = \frac{b'}{2}\bigg[ \big(|\vec{x}|^2+s^2\big)^{-\frac{1}{2}} - \big(|\vec{x}|^2+a^2\big)^{-\frac{1}{2}} \bigg],
\end{equation}

\noindent
where $s$ and $a$ are the core and truncation radii, respectively. The truncated pseudo-Jaffe assumption allows us to explore truncated mass models, but preserves more extended profile options in the limit that the truncation radius ($a$) approaches infinity. The best-fit lens model parameters for all components are listed in Table~\ref{tab:lensmodel}.

\begin{deluxetable*}{llrrrrrrrl}
\tablecaption{Best-fit lens model parameters \label{tab:lensmodel}}
\tablehead{\colhead{Object(s)} & \colhead{Model} & \colhead{$b$} & \colhead{$\Delta$RA} & \colhead{$\Delta$DEC} & \colhead{$e$} & \colhead{$PA$} & \colhead{$s$} & \colhead{$a$} & \colhead{$\alpha$}\\
\colhead{} & \colhead{} & \colhead{} & \colhead{($^{\prime\prime}$)} & \colhead{($^{\prime\prime}$)} & \colhead{} & \colhead{($\degree$)} & \colhead{($^{\prime\prime}$)} & \colhead{($^{\prime\prime}$)} & \colhead{}}
\startdata
Group halo & SPLE & $2.1157$ & $-0.0157$ & $-0.1954$ & $0.331$ & $-82.7$ & {} & {} & $1.51$ \\
\hline
Central galaxies & SIS & $0.7184$ & $0.0585$ & $-0.0147$ & {} & {} & {} & {} & $1.0$ \\
{} & SIS & $0.9551$ & $-0.5820$ & $-0.7580$ & {} & {} & {} & {} & $1.0$ \\
\hline
Southern perturber & p-Jaffe & $1.0833$ & $3.7344$ & $-8.4207$ & $0.244$ & $21.0$ & $0.3295$ & $0.5045$ & {} \\
\hline
Other galaxies & SIS & $0.26420$ & $-2.2917$ & $6.8895$ & {} & {} & {} & {} & $1.0$ \\
{} & SIS & $0.0735$ & $-4.4900$ & $7.5848$ & {} & {} & {} & {} & $1.0$ \\
{} & SIS & $0.1843$ & $-5.6797$ & $6.3216$ & {} & {} & {} & {} & $1.0$ \\
{} & SIS & $0.0290$ & $-4.9115$ & $10.7200$ & {} & {} & {} & {} & $1.0$ \\
{} & SIS & $0.0475$ & $-4.8044$ & $2.3711$ & {} & {} & {} & {} & $1.0$ \\
{} & SIS & $0.3614$ & $-7.4541$ & $-0.2359$ & {} &  {} & {} & {} & $1.0$ \\
{} & SIS & $0.9159$ & $-9.7170$ & $-6.1960$ & {} & {} & {} & {} & $1.0$ \\
{} & SIS & $0.1484$ & $-10.8208$ & $-9.2987$ & {} & {} & {} & {} & $1.0$ \\
{} & SIS & $0.0605$ & $3.5408$ & $7.5583$ & {} & {} & {} & {} & $1.0$ \\
{} & SIS & $0.1418$ & $9.3071$ & $-5.0192$ & {} & {} & {} & {} & $1.0$ \\
{} & SIS & $0.0854$ & $3.1367$ & $-4.9471$ & {} & {} & {} & {} & $1.0$ \\
{} & SIS & $0.0882$ & $0.4415$ & $-7.6127$ & {} & {} & {} & {} & $1.0$ \\
{} & SIS & $0.0344$ & $6.4117$ & $-8.7968$ & {} & {} & {} & {} & $1.0$ \\
{} & SIS & $0.0502$ & $-2.4900$ & $-13.3387$ & {} & {} & {} & {} & $1.0$ \\
\enddata
\tablecomments{From left to right, the columns are: a description of the model component, the assumed model for the shape of the lensing potential (either a softened power law ellipsoid (SPLE), single isothermal sphere (SIS), or pseudo-Jaffe ellipsoid (p-Jaffe)), normalized amplitude (varied), offset in right ascension (from $09^{\rm h}01^{\rm m} 22.^{\rm s}3865$; fixed), offset in declination (from $18\degree14^\prime32.^{\prime\prime}6303$; fixed), ellipticity (only relevant for SPLE and p-Jaffe models; fixed), position angle (only relevant for SPLE and p-Jaffe models; fixed), the core radius (only relevant for p-Jaffe model), the truncation radius (only relevant for p-Jaffe model), and index of the power law (only relevant for SPLE model and assumed to be $1.0$ for SIS models). }
\end{deluxetable*}

\begin{figure*}
\epsscale{1}
\plotone{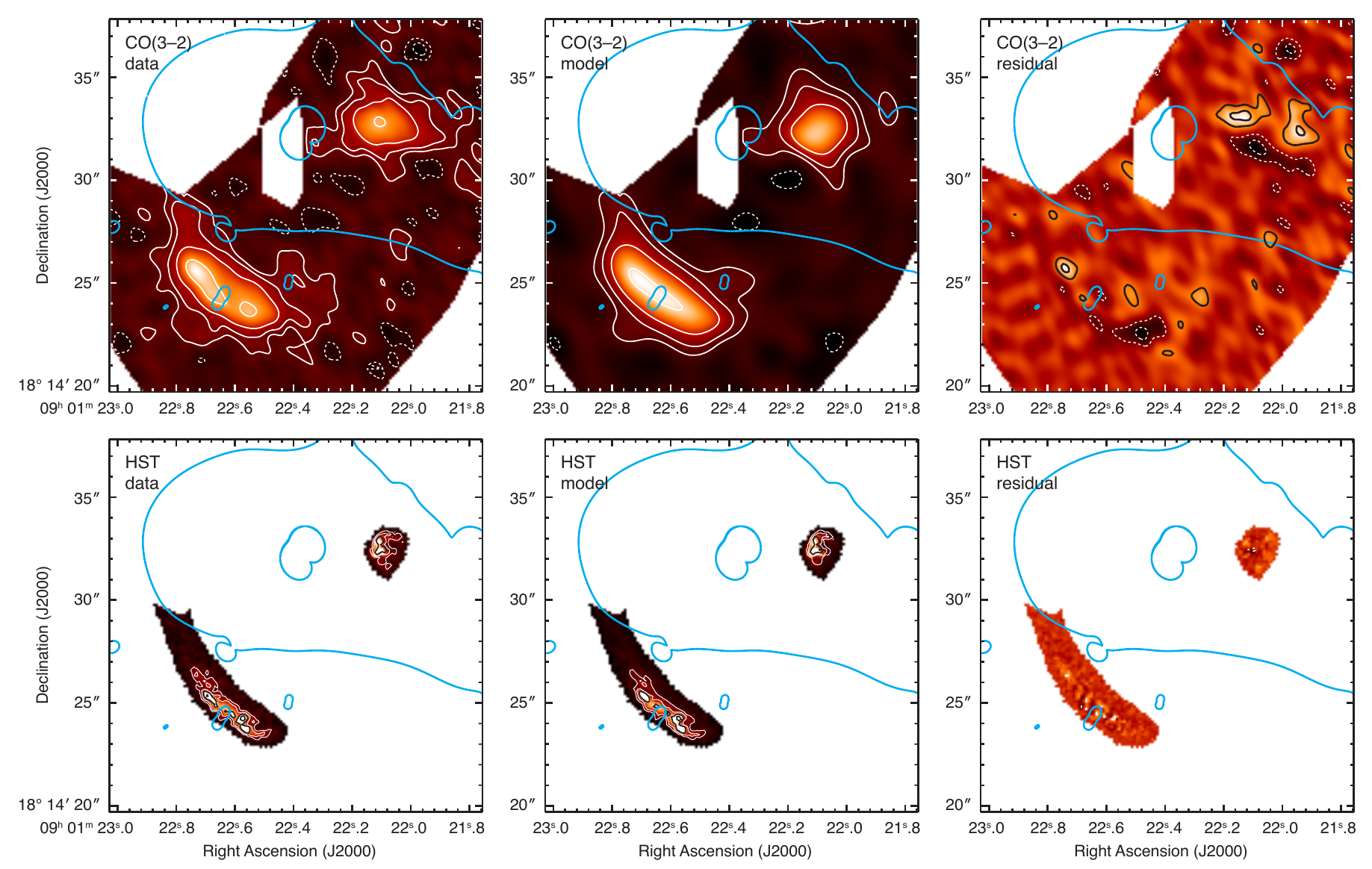}
\caption{Residual differences (right panels) between the observed image-plane data (left panels) and best-fit lensing model image-plane reconstructions (center panels) for the two datasets used to constrain the model: the CO(3--2) map (top row) and {\it HST} F606W image (bottom row). Pixels not used in constraining the data are masked out (most notably the northern image; see text for discussion). Critical curves are shown in blue. Contours are powers of $2\times\pm\sigma$ (i.\/e.\/ $\pm2\sigma$, $\pm4\sigma$, $\pm8\sigma$, etc.); negative contours are dotted. \label{fig:lensmodel}}
\end{figure*}

\begin{figure*}
\epsscale{1}
\plotone{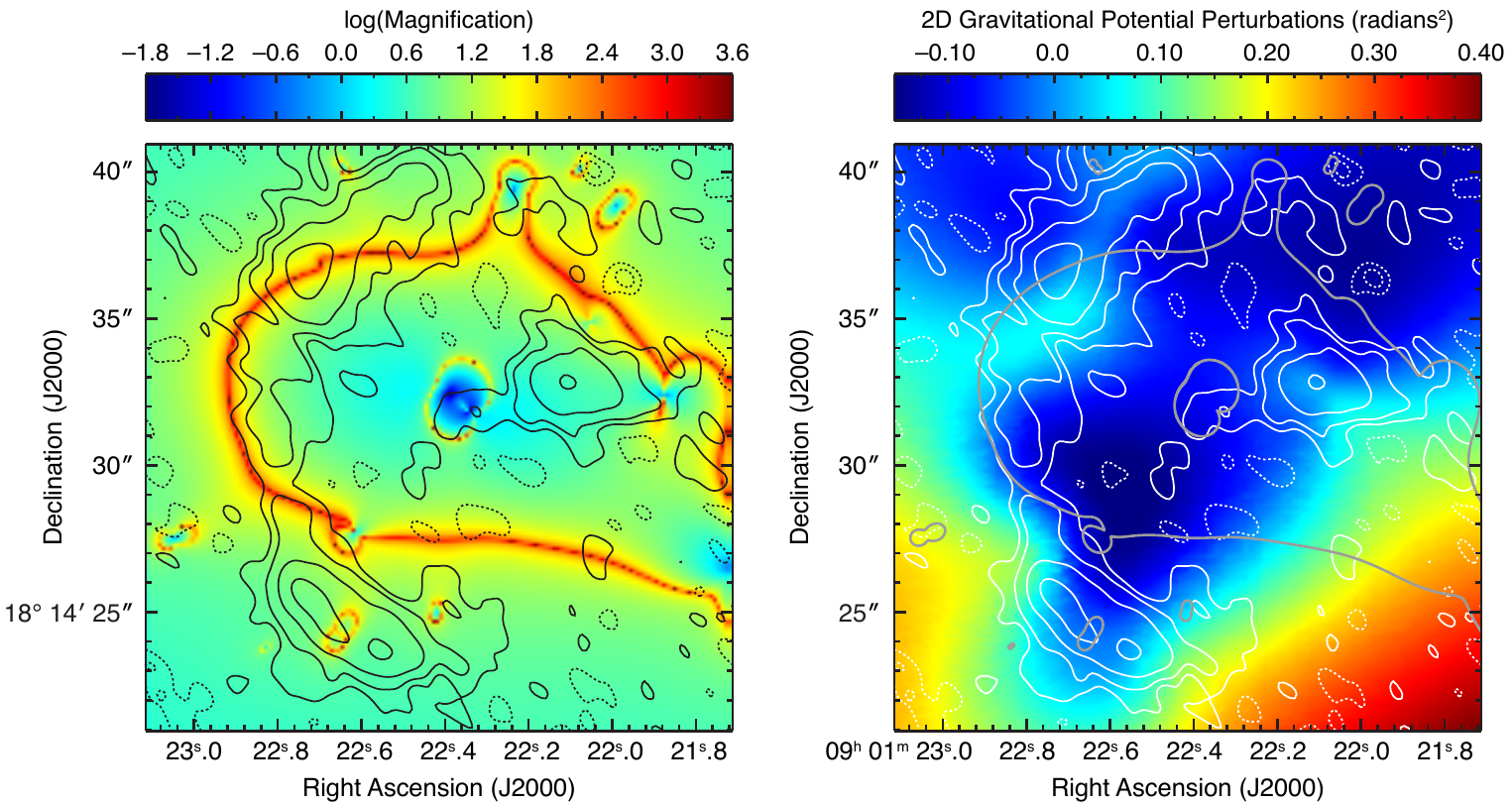}
\caption{The log of the magnification (left) and the 2D projection of the non-parametric perturbations to the lensing potential (right; normalized by the critical lensing density) for the best-fit lens model. Contours for the integrated line CO(3--2) map are shown in black (left) or white (right). Contours are powers of $2\times\pm\sigma$ (i.\/e.\/ $\pm2\sigma$, $\pm4\sigma$, $\pm8\sigma$, etc.); negative contours are dotted. In the right panel we also show the lens model critical curves in grey. \label{fig:mags}}
\end{figure*}

The data used to constrain the model consist of the \textit{HST} F606W imaging and the integrated \mbox{CO(3--2)} intensity map (Figure~\ref{fig:lensmodel}). The pair of merging images comprising the northern arc lie across a critical curve in the image plane and are more highly magnified than the southern and western images (Figure~\ref{fig:mags}). A larger magnification can allow for a more detailed analysis, but only over the fraction of the source that has crossed the caustic. There is also a larger uncertainty associated with the source-plane reconstruction using the northern arc, as the magnification varies rapidly near the critical curve (Figure~\ref{fig:mags}). For these reasons, we do not include the northern arc when constraining the lens model parameters or performing the source-plane reconstructions presented throughout. 

In addition to optimizing the lens model parameters, we include a registration offset between these data sets (referenced to the \mbox{CO(3--2)} data). For each set of lens model parameters and registration offsets, a goodness-of-fit statistic is computed by multiplicatively combining the Bayesian evidence from the optical and radio. We use the framework described in \citet{tagore2014}, \citet{vegetti2009}, and \citet{suyu2006} to reconstruct the pixelated source distribution of J0901 in the source plane, as seen in each band. An irregular, adaptive source grid is used with priors on the sources' surface brightness in the form of curvature regularization; the Bayesian evidence is maximized at each step. After optimization, slight discrepancies between the optical data and the model remain. We add smoothly varying, non-parametric perturbations to the potential to compensate for limitations of the macro-model (Figure~\ref{fig:mags}). These lens potential perturbations are at the 1--2\% level, which correspond to changes in the deflection angle of $100 \,{\rm mas}$ or less. For the optical {\it HST} data, such changes are significant; however, because the beam size is $\sim1^{\prime\prime}$ in the radio bands, the effect on the CO data is negligible. 

Lens modeling of interferometric maps is complicated by the imaging process, which does not conserve surface brightness, can be strongly affected by choices in mapping parameters (e.g., visibility weights), and yields noise that is correlated in the resulting image. All of these effects can potentially cause the lens model and source-plane reconstruction to diverge from reality. While a number of routines have been developed in recent years to constrain lens models using visibility data directly \citep[e.\/g.\/,][]{bussmann2012b,bussmann2013,hezaveh2013,hezaveh2016,rybak2015a,spilker2016a,dye2018}, many rely on parametric source models, which are overly simplistic compared to the resolved observations we have for J0901. Recognizing that lens models derived from visibility data and from deconvolved maps are both fundamentally limited by incomplete sampling in the $uv$ plane, we prefer to exploit the well-resolved structure in our maps of J0901 to derive our lens model. We defer comparisons with source-plane reconstructions inferred from non-parametric visibility-based models to future work.

In order to account for the image-plane correlated noise in our lens modeling, we follow the noise scaling technique of \citet{riechers2008a}. For an individual data set, we scale the noise (for input into the lens modeling code) by some factor greater than unity that could be determined and verified by comparing the statistical properties of noise residuals in areas where lensed features are present and absent. However, because we are comparing source reconstructions across various data sets with different noise properties, we fix the noise scaling. A large noise scaling factor allows the code to under-fit the data in the formal reduced-$\chi^2$ sense, since the code assumes there is more noise in the data, which leads to a higher regularization strength. Qualitatively, this approach smooths the source over a larger physical scale, and the resulting source-plane beam is larger.

\begin{figure}
\plotone{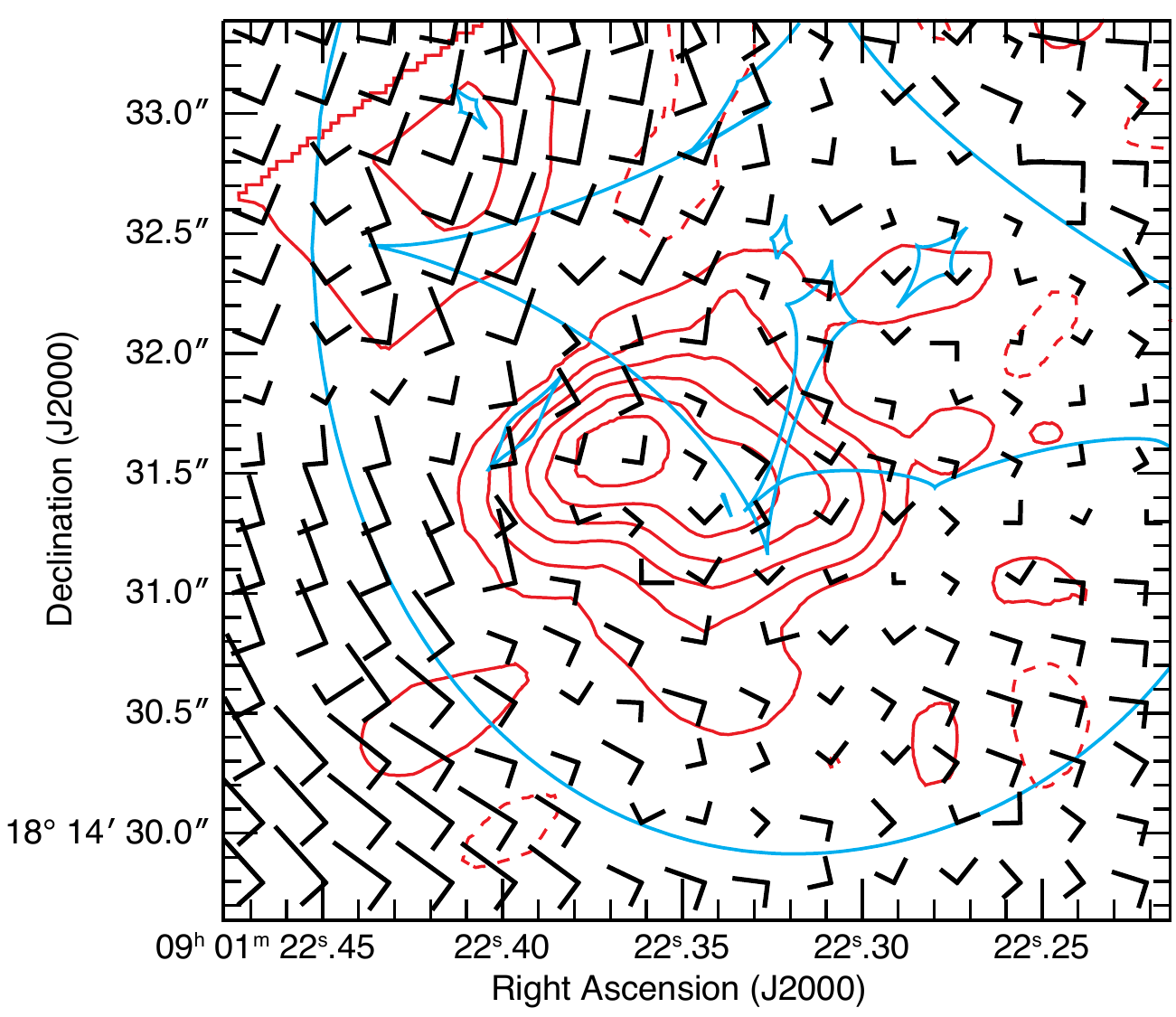}
\caption{The synthesized beam/PSF for the matched datasets as a function of position in the source plane. The black vectors are the beam/PSF HWHM at the pixel for their common origin; every eighth pixel is shown for clarity. The red contours show the source-plane reconstruction of the \mbox{CO(3--2)} integrated line map using the matched resolution data. Contours are multiples of $\pm3\sigma_{obs}$, but note that due to spatial variation in the noise, these surface brightness levels do not correspond to lines of constant significance. Negative contours are dashed. The blue lines indicate the source-plane lensing caustics. \label{fig:beams}}
\end{figure}

Our source-plane reconstructions yield a spatially varying synthesized beam/PSF. In Figure~\ref{fig:beams} we show a grid of the beam HWHMs overlaid on a contour plot of the source-plane reconstruction for the matched \mbox{CO(3--2)} integrated line map as an example of the variation in beam/PSF shape that results from de-lensing. Although the beam shape varies by a factor of a few over the entire reconstruction, the beam is smaller and more consistent in the direction of the emission for J0901. We therefore adopt surface-brightness weighted average beams/PSFs when analyzing the spatial information for J0901; these have FWHMs of $0.2$--$0.3^{\prime\prime}$ (corresponding to physical scales of $1.7$--$2.6\,{\rm kpc}$).

The lens model uncertainties are explored via Markov chain Monte Carlo modeling for the \mbox{CO(3--2)} data only to save computational time. As the \mbox{CO(3--2)} moment map was the primary input used to constrain the lens model, this method accurately captures the uncertainties in lens model parameters. Magnification factors are then derived for the individual maps by de-lensing the emission for the distribution of model parameters. The magnification factor uncertainties thus take into account uncertainties in both the surface brightness of the source and in the lens model parameters.

\subsubsection{Resulting magnification factors and image reconstructions}

\begin{figure*}
\plotone{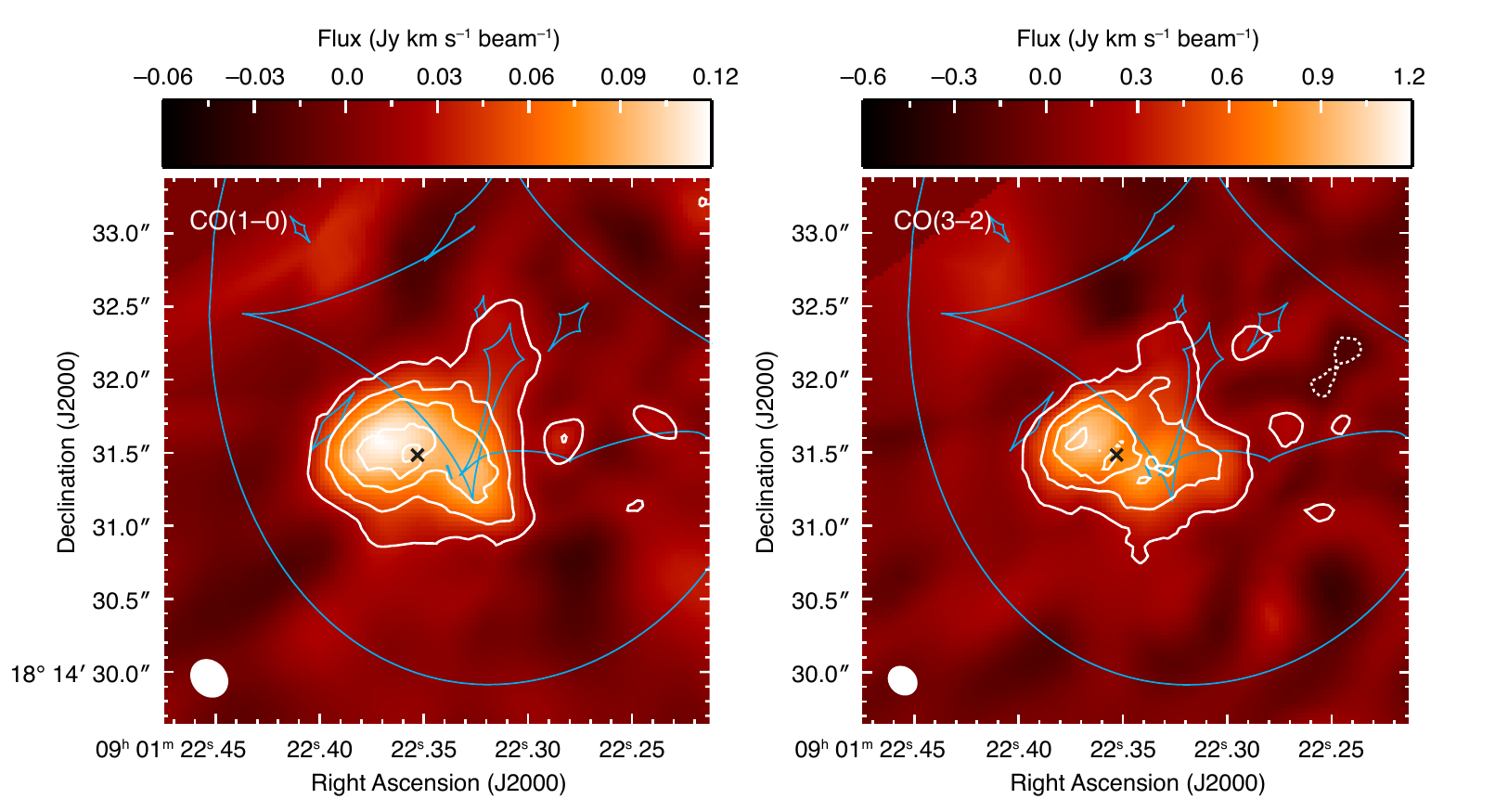}
\caption{Source-plane reconstructions of the integrated \mbox{CO(1--0)} (left) and \mbox{CO(3--2)} (right) intensity maps, derived from the natural resolution observed images with primary beam corrections (shown in Figure~\ref{fig:j0901intline}). Since the reconstructions have spatially varying noise, the contours are generated from the SNR maps and show multiples of $\pm3\sigma$ (negative contours are dotted), which do not strictly follow the surface brightness (color bar; where the minimum and maximum values of the images are shown with vertical dotted lines). The images also have spatially varying resolution, so we show the intensity-weighted average beams at the lower left. Blue lines indicate the image-plane lensing caustics. Black crosses mark the mean dynamical center (see Section~\ref{sec:masses}).\label{fig:delensCOintline}}
\end{figure*}

\begin{figure*}
\epsscale{1.15}
\plotone{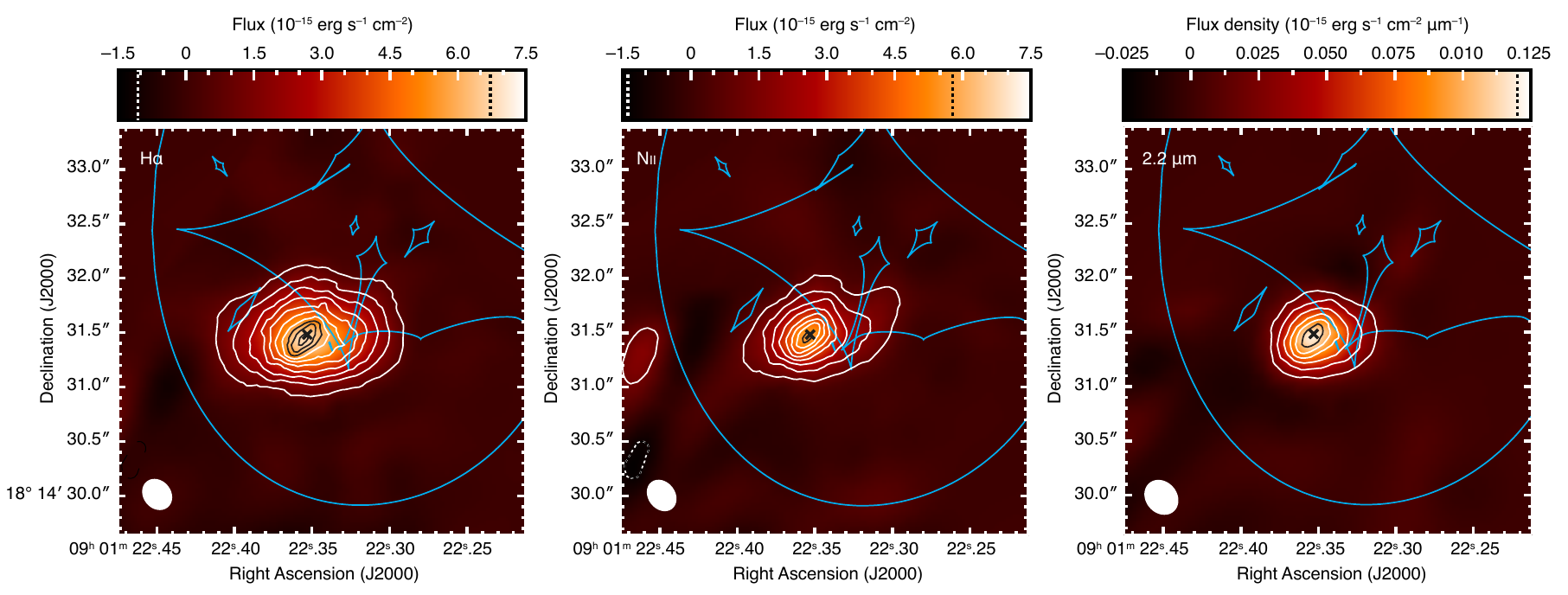}
\caption{Source-plane reconstructions of the ${\rm H\alpha}$ integrated line (left), [N\,{\sc ii}] integrated line (middle), and $2.2\,{\rm \mu m}$ (observed frame) continuum (right) intensity maps, derived the natural resolution observed images (shown in Figures~\ref{fig:j0901halpha} and \ref{fig:j0901cont}). SNR contours, the intensity-weighted average PSF, caustics, and their descriptions are as given in Figure~\ref{fig:delensCOintline}. Black crosses mark the mean dynamical center (see Section~\ref{sec:masses}).
\label{fig:delensVLT}}
\end{figure*}

With the lens model optimized, we perform source reconstructions of the integrated line maps, the individual velocity channel maps, and the $2.2\,{\rm \mu m}$ continuum map. We present the natural resolution source-plane reconstructions of the CO, ${\rm H\alpha}$, and [N\,{\sc ii}] lines for J0901 in Figures~\ref{fig:delensCOintline} and \ref{fig:delensVLT}. In Table~\ref{tab:mag}, we present the 50th percentile magnification factors and 68\% confidence intervals derived using the ``natural" resolution data and the magnification factors derived from the ``matched" resolution data, for each image separately and in aggregate.

\begin{deluxetable*}{llrrrr}
\tablecaption{Magnification factors \label{tab:mag}}
\tablehead{\colhead{Transition} & \colhead{Map} & \colhead{North} & \colhead{South} & \colhead{West} & \colhead{Total} }
\startdata
CO(1--0) & natural & $10.2^{+1.2}_{-0.9}$ & $7.4^{+0.6}_{-0.5}$ & $5.3^{+0.4}_{-0.4}$ & $22.7^{+2.1}_{-1.5}$\\
 & matched  & $20.9^{+4.3}_{-2.9}$ & $15.1^{+2.7}_{-1.9}$ & $11.1^{+2.0}_{-1.5}$ & $47.2^{+8.8}_{-5.7}$\\
 \hline
CO(3--2) & natural & $14.2^{+1.8}_{-1.6}$ & $10.4^{+1.4}_{-1.3}$ & $5.5^{+0.8}_{-0.7}$ & $30.1^{+3.7}_{-3.2}$\\
 & matched  & $14.1^{+1.8}_{-1.5}$ & $10.7^{+1.2}_{-1.1}$ & $5.7^{+0.7}_{-0.6}$ & $30.6^{+3.3}_{-2.9}$\\
 \hline
${\rm H}\alpha$ & natural & $11.8^{+2.2}_{-1.9}$ & $11.4^{+1.7}_{-1.3}$ & $6.3^{+0.9}_{-0.7}$ & $29.6^{+4.0}_{-3.1}$\\
 & matched  & $12.2^{+2.0}_{-1.7}$ & $11.9^{+1.9}_{-1.7}$ & $5.7^{+0.9}_{-0.7}$ & $29.9^{+4.3}_{-3.3}$\\
 \hline
{[N\,{\sc ii}]} & natural & $11.2^{+2.7}_{-2.1}$ & $11.5^{+3.3}_{-2.0}$ & $4.5^{+1.1}_{-0.8}$ & $27.2^{+5.8}_{-4.1}$\\
 & matched  & $9.7^{+1.2}_{-1.1}$ & $8.8^{+1.3}_{-1.0}$ & $3.5^{+0.5}_{-0.4}$ & $21.9^{+2.5}_{-2.1}$\\
 \hline
$2.2\,{\rm \mu m}$ & natural & $11.1^{+5.4}_{-3.7}$ & $16.7^{+9.5}_{-4.2}$ & $8.6^{+3.8}_{-1.9}$ & $37.1^{+16.1}_{-8.1}$\\
 & matched  & $8.9^{+4.6}_{-3.3}$ & $16.9^{+12.1}_{-5.1}$ & $7.9^{+5.3}_{-2.3}$ & $33.7^{+19.2}_{-8.9}$\\
\enddata
\end{deluxetable*} 

While the CO, ${\rm H\alpha}$, and [N\,{\sc ii}] lines all show two emission peaks in the southern arc in the image plane, those peaks do not correspond to one another across all lines. In the source-plane reconstructions, the CO peaks remain distinct but the ${\rm H\alpha}$ and [N\,{\sc ii}] peaks do not. The two peaks seen in the VLT maps are nearly aligned with the positions of the average dynamical center determined from the channelized source reconstructions (see Section~\ref{sec:masses}), and are potentially multiple images of the same region within J0901 caused by a foreground member of the lensing group. However, the two peaks may also have disappeared on reconstruction due to the degree of regularization (i.\/e.\/, the smoothness prior may have ``won" over fitting the data due to noise or flaws in the lens model), and/or because the CO and {\it HST} data used to constrain the model may not have much power over the relatively small region encompassed by the two VLT peaks. 

\begin{figure*}
\plotone{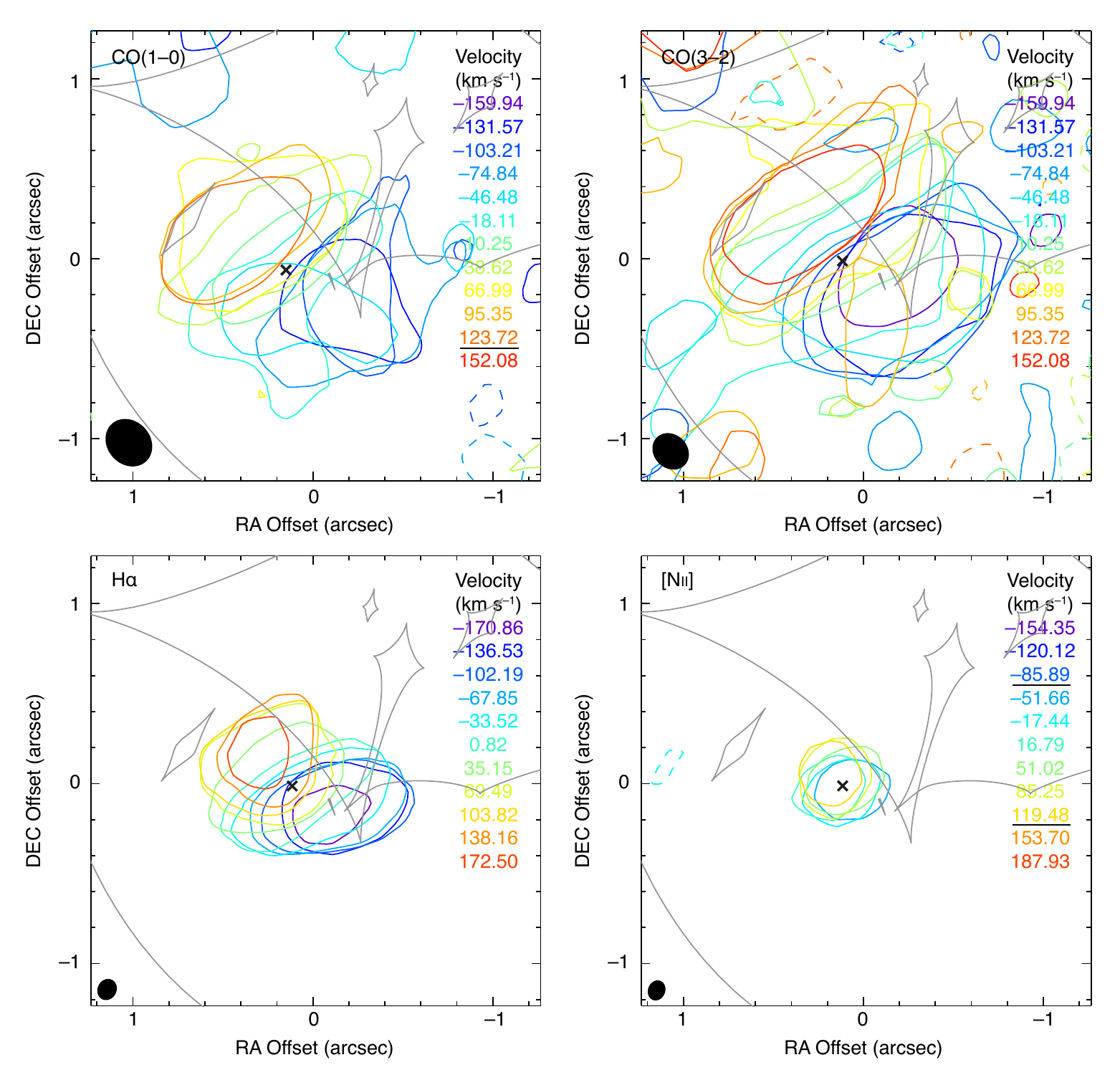}
\caption{Overlaid contours of the source-plane reconstructions for the \mbox{CO(1--0)} (upper left), \mbox{CO(3--2)} (upper right), ${\rm H\alpha}$ (lower left), and [N{\sc ii}] (lower right) channel maps using the natural resolution observed images, colorized by their velocities relative to the $z=2.2586$ ${\rm H\alpha}$ systemic redshift from \citet{hainline2009}. The images are centered at $\alpha {\rm (J2000)} = 09^{\rm h}01^{\rm m}22.^{\rm s}34$ and $\delta {\rm (J2000)} = +18\degree14^{\prime}31.^{\prime\prime}5$. Contours are for the same surface brightness levels as in Figure~\ref{fig:j0901renzos} ($\pm3\sigma_{obs}$ level for the two CO lines, and $\pm2\sigma_{obs}$ for the VLT/SINFONI data), but these surface brightness contours are \emph{not} necessarily at the same significance as for the observed data, since the source-plane reconstructions have spatially varying noise. Negative contours are dashed. Channels that do not have emission above the required surface brightness are separated by black lines in the legends. The images also have spatially varying resolution, so we show the intensity-weighted average beams and PSFs at the lower left. Gray lines indicate the source-plane lensing caustics. Black crosses mark the mean dynamical center (see Section~\ref{sec:masses}). \label{fig:delensrenzos}}
\end{figure*}

\begin{figure*}
\plotone{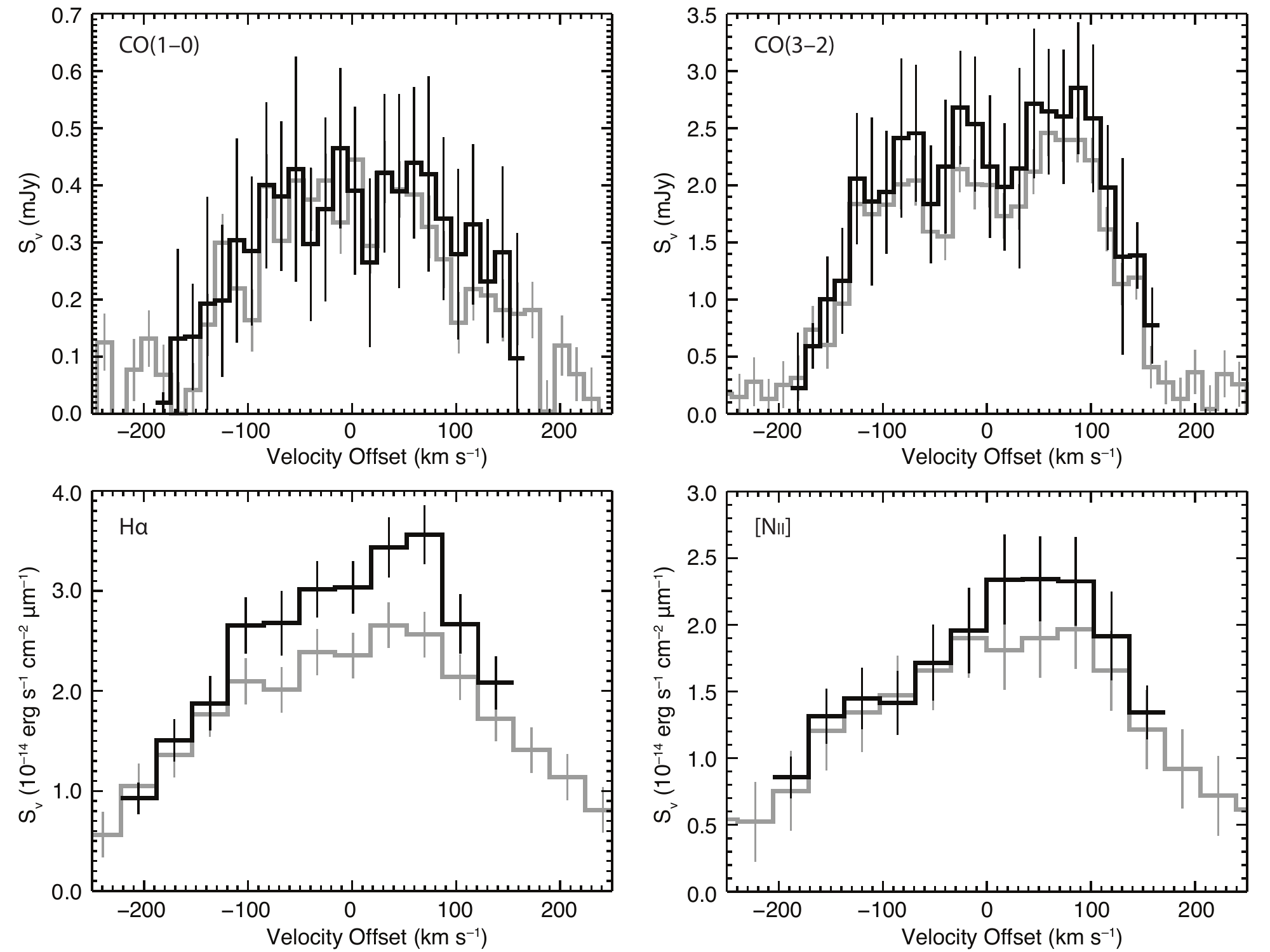}
\caption{Spectral line profiles for CO(1--0) (upper left), CO(3--2) (upper right), H$\alpha$ (lower left) and [N\,{\sc ii}] (lower right) extracted from the channelized image plane lensing reconstructions (black). We also show the observed line profiles (gray) extracted from the southern and western images (the images used to construct the lens model) divided by the average channelized magnification factor (since the per channel magnifications are on average lower than what was determined for the integrated line maps and are noisier/more uncertain). Since the flux is extracted in different ways for the image- and source-plane channel maps (as the image-plane aperture would be warped into different shapes in different source plane channels), the southern+western image-plane comparison spectra are not expected to scatter evenly below and above the source-plane spectra, despite being scaled by the mean per-channel magnification. The vertical bars denote $\pm1\sigma$ uncertainties. Channels far from the systemic redshift were not delensed. \label{fig:j0901delens_spec}}
\end{figure*}

We also reconstruct J0901 in the source plane for the individual channel maps (Figure~\ref{fig:delensrenzos}). The prominent velocity gradient observed in the image plane is also apparent in the reconstructed channel maps. The well-resolved and smooth velocity gradient seen in all lines suggests that J0901 is likely a disk galaxy, despite the two bright peaks seen in the integrated line maps. We extract the spectra from the reconstructed channel maps and compare the line profiles to the observed profiles from the image plane (Figure~\ref{fig:j0901delens_spec}). Since the per-channel magnification factors were not computed to include the northern image, we use the sum of the southern and western observed spectra scaled by the mean per-channel magnification factor in order to understand what effects differential lensing might have on the line profile\footnote{The per-channel magnification factors are, on average, lower than what was determined for the integrated line maps, and they are much noisier. We therefore exclude unphysical magnification factors outside the range of $0$--$100$ when computing the mean magnification factor for this comparison.}. We extracted the source-plane spectra in apertures defined by the ${\rm SNR}>2$ regions in the corresponding integrated line maps. We note that this method is not a perfect match to the procedure used to extract the image-plane spectra; a more perfect match would require de-lensing the image-plane aperture for each channel. Since the area occupied by a channel's emission varies with velocity (as expected, particularly when considering the variation in magnification factor), the aperture defined by the integrated line map reconstruction may miss some emission in individual channels. However, this method is adequate for revealing any dramatic or velocity-correlated differential lensing effects. 

Differential lensing does not appear to strongly affect the shape of the line profile of J0901 in the southern and western images. Since it is the bright northern image that only captures a portion of J0901's source plane structure (and thus only a portion of the velocity structure), one might suspect that any distortions of the line profile are most likely to appear in analyses that include the northern image. However, it is the northern image's CO spectral profile that shows the double-peaked structure typical of rotating disks (Figure~\ref{fig:j0901spectra}), which is perhaps only hinted at in the combined spectrum of the southern and western images and their reconstruction (Figure~\ref{fig:j0901delens_spec}). While the spatial structure of the least-distorted western image best matches the source-plane reconstruction, as expected, the de-lensed ${\rm H\alpha}$ and [N\,{\sc ii}] spectral lines appear to peak at redder wavelengths than seen in the observed spectrum of the western image. In addition, some of the internal structure of J0901 is multiply imaged \emph{within} the Southern arc due to a foreground lensing group member. We are therefore unable to firmly constrain the intrinsic profiles of the ${\rm H\alpha}$ and [N\,{\sc ii}] spectral lines for J0901.

\subsection{Integrated properties: masses and SFR}
\label{sec:masses}

\subsubsection{Gas mass and dust-to-gas ratio}

In order to estimate a gas mass for J0901, we use the magnification-corrected natural \mbox{CO(1--0)} line luminosity derived from all three images, obtaining $M_{\rm gas}=(1.6^{+0.3}_{-0.2})\times10^{11}(\alpha_{\rm CO}/4.6)\,M_\odot$ \citep{solomon1991}. We use the Milky Way CO-to-${\rm H_2}$ conversion factor due to J0901's disk-like ordered rotation (Figure~\ref{fig:delensrenzos}), but it is also the value favored by the \citet{narayanan2012} continuous metallicity and surface-brightness dependent version of the CO-to-${\rm H_2}$ conversion factor. The metallicity-dependent form of the CO-to-${\rm H_2}$ conversion factor presented in \citet{genzel2015} and \citet{tacconi2017} yields a slightly lower value of $\alpha_{\rm CO}=3.8\,{\rm M_\odot\,K^{-1}\,km^{-1}\,s\,pc^{-2}}$. However, the inferred gas mass is consistent with our Milky Way $\alpha_{\rm CO}$-derived mass within the uncertainties. We note that there is also some uncertainty on the metallicity of J0901 (see Section~\ref{sec:metallicity}). As a sanity check on the $\sim30\%$ difference between the \mbox{CO(1--0)} and \mbox{CO(3--2)} lines' magnification factors, we also calculate $M_{\rm gas}$ using the \mbox{CO(3--2)} map and its corresponding magnification (corrected for excitation using our measured global $r_{3,1}$ \emph{without} magnification correction) and find $M_{\rm gas}=(1.2\pm0.3)\times10^{11}(\alpha_{\rm CO}/4.6)\,M_\odot$; this value is consistent with the \mbox{CO(1--0)}-derived gas mass and therefore gives additional credibility to the difference in the two lines' magnification factors (at least for the natural resolution images).

Following \citet{scoville2016}, we also use the $877\,{\rm \mu m}$ (observed frame) SMA continuum detection as an alternative probe of the gas mass. This method relies on the adoption of a dust temperature; \citet{scoville2014,scoville2016} recommend against using dust temperatures derived from multi-band SED fits ($T_{dust}=36\,{\rm K}$ in the case of J0901; \citealt{saintonge2013}), since they are luminosity weighted and thus biased towards the hotter components of the ISM that do not make up the bulk of the mass, and instead recommend the adoption of $T_{dust}=25\,{\rm K}$. Both values result in $\sim2$--$3\times$ lower ISM masses than the CO-derived gas masses ($M_{\rm mol}=(4.8\pm1.3)\times10^{10}\,{M_\sun}$ for $T_{dust}=36\,{\rm K}$ and $M_{\rm mol}=(7.1\pm2.0)\times10^{10}\,{M_\sun}$ for $T_{dust}=25\,{\rm K}$, when corrected by the \mbox{CO(3--2)} ``natural" magnification factor). These continuum-derived ISM masses suggest a lower value of $\alpha_{\rm CO}\sim1.4$--$2$ would be more appropriate for J0901 (closer to values derived for low-metallicity systems, or to the canonical value used for local U/LIRGs). However, since we do not independently derive a magnification factor for the $877\,{\rm \mu m}$ continuum data due to its low angular resolution and S/N, there is some additional uncertainty in the continuum-derived ISM mass and implied CO-to-${\rm H_2}$ conversion factor.

Given the uncertainty in $\alpha_{\rm CO}$ for J0901, we adopt the magnification-corrected, natural resolution \mbox{CO(1--0)}-derived value of $M_{\rm gas}=(1.6^{+0.3}_{-0.2})\times10^{11}(\alpha_{\rm CO}/4.6)\,M_\odot$, carrying the uncertainty in $\alpha_{\rm CO}$ as a free parameter. Even with conversion factor uncertainties, we note that the gas mass of J0901 is comparable to those of other galaxies selected at submillimeter wavelengths, but larger than those of other UV-selected high-redshift galaxies \citep[e.\/g.\/,][]{riechers2010b}.

Adopting the dust mass from \citet{saintonge2013}, corrected to our \mbox{CO(3--2)} magnification factor, we obtain a dust-to-gas mass ratio of $(4.7^{+1.4}_{-1.2})\times10^{-3}(\alpha_{\rm CO}/4.6)^{-1}$ for J0901. This ratio is within the normal range for disk galaxies in the local universe \citep[e.\/g.\/,][]{draine2007b} but is a bit low for those with the same metallicity (as seen for the high-redshift galaxies in \citealt{saintonge2013}). However, the dust-to-gas mass ratio strongly depends on the assumed CO-to-${\rm H_2}$ conversion factor as well as the properties of dust adopted by the \citet{draine2007a} dust models. Lower CO-to-${\rm H_2}$ conversion factors would increase the dust-to-gas mass ratio by a factor of $\sim5$, bringing it more in line with the dust-to-gas ratios of systems where authors tend to adopt those lower values (i.\/e.\/, SMGs and U/LIRGs; e.\/g.\/, \citealt{santini2010}).

\subsubsection{SFR and stellar mass}

Using our new magnification factors and ${\rm H\alpha}$ measurements, we determine improved SFRs for J0901. We use the SFR scaling factor from \citet{hao2011}/\citet{murphy2011} (as compiled in \citealt{kennicutt2012}) scaled to a \citet{kroupa2001} initial mass function. We find ${\rm SFR_{H\alpha}}=14.5^{+2.1}_{-1.7}\,{\rm M_\odot\,yr^{-1}}$ using the total $L_{\rm H\alpha}$ and native magnification factor \emph{without} correction for obscuration. \citet{hainline2009} measured the ${\rm H\alpha}$ and ${\rm H\beta}$ lines for two regions within J0901, finding extreme obscuration corrections from ${\rm H\alpha}$/${\rm H\beta}$ that would increase the SFR by a factor of $\gtrsim20$. However, that ratio could have been affected by the coincidence of a skyline with the ${\rm H\beta}$ emission. Using the total infrared luminosity ($L_{\rm TIR}$ from $8$--$1000\,{\rm \mu m}$) derived from the \citet{draine2007b} fits to J0901's dust SED in \citet{saintonge2013} ($L_{\rm TIR}=1.80^{+0.42}_{-0.41}\times10^{12}\,L_\sun$ assuming our new magnification factor for the native-resolution \mbox{CO(3--2)} data) and our choice in in IMF yields ${\rm SFR_{TIR}}=268^{+63}_{-61}\,{\rm M_\odot\,yr^{-1}}$, comparable to the expected value based on the ${\rm H\beta}$ extinction correction to ${\rm H\alpha}$. \citet{kennicutt2012}/\citet{kennicutt2009} also give an alternative method for correcting ${\rm H\alpha}$ to account for obscured star formation using the observed $L_{\rm TIR}$, but this method yields a much smaller value of ${\rm SFR_{H\alpha +TIR}}=103^{+21}_{-20}\,{\rm M_\odot\,yr^{-1}}$ (where we have corrected the luminosities for the different magnification factors for ${\rm H\alpha}$ and TIR as above). This hybrid method for calculating obscured SFRs involves a number of assumptions that may not apply to galaxies in the early universe, and was calibrated using galaxies with infrared luminosities lower than that of J0901 (albeit with similar $L_{\rm TIR}/L_{\rm H\alpha}$ ratios and attenuation levels). We therefore adopt ${\rm SFR_{TIR}}=268^{+63}_{-61}\,{\rm M_\odot\,yr^{-1}}$ for our subsequent analysis, since it likely accounts for the bulk of the star formation in J0901 and is not likely contaminated by significant emission from the AGN \citep{fadely2010}.

The fraction of the total SFR that can be accounted for by our ${\rm H\alpha}$ measurements is consistent with the ${\rm SFR}_{\rm UV}/{\rm SFR}_{\rm IR}$ derived in \citet{saintonge2013}: ${\rm SFR}_{\rm H\alpha}/{\rm SFR}_{\rm TIR}=0.054^{+0.015}_{-0.014}$ vs. ${\rm SFR}_{\rm UV}/{\rm SFR}_{\rm IR}=0.040\pm0.007$. Since J0901 is known to have an AGN \citep{hainline2009} on the basis of its high [N\,{\sc ii}]/${\rm H\alpha}$ line ratio and large ${\rm H\alpha}$ FWHM, it is possible that the ${\rm H\alpha}$-determined SFR is contaminated by emission from the AGN; IFU observations of the ${\rm H\alpha}$ emission from the nuclear region of J0901 obtained using adaptive optics show signs of a broad low-level outflow once disk rotation is corrected for \citep{genzel2014}. However, for the emission from both the disk and nucleus analyzed here, the ${\rm H\alpha}$ FWHM is no wider than one would expect based on single-Gaussian fits to the double-peaked CO line profiles (at least for the line profile derived from the sum of the three images). It seems likely that most of the ${\rm H\alpha}$ emission is due to star formation, and that some emission from the AGN, near the systemic redshift, masks J0901's double peaked profile (particularly given the slightly poorer $\sim40\,{\rm km\,s^{-1}}$ velocity resolution of the VLT data and $\sim150\,{\rm km\,s^{-1}}$ CO peak separations) but contributes only a small amount to the total ${\rm H\alpha}$ luminosity. Higher S/N would be necessary to do a pixel-by-pixel decomposition of the broad and narrow line emission components to correct for the ${\rm H\alpha}$ emission from the AGN, as done for the nucleus in \citet{genzel2014}.

If we re-scale the stellar mass from \citet{saintonge2013} to use the same Kroupa IMF that we assume for our SFR and apply our ${\rm H\alpha}$-determined magnification factor, we find J0901 has $M_\star=(9.5^{+3.8}_{-2.8})\times10^{10}\,{\rm M_\odot}$. Combined with the TIR-derived SFR, J0901 has a specific star formation rate of ${\rm sSFR}=2.8^{+1.3}_{-1.1}\,{\rm Gyr^{-1}}$. Since we have simply corrected the \citet{saintonge2013}-derived values by our new magnification factors (the CO and ${\rm H\alpha}$ magnification factors are very similar), choice of IMF, and TIR/SFR conversion factor, J0901 still falls along the star-forming main sequence (MS; e.\/g.\/, \citealt{noeske2007,speagle2014}), with an upward offset of just $0.27^{+0.20}_{-0.16}\,{\rm dex}$. We also compare J0901's sSFR to the bi-modal MS and starburst (SB) populations parameterized in \citet{sargent2012}/\citet{rodighiero2011}, who find a MS scatter of 0.188\,dex and a second Gaussian peak for starbursts offset by $\log({\rm \langle sSFR_{SB}\rangle/\langle sSFR_{MS}\rangle})=0.59$ with a 0.243\,dex scatter. In this scheme, J0901 falls between the distributions for MS and starbursts at $0.22^{+0.20}_{-0.16}\,{\rm dex}$, but with considerable uncertainty. Based on these parameterizations of the MS, J0901 appears to be a massive but otherwise ``normal" MS galaxy that falls a little to the high side of the sSFR distribution.

\subsubsection{Dynamical mass}

Using our de-lensed images, we can measure the physical size of J0901 and its dynamical mass. Despite the complications potentially introduced by the spatially varying resolution that results from the de-lensing, the size of J0901 is quite robust. Gaussian fits to the de-lensed integrated CO emission maps (without accounting for beam/resolution effects) are consistent for the two lines, with major and minor axis FWHMs of $1.1^{\prime\prime}\pm0.1^{\prime\prime}$ and $0.85^{\prime\prime}\pm0.05^{\prime\prime}$ respectively (position angle of $82\pm7\degree$). The VLT observations have slightly smaller and more elliptical de-lensed angular sizes, $(1.0^{\prime\prime}\pm0.1^{\prime\prime})\times(0.60^{\prime\prime}\pm0.02^{\prime\prime})$ for ${\rm H\alpha}$ and $(0.68^{\prime\prime}\pm0.06^{\prime\prime})\times(0.22^{\prime\prime}\pm0.02^{\prime\prime})$ for [N\,{\sc ii}], at position angles similar to those of the CO lines. At these angular scales, the adopted beam/PSF values do not significantly affect the source sizes, and both the convolved and de-convolved (reported) source sizes are consistent within their uncertainties. 

\begin{deluxetable*}{clccc}
\tablecaption{Kinematic fit parameters \label{tab:dynmodels}}
\tablehead{ \colhead{Model} & \colhead{Parameter} & \multicolumn{3}{c}{Transition} \\
{} & {} & \colhead{CO(1--0)} & \colhead{CO(3--2)} & \colhead{$\rm{H\alpha}$} }
\startdata
Exponential disk & R.A. & ${\rm 09^{h}01^{m}22.^{s}3518}$ & ${\rm 09^{h}01^{m}22.^{s}3533}$ & ${\rm 09^{h}01^{m}22.^{s}3523}$ \\
{} & Dec. & ${\rm +18\degree14^\prime31.^{\prime\prime}4922}$ & ${\rm +18\degree14^\prime31.^{\prime\prime}4944}$ & ${\rm +18\degree14^\prime31.^{\prime\prime}4387}$ \\
{} & $r_{1/2}$ & $4.83\,{\rm kpc}$ & $4.76\,{\rm kpc}$ & $3.65\,{\rm kpc}$ \\
{} & $i$ & $36\degree$ & $34\degree$ & $23\degree$ \\
{} & P.A. & $51\degree$ & $47\degree$ & $43\degree$ \\
{} & $r_v$ & $0.09\,{\rm kpc}$ & $0.67\,{\rm kpc}$ & $0.56\,{\rm kpc}$ \\
{} & $v_{circ}$ & $188\,{\rm km\,s^{-1}}$ & $230\,{\rm km\,s^{-1}}$ & $345\,{\rm km\,s^{-1}}$ \\
{} & $\sigma_v$ & $38\,{\rm km\,s^{-1}}$ & $39\,{\rm km\,s^{-1}}$ & $60\,{\rm km\,s^{-1}}$ \\
{} & $\chi^2_{red}$ & 1.2 & $1.6$ & 0.97 \\
{} & $M_{dyn}$ & $0.8\times10^{11}\,{\rm M_\odot}$ & $1.2\times10^{11}\,{\rm M_\odot}$ & $2.0\times10^{11}\,{\rm M_\odot}$ \\
\hline
Gaussian & R.A. & ${\rm 09^{h}01^{m}22.^{s}3515}$ & ${\rm 09^{h}01^{m}22.^{s}3527}$ & ${\rm 09^{h}01^{m}22.^{s}3519}$ \\
{} & Dec. & ${\rm +18\degree14^\prime31.^{\prime\prime}4887}$ & ${\rm +18\degree14^\prime31.^{\prime\prime}4925}$ & ${\rm +18\degree14^\prime31.^{\prime\prime}4506}$ \\
{} & $r_{1/2}$ & $4.07\,{\rm kpc}$ & $3.95\,{\rm kpc}$ & $3.14\,{\rm kpc}$ \\
{} & $i$ & $30\degree$ & $27\degree$ & $17\degree$ \\
{} & P.A. & $51\degree$ & $45\degree$ & $42\degree$\\
{} & $r_v$ & $0.05\,{\rm kpc}$ & $0.95\,{\rm kpc}$ & $0.92\,{\rm kpc}$ \\
{} & $v_{circ}$ & $218\,{\rm km\,s^{-1}}$ & $306\,{\rm km\,s^{-1}}$ & $500\,{\rm km\,s^{-1}}$ \\
{} & $\sigma_v$ & $39\,{\rm km\,s^{-1}}$ & $35\,{\rm km\,s^{-1}}$ & $57\,{\rm km\,s^{-1}}$ \\
{} & $\chi^2_{red}$& $1.1$ & $1.5$ & $0.86$ \\
{} & $M_{dyn}$ & $0.9\times10^{11}\,{\rm M_\odot}$ & $1.7\times10^{11}\,{\rm M_\odot}$ & $3.7\times10^{10}\,{\rm M_\odot}$ \\
\enddata
\tablecomments{Since GalPaK$^{\rm 3D}$ does not produce meaningful uncertainties and the assumed models may not accurately reflect the underlying emission and dynamics of J0901, the best-fit values should be treated as approximate. As discussed in the text, we do not consider the fit of the ${\rm H\alpha}$ kinematics for a Gaussian flux profile to be credible.}
\end{deluxetable*}

In order to estimate the dynamical mass, we analyze our de-lensed three dimensional data using the Bayesian Monte Carlo Markov Chain tool GalPaK$^{\rm 3D}$ \citep{bouche2015}, which constrains parametric fits to galaxy morphologies and dynamics while accounting for instrumentation-induced correlations in both the spatial and spectral directions. For the parametric model, we assume either a Gaussian or exponential intensity distribution originating from an inclined thick disk with a rotation profile of $v(r)=v_{circ}\tan^{-1}(r/r_v)$ and intrinsic velocity dispersion $\sigma_v$. In Table~\ref{tab:dynmodels}, we list the best-fit parameters for both models, and the resulting dynamical mass estimates using $M_{dyn} =233.5 (2r_{1/2})v_{circ}^2$ (from the standard $M_{dyn} = rv^2/G$ with units of the dynamical mass, half-light radius, and circular velocity set to solar masses, parsecs, and kilometers per second, respectively). For the radius, we use twice the half-light radius since that is a reasonable approximation for the radius that encompasses 90\% of the emission for both assumptions of Gaussian and exponential flux profiles. We fit dynamical models to the \mbox{CO(1--0)}, \mbox{CO(3--2)}, and ${\rm H\alpha}$ data for both the Gaussian or exponential flux distributions in order to estimate systematic uncertainties caused by model assumptions that may not accurately describe the underlying emission. Attempts to fit the native resolution reconstruction of the [N\,{\sc ii}] maps did not converge. We suspect this failure is due to a combination of factors, including models that poorly describe the observed emission (which might be expected if the [N\,{\sc ii}] emission is mostly associated with the central AGN), reconstructed velocity channels that are limited in number and do not fully trace the broad emission wings, and the lower S/N of these data. The fit to the reconstructed ${\rm H\alpha}$ map for the assumption of a Gaussian intensity distribution converges to circular velocities significantly larger than that of the other emission lines and flux profiles, likely for the same reasons that the [N\,{\sc ii}] does not converge at all. The sub-unity reduced $\chi^2$ values for both ${\rm H\alpha}$ fits are due to the small number of reconstructed velocity channels (11) and the large number of model parameters being fit (10). 

Using the five consistent best-fit models for the three successfully fit lines, we calculate a mean dynamical center for J0901 of R.A. ${\rm 09^{h}01^{m}22.^{s}3523}$ and Dec. ${\rm +18\degree14^\prime31.^{\prime\prime}4813}$. We then use the lens model to project the position of the dynamical center to the image plane; these positions are shown as black crosses in Figures~\ref{fig:j0901intline}, \ref{fig:j0901halpha}, \ref{fig:j0901cont}, and \ref{fig:j0901renzos}. As the coordinates of the mean dynamical center are outside the (primary) lensing caustic, that position only appears in the southern and western images. For the southern image, the foreground member of the lensing group creates two sub-images of the mean dynamical center position. As the two peaks of emission in the VLT ${\rm H\alpha}$, [N\,{\sc ii}], and continuum maps are nearly aligned with the image plane positions of the average dynamical center, these peaks may correspond to multiple images of nuclear emission associated with the central AGN (higher angular resolution observations are necessary to confirm whether these peaks are multiple images or unrelated internal structures).

From these fits, J0901 appears to be consistent with a relatively face-on disk with a half-light radius of $\sim4.25\,{\rm kpc}$ (consistent with sizes from the Gaussian fits we previously derived from the de-lensed integrated line maps). This size is consistent with what has been found for other star-forming galaxies with similar masses and redshifts \citep[e.\/g.\/,][]{vanderwel2014}. The circular velocity is somewhat degenerate with the source size and inclination angle, so the best-fit models either find higher circular velocities with lower inclination angles or lower circular velocities with large inclination angles. On average (neglecting the more questionable fit to the ${\rm H\alpha}$ data), we find $v_{circ}\approx260\,{\rm km\,s^{-1}}$ and $i\approx30\degree$. The \citet{rhoads2014} measurement of $v_{circ}=(120\pm7)/\sin(i)\,{\rm km\,s^{-1}}$ is consistent with our average best-fit circular velocity and inclination angle. Based on the models' best fit circular velocities and velocity dispersions, the molecular gas kinematics appear to be consistent with other high-$z$ disks \citep[e.\/g.\/,][]{tacconi2013}, with $v_{circ}/\sigma_v\sim6$.

These models yield an average dynamical mass estimate of $\sim1.3\times10^{11}\,{\rm M_\sun}$ (again, neglecting the likely unphysical fit to the ${\rm H\alpha}$ data for an assumed Gaussian intensity distribution). All of the five best-fit models' dynamical mass estimates are lower than the total baryonic mass of $2.6^{+0.5}_{-0.3}\times10^{11}\,{\rm M_\odot}$ that we infer from our adopted gas and stellar masses. However, adopting a lower value of the CO-to-${\rm H_2}$ conversion factor significantly alleviates this tension, dropping the total baryonic mass to $1.2^{+0.4}_{-0.3}\times10^{11}\,{\rm M_\odot}$ for $\alpha_{\rm CO}=0.8$. Intermediate values of the CO-to-${\rm H_2}$ conversion factor (favored by metallicity-dependent models, for example) could also be possible if new constraints on the lensing of the stellar mass tracers yield larger magnification factors, or if the dynamical mass is evaluated out to a larger radius (than our assumed value of $2r_{1/2}$) that captures more of the CO emission. Better models of the lensing potential, morphology, and dynamics of J0901 (from data with higher resolution and/or S/N, and/or models that more closely match the true flux distribution and kinematics) may also alleviate some of the tension with the baryonic mass estimates. Models of low inclination systems are particularly sensitive to assumptions of azimuthal symmetry that may not be valid for J0901 or many rotating systems in the early universe; lower inclination angles (which would imply higher circular velocities) may also alleviate tensions between the baryonic and dynamical masses.

\subsection{Spatial variation in CO excitation}
\label{sec:COratios}

In order to understand the gas conditions in J0901, we examine the \mbox{CO(3--2)}/\mbox{CO(1--0)} line ratio in units of brightness temperature (Table~\ref{tab:j0901emission}). We find that the global line ratios of the three images do not differ significantly. Using the matched \mbox{CO(1--0)} image-plane data, we find that J0901 has a global $r_{3,1}=0.79\pm0.12$. This value is comparable to the $r_{3,1}$ found for SMGs and LBGs (\citealt{riechers2010b,sharon2016}; although the sample size is small), and larger than the value implied from excitation analyses of $z\sim1.5$ $BzK$-selected galaxies \citep{dannerbauer2009,daddi2015}. We note that the $r_{3,1}$ value implied by the natural maps is only slightly lower but not significantly different from that of the matched maps, with a global $r_{3,1}=0.75\pm0.11$. It is therefore unlikely that the different observations' $uv$ sampling are leading to a recovery of emission on very different angular scales. However, if we fold in magnification corrections, $r_{3,1}$ significantly decreases for comparisons using the natural resolution data and their corresponding magnification factors ($r_{3,1}=0.56^{+0.13}_{-0.10}$), and increases for the matched resolution data ($r_{3,1}=1.23^{+0.33}_{-0.27}$).

\begin{figure*}
\epsscale{0.9}
\plotone{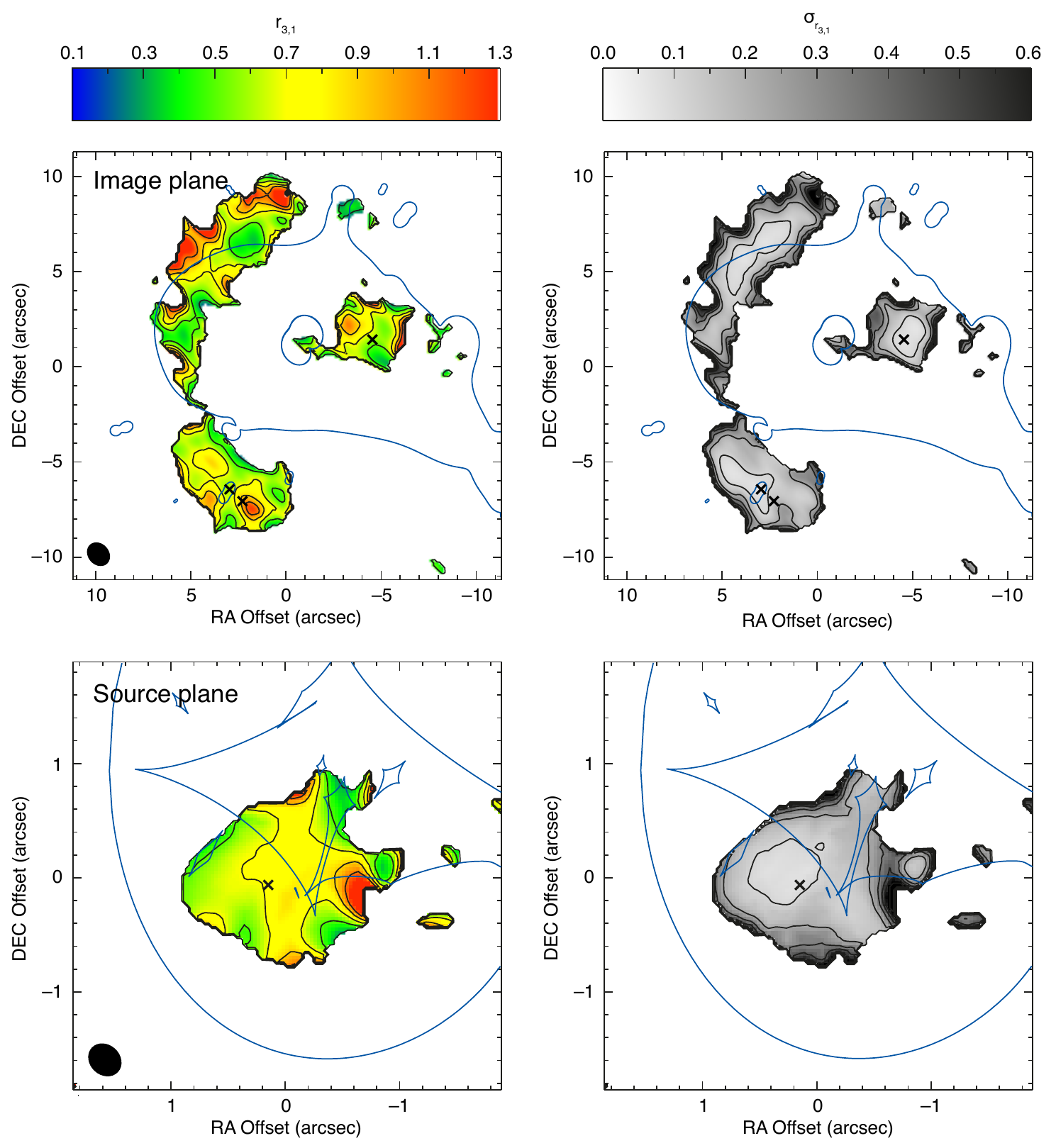}
\caption{Map of the CO(3--2)/CO(1--0) line ratio (left) and statistical uncertainty in the line ratio (right) in units of brightness temperature in the image plane (upper row) and in the de-lensed source-plane reconstruction (lower row). Both ratio maps use the ``matched" datasets with the same spatial resolution and inner $uv$ radius. Negative and $<2\sigma$ significance pixels have been blanked out. For the ratio maps, contours are in steps of $\Delta r_{3,1}=0.2$, and the color mapping is saturated at $r_{3,1}=1.3$. For the uncertainty maps, contours are in steps of $\Delta \sigma_{r_{3,1}}=0.1$, and the color mapping is saturated at $\sigma_{r_{3,1}}=0.6$. Blue lines indicate the image-plane lensing critical curves or source-plane caustics (Section~\ref{sec:lensing}). Black crosses mark the mean dynamical center determined from the source-plane reconstructions and lens modeling (see Section~\ref{sec:masses}).\label{fig:j0901intlineratio}}
\end{figure*}

\begin{figure*}
\plotone{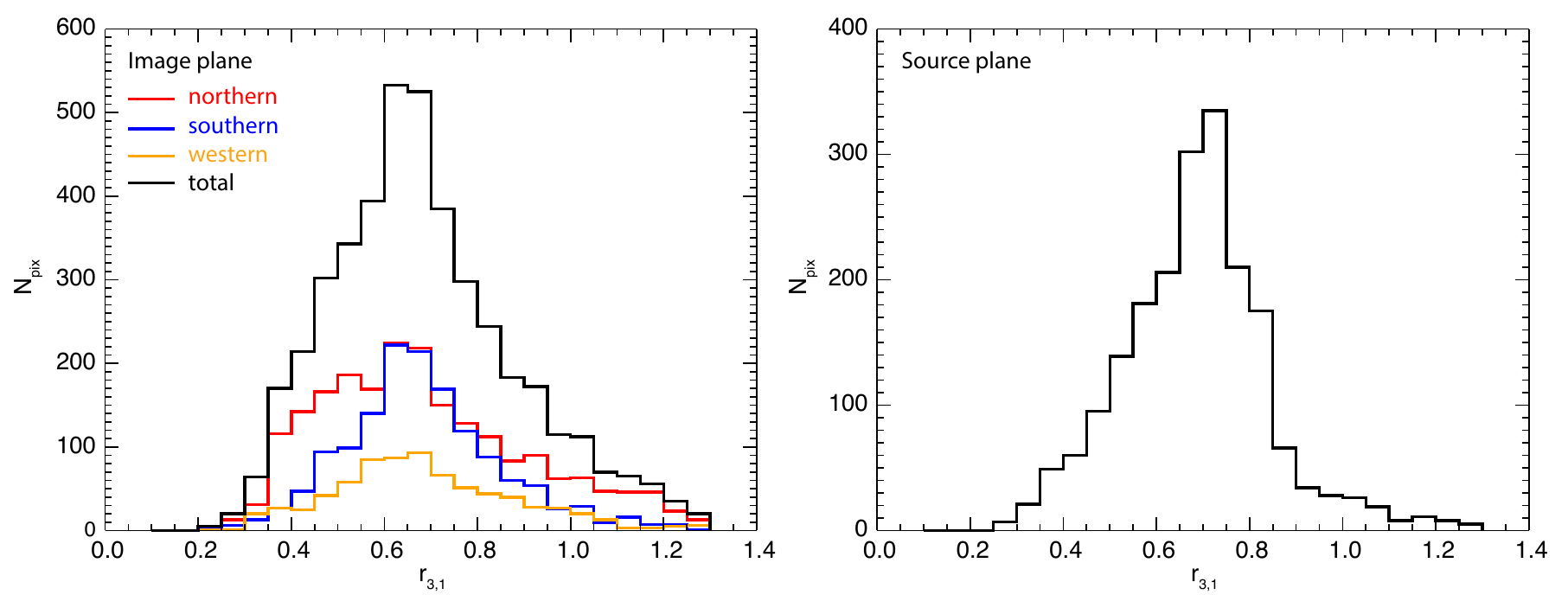}
\caption{Distribution of CO(3--2)/CO(1--0) pixel values (in units of brightness temperature) in both the image plane (left) and reconstructed source plane (right). The pixels used in these distributions are the same as in Figure~\ref{fig:j0901intlineratio}, which are clipped at the $2\sigma$ level. For the image plane maps, we show the pixel distribution for the northern (red), southern (blue), and western (gold) images separately as well as in aggregate (black). \label{fig:j0901ratiohisto}}
\end{figure*}

The strong gravitational lensing of J0901 yields additional angular resolution, which allows us to examine spatial variation in the CO excitation. For comparisons to the \mbox{CO(3--2)} map, we used the matched \mbox{CO(1--0)} map. Figure~\ref{fig:j0901intlineratio} shows the integrated line ratio map for J0901. The average value of $r_{3,1}$ in the line ratio map is $\sim0.8$, in line with the $r_{3,1}$ calculated from the integrated line flux of the $uv$-clipped \mbox{CO(1--0)} map. However, if we look at distribution of $r_{3,1}$ values in the map (Figure~\ref{fig:j0901ratiohisto}), we see that the distributions peak at slightly lower values of $r_{3,1}\sim0.6$--$0.7$ for all images and for the source-plane reconstruction. Given this lower peak $r_{3,1}$ in the source-plane reconstruction, we do not trust the large magnification factor derived for the matched-resolution \mbox{CO(1--0)} data that yields the unusually large global $r_{3,1}\approx1.2$. For the image-plane $r_{3,1}$ distributions, a strong tail out to higher excitations biases the average $r_{3,1}$ value, and most of the gas has a lower \mbox{CO(3--2)}/\mbox{CO(1--0)} line ratio. While the image plane $r_{3,1}$ distribution appears roughly log-normal, which may hint at emission from higher density gas phases, we do not ascribe much significance to this shape, given the underlying noise in the two maps and the $2\sigma$ significance clipping that is applied. Given the Gaussian noise in the individual CO maps, the ratio map noise should follow a Cauchy distribution, which could skew the distribution of per-pixel $r_{3,1}$ values if it is not properly accounted for. However, the noise distribution is further complicated by the primary beam corrections required to accurately measure the flux in an extended source such as J0901. We therefore trust only the peak values of the $r_{3,1}$ distributions. 

For the integrated line ratio map, the lower-excitation gas (areas in the map with lower values of $r_{3,1}$) appears to be more spatially extended than the higher excitation gas, especially on the basis of the southern image and reconstructed source plane maps. The line ratio map for the source-plane reconstruction looks similar to that of the western image, which we expect since the western image is the least distorted. For the northern image, it is difficult to determine whether the large $r_{3,1}$ values near the image's edge are caused by noise and weak emission or by genuine differences between the CO emission in the two maps (potentially amplified by lensing). Examining the line ratio maps as a function of channel does not reveal any significant velocity trend, in either the image or the source plane, due to the lower SNR of individual channel maps (which is then amplified when taking their ratio). 

\begin{figure}
\plotone{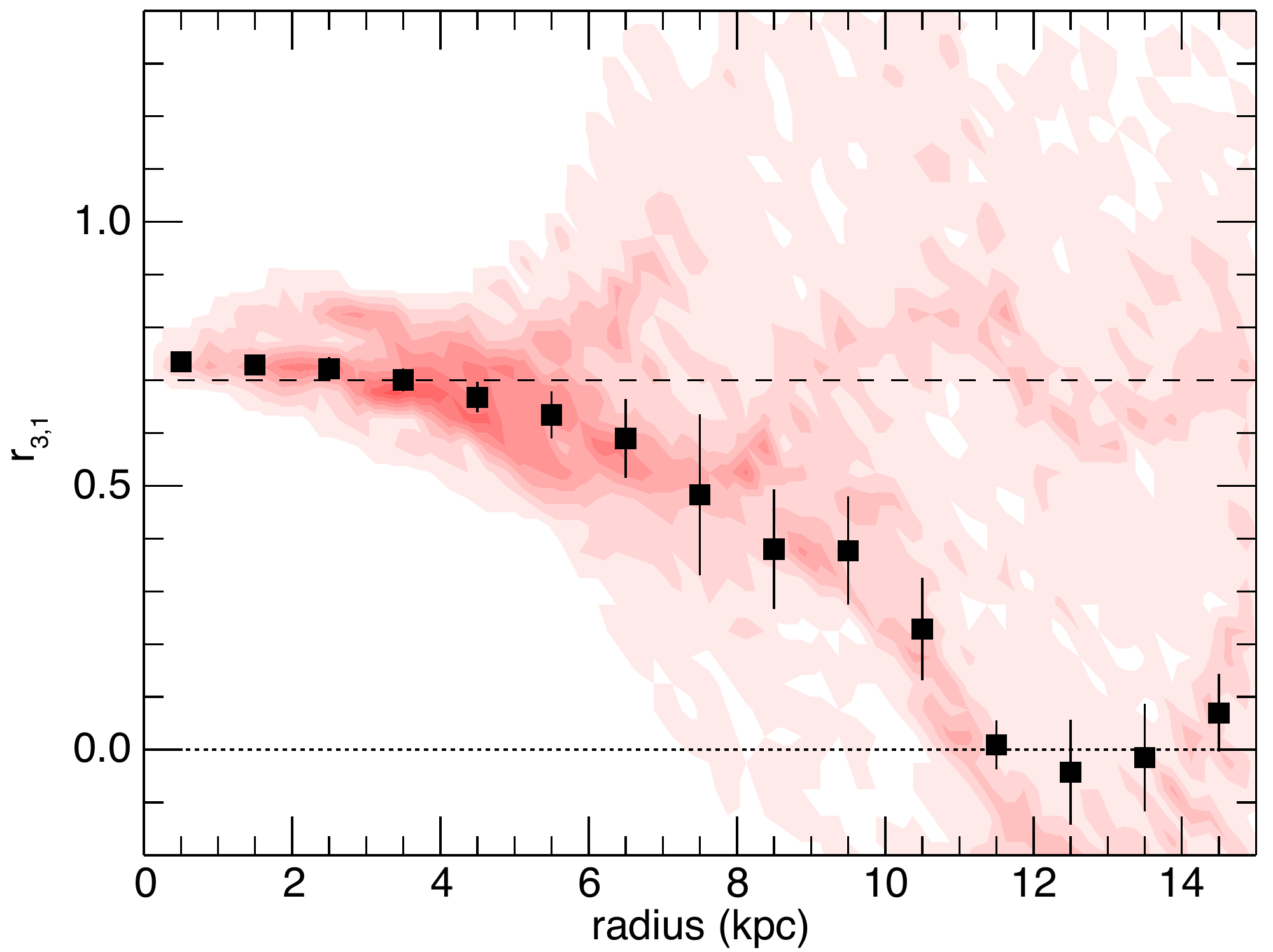}
\caption{The distribution of CO(3--2)/CO(1--0) line ratios for pixels in the matched-resolution source-plane reconstructions as a function of radius relative to the dynamical center of J0901. For each bin (with width $\Delta r_{3,1}=0.05$ and $\Delta r=0.25\,{\rm kpc}$), one of the eight red tones is assigned, starting at one pixel per bin, and in steps of three pixels per bin thereafter. We include all pixels regardless of their statistical significance. Radial positions account for the inclination of the source. We use the mean dynamical center, position angle, and inclination angle from the best-fit models in Table~\ref{tab:dynmodels}, omitting the model for the ${\rm H\alpha}$ data using a Gaussian flux profile since that model does not converge to sensible values. The black squares are the variance-weighted mean $r_{3,1}$ values for pixels in bins of $1\,{\rm kpc}$. Associated uncertainties are calculated from a bootstrap analysis (with replacement) in which we calculate the dispersion from the variance-weighted mean for $10^4$ iterations of the underlying \mbox{CO(1--0)} and \mbox{CO(3--2)} pixels, after randomly perturbing the pixels' fluxes in each iteration by their uncertainties as determined from the lens reconstructions. Since the pixels are correlated, adjascent binned average $r_{3,1}$ values are also correlated; however, the intensity-weighted average PSF's major axis FWHM is $\sim2\,{\rm kpc}$ (when tilted by J0901's inclination angle), so every other bin is approximately uncorrelated. The dashed line corresponds to the approximate peak value in the $r_{3,1}$ histogram for the reconstructed source as shown in Figure~\ref{fig:j0901ratiohisto} ($r_{3,1}=0.7$). The dotted line corresponds to $r_{3,1}=0$ for reference. \label{fig:radialratio}}
\end{figure}

For the source-plane reconstruction maps using the matched-resolution data, in Figure~\ref{fig:radialratio} we show $r_{3,1}$ as a function of the physical radius from J0901's dynamical center. Unlike the mapped values of $r_{3,1}$ in Figure~\ref{fig:j0901intlineratio}, we include all pixels, regardless of their statistical significance. In order to calculate each pixel's distance from the center, including inclination corrections, we use the mean dynamical center, position angle, and inclination angle from the best-fit models in Table~\ref{tab:dynmodels}, omitting the model for the ${\rm H\alpha}$ data using a Gaussian flux profile since that model does not converge to sensible values. The distribution of $r_{3,1}$ values decreases as a function of radius, which is clearest in the variance-weighted mean $r_{3,1}$ values calculated in bins of $1\,{\rm kpc}$. Since the pixels are correlated, the binned average $r_{3,1}$ values are also correlated. However, since the intensity-weighted average PSF's major axis FWHM (which approximately gives the resolution and thus correlation length of the data) is $\sim2\,{\rm kpc}$ when tilted by J0901's inclination angle, every other bin is approximately uncorrelated. By using the variance-weighted means in our radial bins, we can retrieve average values that are not biased by noise-dominated pixels that scatter to large $r_{3,1}$ or have unphysical negative $r_{3,1}$ values. The spatial distribution of line ratios in J0901 is consistent with a picture of multi-phase gas in which the bulk of the molecular ISM is in an extended cool/low-density phase, containing smaller embedded regions of gas in a warm/high-density phase \citep[e.\/g.\/,][]{ivison2011,thomson2012} that is somewhat more centrally concentrated.

\subsection{Spatial variation in metallicity}
\label{sec:metallicity}

\begin{figure*}
\plotone{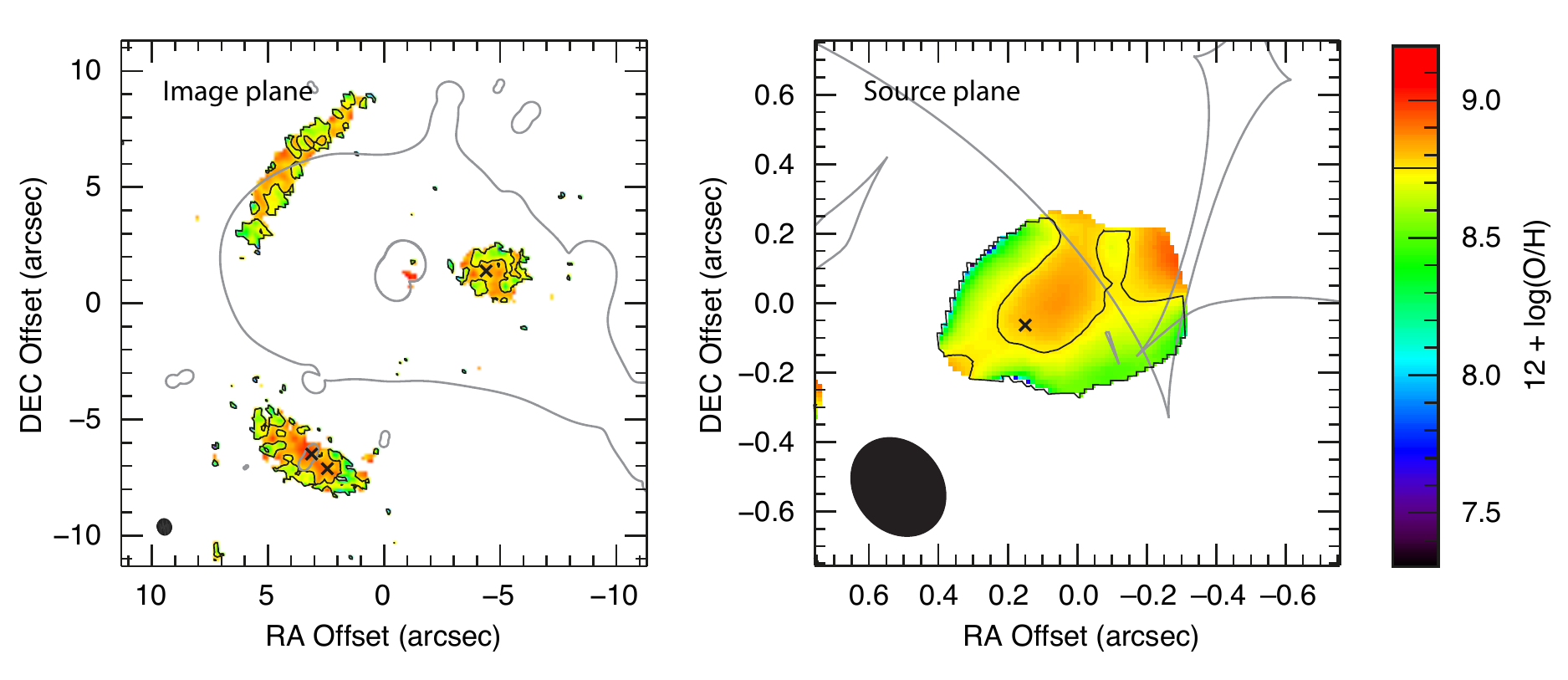}
\caption{Map of the metallicity as estimated from the [N\,{\sc ii}]/${\rm H\alpha}$ line ratio \citep{pettini2004} in both the image plane (left) and source-plane reconstruction (right). The black contour at $12+\log({\rm O/H})=8.75$ represents the upper limit on the range for which [N\,{\sc ii}]/${\rm H\alpha}$ accurately estimates the metallicity. Pixels with $<2\sigma$ significance in the ${\rm H\alpha}$ line have been blanked out (excluding $<2\sigma$ significance pixels in the image plane [N\,{\sc ii}] map would remove nearly all pixels below $12+\log({\rm O/H})=8.75$). PSFs are shown at lower left. Gray lines indicate the image-plane lensing critical curves or source-plane caustics (Section~\ref{sec:lensing}). Black crosses mark the mean dynamical center determined from the source-plane reconstructions and lens modeling (see Section~\ref{sec:masses}). \label{fig:j0901metal}}
\end{figure*}

Using the [N\,{\sc ii}] and ${\rm H\alpha}$ maps, we also examine spatial variations in the metallicity of J0901. We estimate the metallicity using

\begin{equation}\label{eq:metallicity}
12+\log({\rm O/H}) = 8.90+0.57\log(\text{[N\,{\sc ii}]}/{\rm H\alpha})
\end{equation}

\noindent
from \citet{pettini2004}, which is valid for $7.5>12+\log({\rm O/H})>8.75$ (using 8.66 as the solar abundance; \citealt{asplund2004}). In our map of the metallicity (Figure~\ref{fig:j0901metal}) we blank out any pixels with $<2\sigma$ significance in the ${\rm H\alpha}$ map. We find that a substantial fraction of the source has $12+\log({\rm O/H})$ values larger than the range where the [N\,{\sc ii}]/${\rm H\alpha}$ accurately traces the metallicity (although it has been suggested that at high redshift, the threshold at which the [N\,{\sc ii}]/${\rm H\alpha}$ ratio becomes affected by the AGN is higher; e.\/g.\/, \citealt{kewley2013a,kewley2013b}); the average pixelized value is $12+\log({\rm O/H})=8.73\pm0.21$ vs.~ $12+\log({\rm O/H})=7.3\pm1.1$ calculated from the ratio of the total luminosities (without magnification correction). Larger values of [N\,{\sc ii}]/${\rm H\alpha}$ cannot be produced in the photoionization regions of massive stars, indicating potential heating or shocked excitation by a central AGN or its winds \citep[e.\/g.\/,][]{baldwin1981,kauffman2003}. The high central [N\,{\sc ii}]/${\rm H\alpha}$ ratio seen in the source-plane reconstruction, least-distorted western image, and southern image is in line with previous evidence of an AGN in J0901 \citep{hainline2009, diehl2009, genzel2014}. However, we note that the average pixelized metallicity is also much closer to the metallicity predicted by the mass-metallicity relation for high-$z$ galaxies, which implies $12+\log({\rm O/H})=8.5$--$8.7$ for J0901's new magnification-corrected stellar mass (depending on which relation we use; \citealt{genzel2012,wuyts2014,sanders2018}).

\begin{figure*}
\plotone{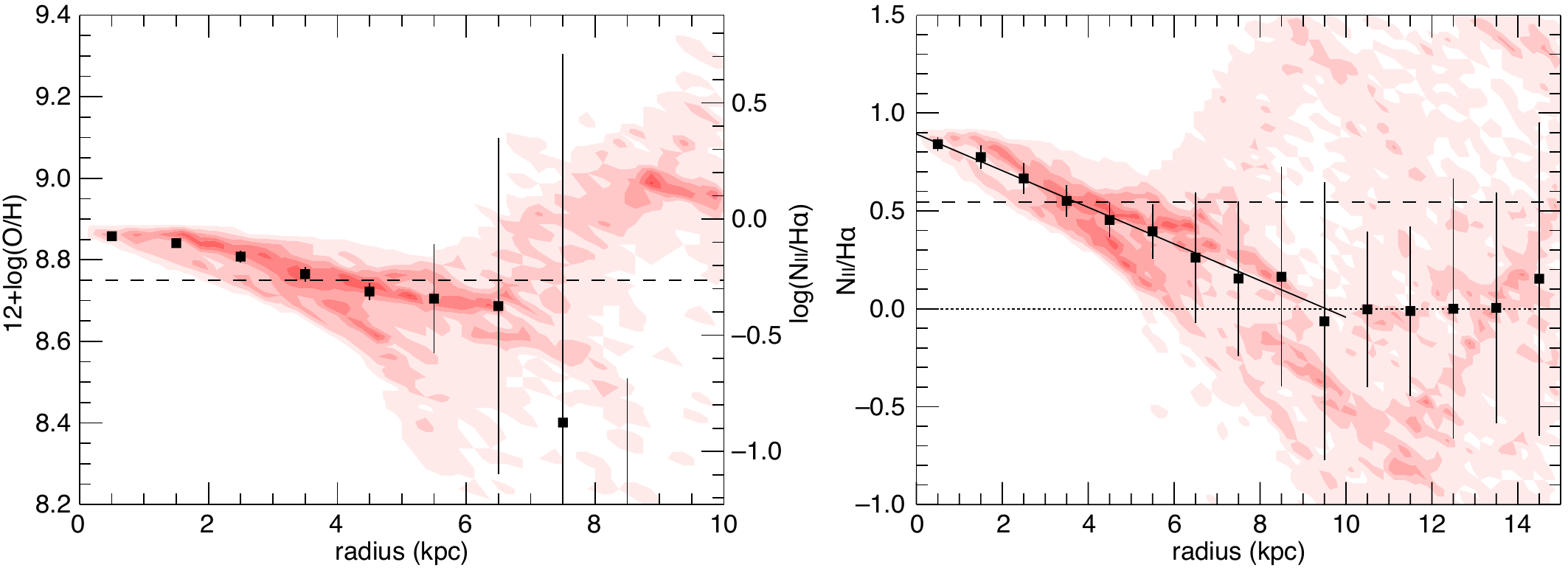}
\caption{The metallicity (or log([N\,{\sc ii}]/H$\alpha$); left) and [N\,{\sc ii}]/H$\alpha$ ratio (right) for individual pixels in the matched-resolution source-plane reconstructions as a function of radius relative to the dynamical center of J0901. For each bin (with width $\Delta r=0.25\,{\rm kpc}$ and either $\Delta Z=0.025$ or $\Delta$([N\,{\sc ii}]/H$\alpha$)$=0.05$), one of the six (left) or five (right) red tones is assigned, starting at one pixel per bin, and in steps of three pixels per bin thereafter. We include all pixels regardless of their statistical significance. Radial positions account for the inclination of the source. We use the mean dynamical center, position angle, and inclination angle from the best-fit models in Table~\ref{tab:dynmodels}, omitting the model for the ${\rm H\alpha}$ data using a Gaussian flux profile since that model does not converge to sensible values. The black squares are the variance-weighted mean values for pixels in bins of $1\,{\rm kpc}$. Associated uncertainties are calculated from a bootstrap analysis (with replacement) in which we calculate the dispersion from the variance-weighted mean for $10^4$ iterations of the underlying ${\rm H\alpha}$ and [N\,{\sc ii}] pixels; the pixels' fluxes in each iteration are randomly perturbed by their uncertainties as determined from the lens reconstructions. Since the pixels are correlated, adjascent binned average values are also correlated; the intensity-weighted average PSF's major axis FWHM is $\sim2\,{\rm kpc}$ (when tilted by J0901's inclination angle), so every other bin is approximately uncorrelated. The dashed lines correspond to the value above which the [N\,{\sc ii}]/H$\alpha$ ratio is no longer believed to be an accurate tracer of the metallicity (at least in the local universe). For the right panel, we also show [N\,{\sc ii}]/H$\alpha=0$ for clarity (dotted line; negative values are caused by noise), and the best-fit linear relation for $r\leq10\,{\rm kpc}$ (solid line). \label{fig:j0901metalrad}}
\end{figure*}

Caveats on the validity of using [N\,{\sc ii}]/${\rm H\alpha}$ to trace metallicity aside, in Figure~\ref{fig:j0901metalrad}, we examine the radial decrease in metallicity in more detail. Like the radial $r_{3,1}$ plot, we calculate $12+\log({\rm O/H})$ for each pixel in the matched-resolution source-plane reconstructions regardless of SNR. We calculate each pixel's radial distance from the average dynamical center, corrected for inclination angle, using the best-fit models in Table~\ref{tab:dynmodels} (again, omitting the model for the ${\rm H\alpha}$ data using a Gaussian flux profile). While there is a weak radial gradient in metallicity out to $\sim5\,{\rm kpc}$, any trends at larger radii are lost in the noise. However, the roughly linear radial gradient in [N\,{\sc ii}]/${\rm H\alpha}$ (rather than its log $\leftrightarrow$ the metallicity) may extend to $\sim10\,{\rm kpc}$ with a slope of $\sim-0.1\,{\rm kpc^{-1}}$ (from a linear best-fit to the binned values with no correction for beam smearing). The radial metallicity gradient of $\sim-0.03\,{\rm dex\,kpc^{-1}}$ (from a linear best-fit to the binned values with $r\leq5\,{\rm kpc}$ and no correction for beam smearing) is on the flatter end of (albeit consistent with) the distribution for disk galaxies in the local universe \citep[e.\/g.\/,][]{rupke2010b}. However, high-redshift galaxies appear to have a wide range of metallicity gradients \citep[e.\/g.\/,][and references therein]{wuyts2016a}, within which J0901 falls, making the physical interpretation of the gradient difficult even without accounting for the potential influence of the central AGN.

\subsection{Spatially resolved Schmidt-Kennicutt relation}
\label{sec:KSlaw}

\subsubsection{Methods}

We examine the spatially resolved Schmidt-Kennicutt relation \citep{schmidt1959,kennicutt1998} for J0901 using the ${\rm H\alpha}$ and CO maps smoothed to the same spatial resolution. We use the $L_{\rm H\alpha}$-SFR conversion factor given in \citet{kennicutt2012}, which assumes a \citet{kroupa2001} initial mass function (IMF). The ${\rm H\alpha}$ brightness has \emph{not} been corrected for extinction. Properly accounting for spatially varying extinction can significantly affect the slope of the Schmidt-Kennicutt relation \citep{genzel2013}, but our current dust continuum data lack the spatial resolution for us to effectively perform such a correction. A global correction for the extinction (as in \citealt{sharon2013}) would simply offset the relation to higher SFR surface densities (discussed further below).

In order to fit the Schmidt-Kennicutt relation in J0901 to a power law, we follow the methodology of \citet{blanc2009} and \citet{leroy2013} and perform a Bayesian analysis, since standard orthogonal least squares regression fits are biased by clipping of the molecular gas and star formation surface densities at a chosen significance level. While the full methodology is presented in \citet{blanc2009} and \citet{leroy2013}, in short, we iteratively calculate the SFR surface density for a random sample (with replacement) of the observed pixelized molecular gas surface densities in J0901 for a grid of potential normalization factors ($A$), indices ($n$), and intrinsic scatter values ($\sigma$) in the equation

\begin{equation}\label{eq:SKlaw}
\frac{\Sigma_{\rm SFR}}{1\,M_\sun\,{\rm yr^{-1}\,kpc^{-2}}}=A\left(\frac{\Sigma_{\rm gas}}{1000\,M_\sun\,{\rm pc^{-2}}}\right)^{n}\times10^{{\mathcal N}(0,\sigma)},
\end{equation}

\noindent
where ${\mathcal N}(0,\sigma)$ is a normal distribution with mean zero and standard deviation $\sigma$. For each possible combination of Schmidt-Kennicutt relation parameters, we grid the resulting model values of $\Sigma_{\rm SFR}$ and $\Sigma_{\rm gas}$ and compare them to a grid of the measured values to calculate a $\chi^2$ value. As in \citet{leroy2013}, we apply a $2\sigma$ cut in gas mass surface density before comparing the grids of the observed and model data points in order to define a clear y-axis; since this cut is applied \emph{after} the data are simulated, it does not bias the selection of the best-fit model in the same way as more conventional linear fitting algorithms. For each of the three Schmidt-Kennicutt parameters, we fit polynomials to the shape of their $\chi^2$ values (taking the minimum $\chi^2$ along the complementary parameters' axes, collapsing the model grid to a distribution of $\chi^2$ values for each parameter separately), and use the polynomials' minima as the best-fit values of the parameters. We then perform this comparison multiple times, each time removing a pixel at random, perturbing the grid on which we compare the source and model, and perturbing the emission for both tracers by both the additive statistical uncertainty (on a per-pixel basis) and the multiplicative flux calibration uncertainty (applied to all pixels). Our best-fit values for the Schmidt-Kennicutt relation and their uncertainties (both statistical and systematic) are given by the mean and standard deviation of the resulting distribution of fitted parameters. 

In addition to the different denominator of Eqn.~\ref{eq:SKlaw} (which must be accounted for in comparisons to other Schmidt-Kennicutt studies and is chosen to reduce the fitting covariance), our implementation of the algorithm differs from that of \citet{blanc2009} and \citet{leroy2013} in the following ways: (1) We randomly draw $10^4$ values of $\Sigma_{\rm gas}$ for calculating model values of $\Sigma_{\rm SFR}$ and allow repeats, but we only perform the iterative fitting routine for 100 perturbations of the model/source (due to computational/time limits). (2) Since our $\Sigma_{\rm gas}$ and $\Sigma_{\rm SFR}$ uncertainties are both dominated by measurement uncertainties, we additively perturb both the model gas and SFR surface densities. (3) We sample the Schmidt-Kennicutt parameters in $\Delta \log(A)=0.03$ from $-1.25\geq\log(A)\geq-0.5$, $\Delta n=0.05$ from $0.8\geq n\geq2.3$, and $\Delta\sigma=0.015$ from $0\geq\sigma\geq0.3$. (4) We assume a $10\%$ flux calibration uncertainty for the ${\rm H\alpha}$ and CO data. (5) Our minimum $\chi^2$ curves are fit by a third order polynomial rather than a second order polynomial as in \citet{leroy2013} in order to better fit our skewed $\chi^2$ curves.

\subsubsection{Results for J0901}

\begin{figure*}
\epsscale{1.2}
\plotone{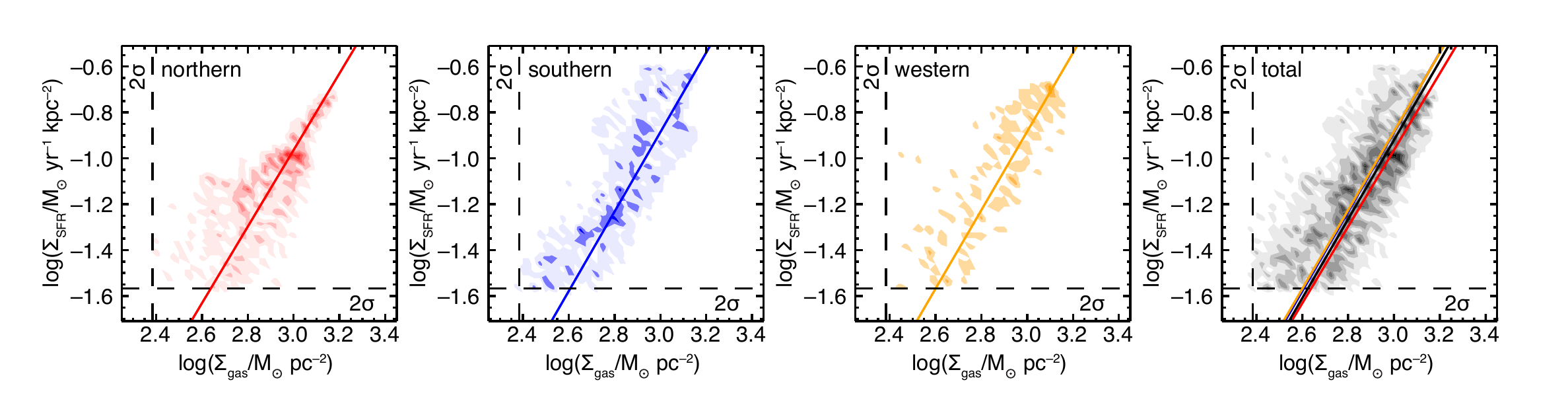}
\caption{Star formation rate surface density as measured by ${\rm H\alpha}$ surface brightness (uncorrected for extinction) vs.~\mbox{CO(1--0)}-determined molecular gas mass surface density of J0901 (using the natural resolution/weighted data). From left to right, the four panels plot the density of pixels in 0.025 dex bins in gas mass and SFR surface density for the northern image (red), southern image (blue), western image (gold), and all images combined (gray). The color tones start at one pixel per bin and are in steps of two pixels per bin thereafter; there are six, three, three, and seven color steps in the northern, southern, western, and total panels, respectively. The two dashed lines mark $2\sigma$ surface brightness cuts, but only the gas surface density cut was applied during the linear fit (as described in the text). In the rightmost plot we include the linear fits for the individual images for easier comparisons; note that the blue line for the southern image falls nearly directly under the gold line for the western image. \label{fig:j0901kslaw10}}
\end{figure*}

\begin{figure*}
\epsscale{1.2}
\plotone{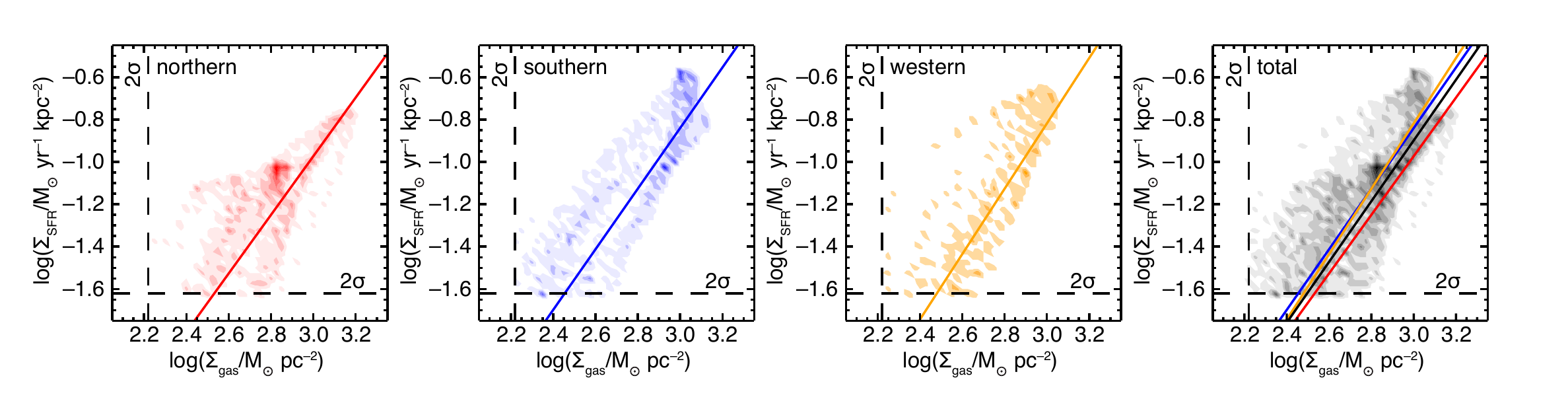}
\caption{Star formation rate surface density as measured by ${\rm H\alpha}$ surface brightness (uncorrected for extinction) vs.~\mbox{CO(3--2)}-determined molecular gas mass surface density of J0901 (using the natural resolution data). All lines and colors are as described in Figure~\ref{fig:j0901kslaw10}, but there are seven, six, three, and eight steps in the northern, southern, western, and total panels, respectively, and the axis ranges differ as well. \label{fig:j0901kslaw32}}
\end{figure*}

\begin{figure}
\plotone{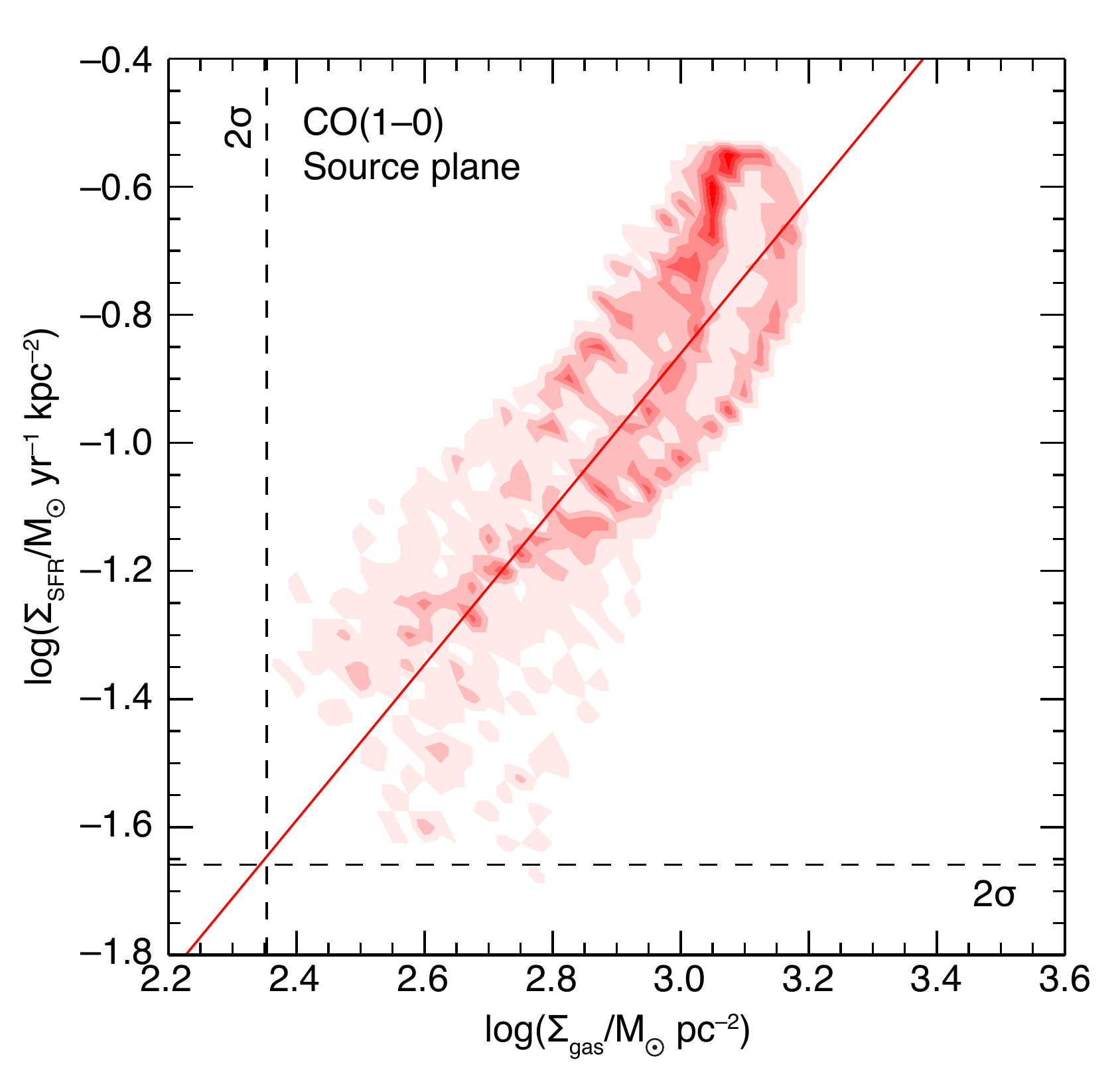}
\plotone{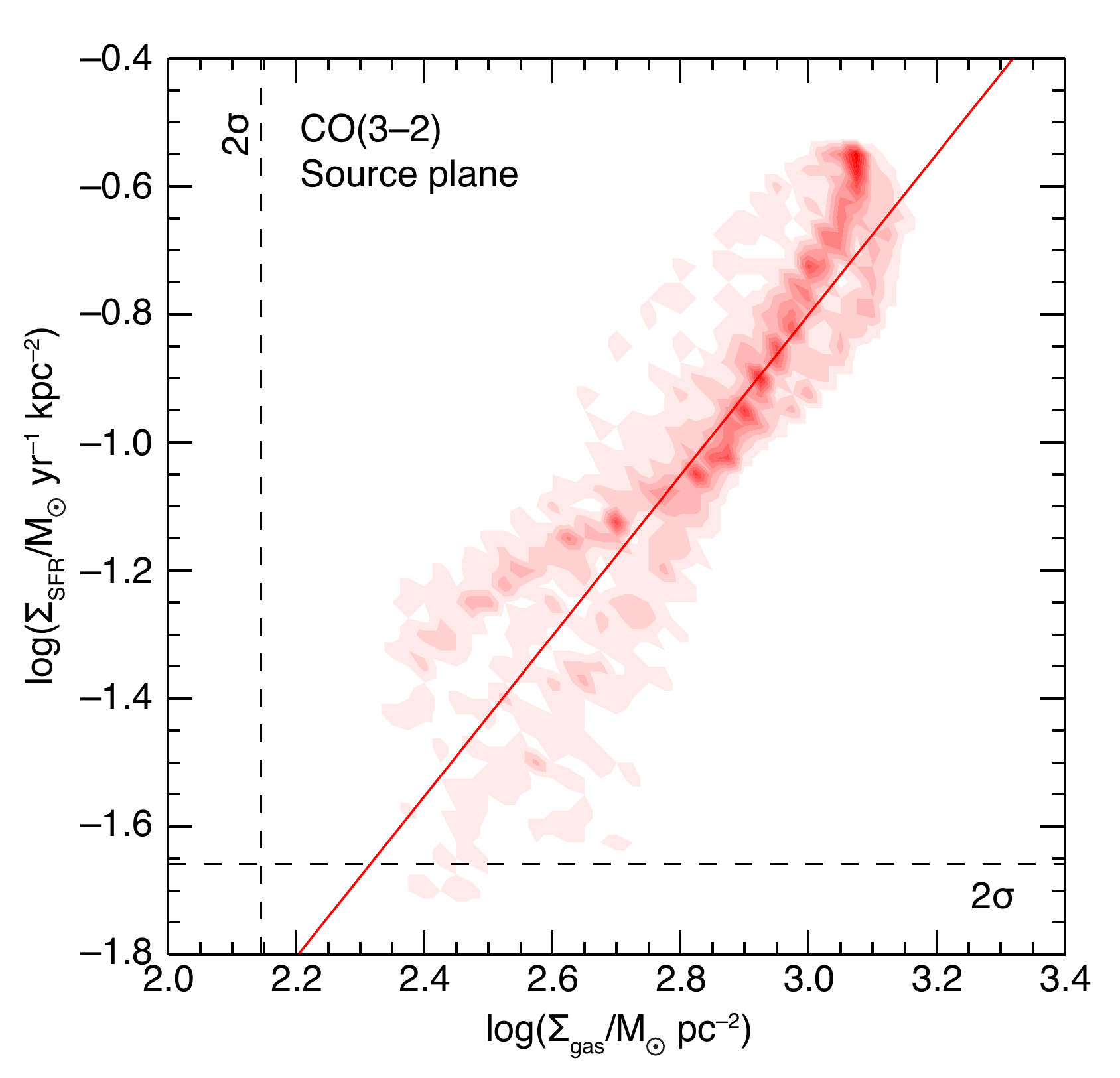}
\caption{Star formation rate surface density as measured by ${\rm H\alpha}$ surface brightness (uncorrected for extinction) vs.~\mbox{CO(1--0)}-determined molecular gas mass surface density (top) and \mbox{CO(3--2)}-determined molecular gas mass surface density (bottom) using the de-lensed CO images of J0901 derived from maps with matched beams/PSFs and inner $uv$ radii. For each $0.025\,{\rm dex}$ bin in gas mass and SFR surface density, one of six or ten red tones is assigned (for the upper or lower panels, respectively), starting at one pixel per bin, and in steps of two pixels per bin thereafter. The two dashed lines mark $2\sigma$ surface brightness cuts applied to the image-plane data, which are not applied here since pixels with at least $2\sigma$ significance correspond to different surface brightnesses in the de-lensed data. \label{fig:kslawdelens}}
\end{figure}

In Figures~\ref{fig:j0901kslaw10} and \ref{fig:j0901kslaw32} we show the Schmidt-Kennicutt relation using the native resolution integrated \mbox{CO(1--0)} and \mbox{CO(3--2)} maps. In order to determine whether differential lensing affects the observed Schmidt-Kennicutt relation in J0901, we analyze each image of J0901 separately and all three images combined. We also compare these results to a Schmidt-Kennicutt analysis using the matched maps (not shown) and the source-plane reconstructions of the matched maps for both CO lines (Figure~\ref{fig:kslawdelens}). Table~\ref{tab:sk} lists the best-fit parameters of the Schmidt-Kennicutt relation in Equation~\ref{eq:SKlaw}. The power law fits are roughly consistent with super-linear indices of $n\approx1.5$ for both CO transitions, with a mean value of $\bar n=1.54\pm0.13$, although individual fits for the image plane analyses range from $n=1.38$--$1.73$. 

\begin{deluxetable*}{llrrccc}
\tablewidth{0pt}
\tablecaption{J0901 Schmidt-Kennicutt fit parameters \label{tab:sk}}
\tablehead{ \colhead{Image} & \colhead{line} & \colhead{$N_{pix}$} & \colhead{$N_{ind}$} & \colhead{$A$} & \colhead{$n$} & \colhead{$\sigma$}}
\startdata
North & CO(1--0) & $1891$ & $23.5$ & $-0.96\pm0.08$ & $1.67\pm0.06$ & $0.21\pm0.02$ \\
{} & CO(1--0)$_{m}$ & $1785$ & $17.7$ & $-0.97\pm0.06$ & $1.40\pm0.08$ & $0.24\pm0.01$ \\
{} & CO(3--2) & $2278$ & $24.1$ & $-0.97\pm0.06$ & $1.38\pm0.04$ & $0.23\pm0.01$ \\
{} & CO(3--2)$_{m}$ & $2366$ & $22.1$ & $-0.96\pm0.07$ & $1.41\pm0.04$ & $0.23\pm0.01$ \\
\hline
South & CO(1--0) & $1478$ & $18.4$ & $-0.89\pm0.07$ & $1.73\pm0.08$ & $0.23\pm0.01$ \\
{} & CO(1--0)$_{m}$ & $1570$ & $14.7$ & $-0.86\pm0.07$ & $1.70\pm0.09$ & $0.24\pm0.01$ \\
{} & CO(3--2) & $1757$ & $18.6$ & $-0.84\pm0.05$ & $1.43\pm0.04$ & $0.22\pm0.01$ \\
{} & CO(3--2)$_{m}$ & $1851$ & $17.3$ & $-0.83\pm0.07$ & $1.42\pm0.04$ & $0.21\pm0.01$ \\
\hline
West & CO(1--0) & $858$ & $10.7$ & $-0.88\pm0.08$ & $1.71\pm0.10$ & $0.24\pm0.02$ \\
{} & CO(1--0)$_{m}$ & $793$ & $7.4$ & $-0.82\pm0.07$ & $1.56\pm0.10$ & $0.21\pm0.02$ \\
{} & CO(3--2) & $954$ & $10.1$ & $-0.82\pm0.07$ & $1.54\pm0.07$ & $0.26\pm0.01$ \\
{} & CO(3--2)$_{m}$ & $982$ & $9.2$ & $-0.80\pm0.08$ & $1.51\pm0.07$ & $0.25\pm0.01$ \\
\hline
Total & CO(1--0) & $4227$ & $52.6$ & $-0.92\pm0.07$ & $1.72\pm0.06$ & $0.23\pm0.01$ \\
{} & CO(1--0)$_{m}$ & $4148$ & $38.8$ & $-0.91\pm0.05$ & $1.55\pm0.08$ & $0.24\pm0.01$ \\
{} & CO(3--2) & $4989$ & $52.8$ & $-0.90\pm0.04$ & $1.43\pm0.03$ & $0.239\pm0.004$ \\
{} & CO(3--2)$_{m}$ & $5199$ & $48.6$ & $-0.89\pm0.06$ & $1.44\pm0.03$ & $0.237\pm0.004$ \\
\hline
De-lensed & CO(1--0)$_{m}$ & $1688$ & $15.8$ & $-0.86\pm0.06$ & $1.22\pm0.03$ & $0.18\pm0.01$ \\
{} & CO(3--2)$_{m}$ & $1673$ & $15.7$ & $-0.80\pm0.06$ & $1.25\pm0.03$ & $0.13\pm0.01$ \\
\enddata
\tablecomments{The lines with subscript $m$ use the matched maps with the same smoothed resolution and the same inner $uv$ radius. $N_{pix}$ lists the total number of pixels used in the fitting procedure, which are not all independent from one another due to the beam/PSF size, and $N_{ind}$ lists the number of independent resolution elements (beams/PSFs) to which those pixels correspond.}
\end{deluxetable*}

We note that in Figures~\ref{fig:j0901kslaw10}--\ref{fig:kslawcompare}, we show much smaller $\Sigma_{\rm SFR}$ and $\Sigma_{\rm gas}$ bins in our Schmidt-Kennicutt plots ($0.025\,{\rm dex}$) than we use in the fitting analysis ($0.05$--$0.2\,{\rm dex}$, randomly assigned for each iteration of the Schmidt-Kennicutt fitting routine). These smaller bins show some structure in the pixel densities due to differences in the brightness distributions of our ${\rm H\alpha}$ and CO maps that are amplified by the beams/PSFs (causing neighboring pixels to be correlated and blending real source structure together, in some cases causing misaligned peaks of emission). The pixel density structures are particularly apparent in the analysis of the source plane reconstructed images since correlations in the observed images can be warped and exaggerated by the de-lensing process. The larger bins we use in the fitting analysis (plus the Monte Carlo perturbations in the model realizations) do not show these structures, and therefore these structures do not bias our power-law fits. The numbers of correlated pixels and corresponding numbers of independent beams/PSFs used in the fits are listed in Table~\ref{tab:sk} as $N_{pix}$ and $N_{ind}$, respectively.

The smoothing and inner-radius $uv$ clipping used to create the ``matched" dataset lowers the best-fit Schmidt-Kennicutt index for the \mbox{CO(1--0)} line only (except for the southern image where the index is unchanged). We suspect that the best-fit index for the \mbox{CO(3--2)} data is unchanged between the ``native" and ``matched" resolution maps because considerably less smoothing (and no $uv$ clipping) is required to create its matched map. Therefore, while the \mbox{CO(1--0)} line might better represent the true distribution of the total molecular gas mass, one must compare the matched maps in order to make fair comparisons between how the choice of molecular gas tracer affects the Schmidt-Kennicutt relation.

We find significant differences between the Schmidt-Kennicutt indices determined from the image-plane data and from the source-plane reconstructions; the best-fit indices for the source-plane reconstructions are much lower with $\bar n=1.24\pm0.02$. It is difficult to explain this difference since lensing conserves surface brightness. We do expect the distributions of pixels in the parameters space of SFR vs.~gas mass to change due both to the existence of multiple images of the same region within the source and to the larger number of samples per region within the source given the uniform image-plane pixel size. These effects should be particularly strong for the northern and southern images since they cross critical curves where the magnification is highest. The western image is the least distorted of the three images, and although it suffers from the lowest S/N, it shows a similar distribution of pixels in the parameters space of SFR vs.~gas mass to the source-plane reconstruction: the highest concentration of pixels is at the highest SFR and gas mass surface densities. The lack of difference in the Schmidt-Kennicutt index between the three images despite these lensing effects is reassuring since it indicates that Schmidt-Kennicutt relation does not change much between different regions within J0901 (at least at the spatial resolutions probed here). The similar pixel distribution for the western image and source plane reconstruction is similarly reassuring since it indicates that our lens modeling is working correctly. However, neither of these points explains why the best-fit index differs between the image plane and source plane analyses. 

We note that there is one significant methodological difference between the image-plane and source-plane Schmidt-Kennicutt analyses that does influence the best-fit index. The source-plane reconstructions for the ${\rm H\alpha}$ data are more compact than for the CO data, and thus the gas and SFR surface densities are largely uncorrelated beyond the edges of the ${\rm H\alpha}$ emission. The uncorrelated pixels, if left in the Schmidt-Kennicutt analysis, result in indices of $n\sim2$, which is even steeper than found in the image plane. Clipping the data at a fixed value of the gas mass or SFR surface density to remove uncorrelated data would both bias the fits and leave too few pixels to support a robust fit. Instead, we only include pixels that have SNR$\geq2$ for both the gas and SFR maps. Due to the spatially varying noise in the source-plane reconstructions, this significance cut does not result in a constant surface brightness cut. We re-derived the fits with SNR cuts between 2 and $4\sigma$ and found no difference in the best-fit values of the Schmidt-Kennicutt parameters, so we believe this method is reliable for the source-plane reconstructions. Therefore we do not think this methodological difference causes the difference in the Schmidt-Kennicutt index between the image and source planes.

One potential way to resolve the difference in the Schmidt-Kennicutt index between the source and image planes would be to observe J0901 at higher angular resolution. While the fit to the western image is determined by a large number of pixels ($\sim800$--$1000$), those pixels only correspond to a small number of independent resolution elements ($\lesssim10$). If the western image is expected to better represent the image-plane structure and have a different distribution of pixels in the parameter space of SFR vs.~gas mass relative to the other images, then more high-SNR independent data points would be helpful. While our calculated uncertainties for the best-fit correlations fold in the effects of excluding individual pixels, they do not fold in the effects of excluding entire resolution elements. Therefore, the uncertainties quoted in Table~\ref{tab:sk} are likely underestimates of the true uncertainties, particularly for the western image.

We also see a significant difference in the observed scatter of the Schmidt-Kennicutt relation, $\sigma$, between the source-plane and image-plane data. While the uncertainty in the scatter is likely an underestimate as described above, we suspect that the difference in the scatter is an artifact of the lens modeling noise scaling and regularization. As discussed in Section~\ref{sec:lensing}, we scale up the assumed image noise during the lens modeling in order to allow the code to under-fit the flux distributions and thus minimize the effects of the correlated noise in the interferometric data. Since this process increases the regularization strength, it effectively smooths the source-plane emission, and thus increases the source-plane SNR while decreasing the scatter of the SFR and gas mass surface densities.

Lastly, we find no significant systematic differences in the best-fit indices between the matched datasets for the two CO lines using either the image-plane data or the source-plane reconstructions, except in the case of the southern image. However, given that the native resolution \mbox{CO(1--0)} and \mbox{CO(3--2)} data produce significantly different indices ($n\sim1.7$ using the \mbox{CO(1--0)} data and $n\sim1.4$ for the \mbox{CO(3--2)} data) and that the source-plane reconstruction produce yet a different index ($n\sim1.2$), the potential differences between indices for the two CO lines are inconclusive. Since we apply a global excitation correction in determining $\Sigma_{gas}$ based on our measured $r_{3,1}$ values, we find consistent offsets ($A$) in the linear fits between the two CO transitions, regardless of map (native, matched, or reconstructed) or lens-plane image (north, south, west, or total).

\subsubsection{Comparisons to other spatially resolved galaxies and important caveats}

\begin{figure*}
\epsscale{0.85}
\plotone{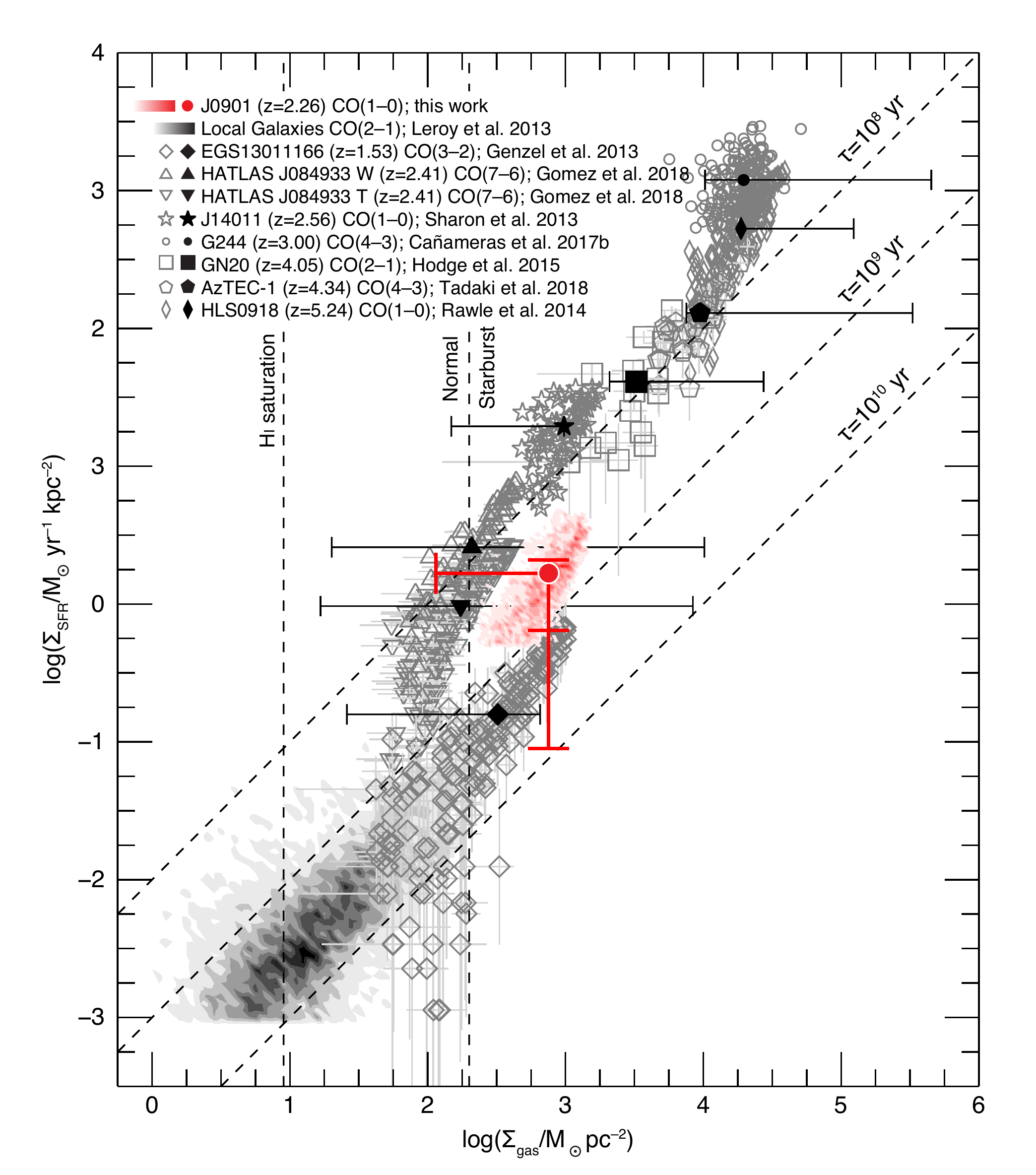}
\caption{Comparison between the star formation rate surface density and gas mass surface density for high-redshift galaxies with resolved pixel-by-pixel analyses: the modestly lensed SMG J14011 (stars; \citealt{sharon2013}), the $z=1.5$ normal disk galaxy EGS13011166 (wide diamonds with error bars; \citealt{genzel2013}), the strongly lensed SMG HLS0918 (narrow diamonds; \citealt{rawle2014}), the SMG GN20 (squares with error bars; \citealt{hodge2015}), the strongly lensed SMG G244 (circles; \citealt{canameras2017b}), the SMG AzTEC-1 (pentagons with error bars; \citealt{tadaki2018}), and the two components of the HyLIRG HATLAS\,J084933 (upward and downward pointing triangles with error bars; \citealt{gomez2018}). The red region shows the density of pixels for all three images of J0901 using the natural resolution \mbox{CO(1--0)} data, but the SFR has been scaled to match the TIR-derived SFR. The gray shaded region shows the \cite{leroy2013} sample of local galaxies. For all high-redshift galaxies, the star formation rates have been converted to the  Kroupa initial mass function (the same IMF as used for the local sample). However, we respect the different authors' choices of $\alpha_{\rm CO}$, and instead show how the locus of points (centered at the mean surface density; black/red symbols) would translate for $0.7\le\alpha_{\rm CO}\le4.6$ using the horizontal lines. Since the molecular gas for GN20, EGS13011166, G244, AzTEC-1, and HATLAS\,J084933 was not observed in the \mbox{CO(1--0)} line, we include an additional excitation uncertainty for a range of possible line ratios, using that of the Milky Way as a lower limit ($r_{2,1}=0.50$, $r_{3,1}=0.26$, $r_{4,1}=0.15$, $r_{7,1}=0.015$; \citealt{fixsen1999}), and thermalized excitation as an upper limit ($r_{2,1}=r_{3,1}=r_{4,1}=r_{7,1}=1.0$). For J0901, the vertical bar denotes how the SFR surface density would scale for different global extinction corrections. From bottom to top, we mark the average SFR surface density if the SFR was determined from the ${\rm H}\alpha$ data without extinction correction (as in Fig.~\ref{fig:j0901kslaw10}), from the ${\rm H}\alpha$ luminosity corrected to include obscured star formation traced by the TIR luminosity following \citet{kennicutt2012}, from the TIR luminosity only (current location; red circle), and from the ${\rm H}\alpha$ luminosity corrected for extinction using the ${\rm H}\alpha$/${\rm H}\beta$ value from \citet{hainline2009}. Dashed lines are as in \citet{bigiel2008}, and include diagonal lines of constant SFE (or the inverse of the gas consumption timescale), the threshold at which atomic gas converts entirely to molecular gas (left vertical line), and a proposed threshold for the transition between ``normal" and ``starburst" modes of star formation (right vertical line; \citealt{bigiel2008}).
\label{fig:kslawcompare}}
\end{figure*}

In Figure~\ref{fig:kslawcompare} we compare the results for J0901 to other high-redshift galaxies in which resolved pixel-by-pixel Schmidt-Kennicutt analyses have been performed. Only eight high-redshift galaxies besides J0901 have been analyzed on a pixel-by-pixel basis on the Schmidt-Kennicutt relationship: seven SMGs and one normal disk galaxy (\citealt{sharon2013, genzel2013, rawle2014, hodge2015, canameras2017b, tadaki2018, gomez2018}; but see also \citealt{freundlich2013,sharda2017}). We also compare J0901 to the sample of local disk galaxies from \citet{leroy2013}. For these comparisons, we convert all measurements to the same \citep{kroupa2001} initial mass function. The position of J0901 (or any galaxy) on the star formation relation strongly depends on the assumed value of the CO-to-${\rm H_2}$ conversion factor. While $\alpha_{\rm CO}$ is still uncertain for high-redshift galaxies, different authors' choices in CO-to-${\rm H_2}$ conversion factors are often justified based on metallicity or dynamical arguments. Therefore we do not correct gas measurements to the same $\alpha_{\rm CO}$, but instead show a horizontal bar to indicate how gas mass surface density measurements would scale for the range of possible conversion values ($0.7\leq\alpha_{\rm CO}\leq4.6$). For GN20, EGS13011166, PLCK G244.8+54.9, and HATLAS\,J084933, where the CO measurements were not made for the $J=1$--$0$ transition, we include additional factors in this rough systematic molecular gas uncertainty to account for the unknown gas excitation; we allow the ratios to the ground state to be as low as those found for the Milky Way ($r_{2,1}=0.50$, $r_{3,1}=0.26$, $r_{4,1}=0.15$, or $r_{7,1}=0.015$; \citealt{fixsen1999}) and as high as thermalized ($r_{2,1}=r_{3,1}=r_{4,1}=r_{7,1}=1.0$).

One significant source of uncertainty in the SFE for J0901 is that we do not have comparable spatial resolution tracers of obscured star formation. Our dust map from the SMA is not sufficiently resolved to map local variations of the dust column, and we do not have ${\rm H}\beta$ maps or enough resolved multi-band optical/UV data to perform a spatially resolved SED analysis as in \citet{genzel2013}. In Figure~\ref{fig:kslawcompare}, the contours are for star formation traced by ${\rm H}\alpha$ but scaled to account for the total SFR as inferred from $L_{\rm TIR}$; this rescaling is analogous to applying a uniform \emph{global} extinction correction (the same technique used to correct the SFR surface density for J14011 in \citealt{sharon2013}). The obscured star formation as probed by the total infrared luminosity is significantly higher than the ${\rm H}\alpha$-traced star formation, and moves J0901 to higher star formation efficiencies, making J0901 significantly offset from the Schmidt-Kennicutt relation found for local normal disk galaxies. For comparison, we also show how the locus of points would change for different extinction corrections to the ${\rm H\alpha}$-derived SFR surface density in Fig.~\ref{fig:kslawcompare}, including no extinction corrections, the total infrared-corrected ${\rm H}\alpha$ emission to measure the SFR as in \citet{kennicutt2012}, and the global extinction value calculated from the ${\rm H}\alpha$/${\rm H}\beta$ ratio in \citet{hainline2009} (using the standard assumption of Case B recombination). Using the total infrared-corrected ${\rm H}\alpha$ emission to measure the SFR moves J0901 to higher star formation efficiencies, but not as high as using $L_{\rm TIR}$ alone. The global extinction value calculated from the ${\rm H}\alpha$/${\rm H}\beta$ ratio is highly uncertain since the ${\rm H}\beta$ line was coincident with a sky line; however, using this line ratio to correct for extinction (and thus obscured star formation) produces a large SFR, comparable to that calculated using the infrared. 

Global extinction corrections preserve the index of the Schmidt-Kennicutt relationship, but different extinction laws and patchy/localized extinction could significantly change the correlation's slope. \citet{genzel2013} find that the index of the Schmidt-Kennicutt relation for EGS13011166 varied between $0.8\leq n\leq1.7$ for the different extinction corrections they explore. Such extinction corrections are particularly challenging for starburst SMGs where nearly all of the star formation is expected to be obscured. For GN20, HLS0918, AzTEC-1, and J084933, their SFR surface densities are inferred from maps of the continuum emission near the peak of the dust SED ($\sim 170\,{\rm \mu m}$ rest frame) scaled to their $L_{\rm TIR}$-determined SFRs \citep{rawle2014,hodge2015}; a similar $L_{\rm TIR}$-determined SFR scaling is used to infer $\Sigma_{\rm SFR}$ for G244, but they scale continuum emission from further down the Rayleigh-Jeans tail of the dust SED ($\sim750\,{\rm \mu m}$ rest frame; \citealt{canameras2017b}), which may better trace dust mass than the SFR. Accounting for additional obscured star formation in J0901 (or unobscured star formation in the case of SMGs) may change the index of the Schmidt-Kennicutt relation, but J0901 would remain at elevated SFE relative to the local relation (regardless of the choice of CO-to-${\rm H_2}$ conversion factor). Resolved dust maps or extinction corrections are necessary to more firmly pin down the index of the Schmidt-Kennicutt relationship for UV-bright high-redshift galaxies and for J0901 specifically.

A second source of uncertainty in the position of J0901 relative to the star formation law is that we do not correct for contaminating AGN emission. While \citet{fadely2010} determine that the AGN in J0901 is not a significant contributor to its FIR luminosity, the AGN may contribute to the ${\rm H\alpha}$ luminosity used in our pixelized analysis. If the AGN is producing ${\rm H\alpha}$ emission in excess of that expected from star formation (\citet{genzel2014} suggest the AGN is responsible for $\sim10\%$ of the ${\rm H\alpha}$ emission), then some regions of J0901 would have an unexpectedly large SFE, which could either globally bias the $\Sigma_{\rm SFR}$-$\Sigma_{\rm gas}$ relation to higher SFEs (e.\/g.\/, if the AGN emission were uncorrelated with the molecular gas) or bias the fit to steeper slopes (if the AGN were fueled by molecular gas, the excess ${\rm H\alpha}$ emission could correspond to high molecular gas surface brightness). In order to determine if there are regions in J0901 that might be affected by the AGN, we examined SFE as a function of $r_{3,1}$ and the [N\,{\sc ii}]/${\rm H\alpha}$ ratio, since gas fueling the AGN may be at higher density, in a higher excitation state, and/or shocked. While the distribution of SFE has a tail towards higher values, we found no significant correlation between SFE and $r_{3,1}$ or SFE and metallicity. Absent some additional tracer of AGN-affected ${\rm H\alpha}$ emission in J0901, we err on the side of using all pixels in the Schmidt-Kennicutt analysis. An alternative method for identifying and excluding ${\rm H\alpha}$ emission from the AGN would be to perform a velocity decomposition for each SINFONI pixel; the broader ${\rm H\alpha}$ line profile could be associated with the AGN rather than star formation, although a narrow-line AGN component might still masquerade as star formation. However, our current data lack the S/N (and likely the spatial resolution) to do such a decomposition.

Adopting $\alpha_{\rm CO}=4.6\,M_\odot\,{\rm (K\,km\,s^{-1}\,pc^2)}^{-1}$ and scaling to the IR-determined SFR, we find J0901 appears slightly offset to higher SFEs relative to the ``sequence of disks" \citep[e.\/g.\/,][]{daddi2010b,genzel2010}, while lower values of $\alpha_{\rm CO}$ move J0901 to the ``sequence of starbursts," in line with the scenario that the distinction between such sequences is at least partly a product of the assumed $\alpha_{\rm CO}$ factors. If we assume the CO-to-${\rm H_2}$ conversion factors are correct for all of the resolved high-redshift sources, J0901 appears to fall along or slightly below a track of high-redshift starbursts with a net index that is potentially super-unity (Figure~\ref{fig:kslawcompare}). However, given the variety of assumptions involved (extinction corrections, excitation corrections, and $\alpha_{\rm CO}$), J0901 and the other eight individual galaxies studied to date may lie within the normal scatter of SFEs.

Individual Schmidt-Kennicutt indices range from $1.0\lesssim n\lesssim2.0$ for these high-redshift sources. Given that many of the sources explored here use low-$J$ CO lines (\mbox{CO(1--0)} or \mbox{CO(2--1)}), and that we find there is no clear excitation difference for J0901 up to \mbox{CO(3--2)}, we do not think the range of indices reflects the critical densities of different observed gas tracers. For the local disk galaxies in \cite{leroy2013}, there is also a wide range of Schmidt-Kennicutt indices (all mapped in the \mbox{CO(2--1)} line), and it is only the distribution of indices that peaks at $n\sim1$. \citet{wei2010} also find a range of Schmidt-Kennicutt indices ($n\sim1.6$--$1.9$, mapped in the \mbox{CO(1--0)} line) for nearby low-mass E/S0 galaxies with a median index of $n\sim1.2$. While some of the apparent variation in local galaxies is due to other factors (like spatially varying CO-to-${\rm H_2}$ conversion factors), some of the scatter is real, and we suspect this is also the case for high-redshift galaxies. Therefore, larger samples of high-redshift galaxies need to be analyzed in a uniform way if we are to say conclusively whether they have a different Schmidt-Kennicutt index or higher SFE, or if there are differences between galaxy populations (like starbursts vs.~normal galaxies). 

While different choices of molecular gas tracers can complicate comparisons between studies of the Schmidt-Kennicutt relation, these tracers' density sensitivities are valuable tools for probing the underlying volumetric ``Schmidt law" (${\rm SFR}\propto\rho_{\rm gas}^n$; \citealt{schmidt1959}). In a series of hydrodynamic galaxy simulations with 3-D non-LTE radiative transfer modeling, \citet{narayanan2008d} and \citet{narayanan2011} demonstrate that the change in the Schmidt-Kennicutt index with CO rotational line (for the surface density or integrated versions of the relation) differs depending on the index of the underlying volumetric star formation law. They argue that the cold gas less directly involved in star formation will be under-luminous in higher-excitation emission lines; therefore, while the intrinsic star-formation relation using a cold gas tracer might have an index of $n=1.5$, higher excitation emission lines would trace less mass per unit of star formation, resulting in observed indices closer to $n=1$. 

Variation in the power law index with gas tracer has been seen in the integrated form of the Schmidt-Kennicutt relation (e.\/g.\/, \citealt{sanders1991,yao2003,gao2004,narayanan2005,bussmann2008,gracia-carpio2008,bayet2009,iono2009,juneau2009}; cf. \citealt{tacconi2013,sharon2016}), but these studies do not observe all tracers for the same galaxies, nor do they examine spatially resolved star formation properties. \citet{greve2014} analyze the integrated CO and FIR properties for a comprehensive sample of local U/LIRGs and high-redshift SMGs (although not every CO line is detected in every galaxy) and find strong trends in the integrated form of the Schmidt-Kennicutt index with critical density. They find $n\sim1$ for CO rotational transitions $\lesssim J_{upper}=6$ and decreasing indices for higher-excitation lines, a pattern that does not match the predictions of \citet{narayanan2008d}. However, this discrepancy is not entirely robust: \citet{kamenetzky2016} perform a similar analysis using largely the same sample, and find near-unity indices for all CO lines up to \mbox{CO(13--12)}, unless AGN host galaxies were included in the analysis. For high-redshift galaxies with surface density measurements, neither J0901 nor the (current) high-redshift sources exhibit the change in index with CO rotational transition predicted by \citet{narayanan2011} for any of the underlying potential Schmidt laws. Given the variation in indices seen for local disk galaxies \citep{leroy2013}, it seems likely that intrinsic variations between galaxies will mask the population average when only small numbers of observations are available at high redshift. Larger samples of galaxies will therefore need to be mapped in multiple gas tracers (in addition to efforts addressing the extinction corrections mentioned previously) in order to properly test the \citet{narayanan2011} surface density predictions and determine the underlying volumetric star formation law.

\subsection{Correlation between CO excitation and SFR}
\label{sec:sfrco}

\begin{figure*}
\epsscale{1.2}
\plotone{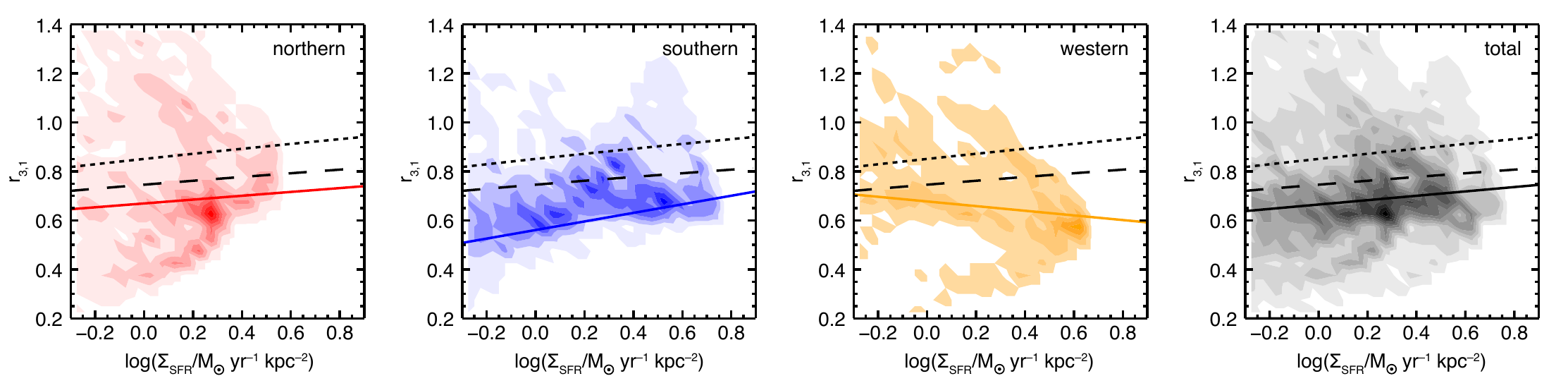}
\caption{$r_{3,1}$ vs.~star formation rate surface density as measured by ${\rm H\alpha}$ surface brightness (corrected globally for extinction) for J0901 (using the $uv$- and resolution-matched maps). From left to right, the four panels plot the density of pixels in 0.05 dex bins for the northern image (red), southern image (blue), western image (gold), and all images combined (gray). The color tones start at one pixel per bin and are in steps of three pixels per bin thereafter; there are eight, six, five, and twelve color steps in the northern, southern, western, and total panels, respectively. The best fit linear relations are shown as solid lines. The black dashed and dotted lines show the correlations between $r_{3,1}$ and $\Sigma_{\rm SFR}$ from \citet{narayanan2014} for the unresolved and resolved cases, respectively. \label{fig:j0901sfrvsr31}}
\end{figure*}

\begin{figure}
\plotone{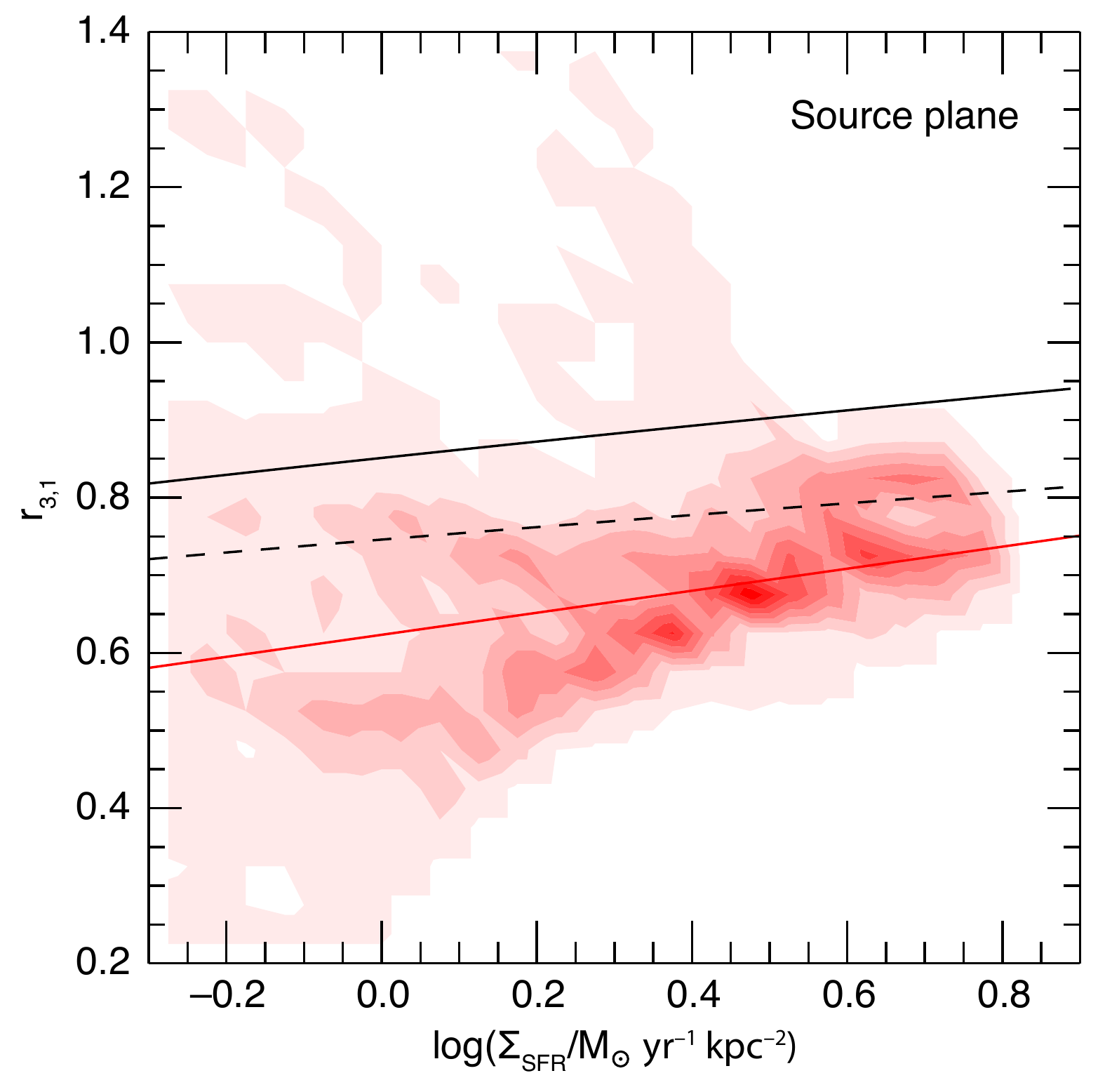}
\caption{$r_{3,1}$ vs.~star formation rate surface density as measured by ${\rm H\alpha}$ surface brightness (corrected globally for extinction) for J0901 using the de-lensed $uv$- and resolution-matched maps. The color tones plot the density of pixels in 0.05 dex bins, starting at one pixel per bin and in steps of five pixels per bin thereafter; there are nine color steps. The best fit linear relation is shown as the solid red line. The black dashed and dotted lines show the correlations between $r_{3,1}$ and $\Sigma_{\rm SFR}$ from \citet{narayanan2014} for the unresolved and resolved cases, respectively. \label{fig:j0901sfrvsr31delens}}
\end{figure}

We also compare our observations of J0901 to the correlation between the shape of the CO spectral line energy distribution (SLED) and $\Sigma_{\rm SFR}$ predicted from a suite of galaxy simulation by \citet{narayanan2014}. Since the correlation is for the \emph{total} SFR surface density, we scale the individual pixels of the ${\rm H\alpha}$ map such that their sum is equal to the TIR-determined SFR (effectively a global extinction correction as described above). In Figures~\ref{fig:j0901sfrvsr31} and \ref{fig:j0901sfrvsr31delens} we plot the $r_{3,1}$ value of each pixel (determined from the $uv$-matched CO maps) vs.~its ${\rm H\alpha}$-determined SFR surface density, scaled to the TIR-determined SFR. We also show the $r_{3,1}$ values predicted for our range of SFR surface densities for both the ``resolved" and ``unresolved" parameterizations of \citet{narayanan2014}; the resolution of our observations, while good for $z\sim2$, is still significantly worse than the $\sim70\,{\rm pc}$ resolution of their simulations. While the \emph{average} $\Sigma_{\rm SFR}$-predicted value of $r_{3,1}=0.77\pm0.02$ is consistent with our measured global average $r_{3,1}=0.79\pm0.12$, the pixelized values of $r_{3,1}$ are generally offset by $\Delta r_{3,1}\approx0.2$.

Since density plots of the southern image, combination of all three images, and the de-lensed source plane reconstruction are all suggestive of correlation between $r_{3,1}$ and the SFR surface density, we attempt a more detailed comparison to the results of \citet{narayanan2014} in order to test if their models can be extended to predict the range of CO excitation variations \emph{within} galaxies. Since any correlation between these two parameters is weak, and the \citet{narayanan2014} relation for $r_{3,1}$ is effectively linear over the relatively narrow range of $\Sigma_{\rm SFR}$ probed by J0901, we attempt a linear fit between $r_{3,1}$ and $\log(\Sigma_{\rm SFR})$. Since, as in fitting the Schmidt-Kennicutt relation, any significance cut on the included pixels could bias the fit (although, for the reasons presented above, we do exclude pixels with fluxes $<2\sigma$ for the source-plane reconstructions), we follow a similar procedure as in our Schmidt-Kennicutt fitting: we randomly sample $\Sigma_{\rm SFR}$ $10^5$ times, use those values to predict $r_{3,1}$ for a specific choice of linear model parameters, perturb $r_{3,1}$ and $\Sigma_{\rm SFR}$ by the noise, compare the binned grid of the model results to that of the observations to calculate a $\chi^2$ value, and repeat for a range of model parameters, ultimately finding the model parameters that produce the lowest $\chi^2$ value. We repeat this procedure 100 times, each time choosing a random bin size (between $0.025$--$0.1$ dex for both $\Sigma_{\rm SFR}$ and $\Delta r_{3,1}$), perturbing the gridding (by up to half a bin width in any direction), perturbing $\Sigma_{\rm SFR}$ and $r_{3,1}$ values by their flux calibration uncertainties ($\sim10\%$), and perturbing individual values of $\Sigma_{\rm SFR}$ and $r_{3,1}$ by their statistical uncertainties. We use the variation in the best-fit values for these 100 iterations to estimate uncertainties. We note that in this case, the choice of noise for $r_{3,1}$ is not trivial, since it is a ratio of two values, and thus the uncertainty depends on the measured values of both the \mbox{CO(1--0)} and \mbox{CO(3--2)} lines (see Figure~\ref{fig:j0901intlineratio}), which are not separately predicted by the model. We therefore produce $10^5$ iterations of the $r_{3,1}$ map in which both image-plane integrated line maps have been perturbed by their own Gaussian noise, and then produce corresponding $r_{3,1}$ uncertainty maps for each iteration.\footnote{We did not scale the noise by the primary beam corrections, which were already applied to the integrated line maps, so the $r_{3,1}$ iterations' scatter is only approximately correct.} We then collect all values of $\sigma_{r_{3,1}}$ (from the $10^5$ uncertainty maps) that correspond to values of $r_{3,1}$ in $\Delta r_{3,1}=0.05$ bins. When perturbing our $\Sigma_{\rm SFR}$-predicted values of $r_{3,1}$ for each model, we randomly choose a $\sigma_{r_{3,1}}$ from the appropriate $\Delta r_{3,1}$ bin.

As for the Schmidt-Kennicutt relation, we include a Gaussian component in case there is intrinsic scatter in addition to the noise. The resulting relationship we attempt to fit is thus

\begin{equation}\label{eq:NK14}
r_{3,1}=A+B\times\log\left(\frac{\Sigma_{\rm SFR}}{2\,{\rm M_\sun\,yr^{-1}\,kpc^{-2}}}\right)+{\mathcal N}(0,\sigma).
\end{equation}

\noindent
For the \citet{narayanan2014} model of $r_{3,1}$, a first order Taylor expansion about a typical $\log(\Sigma_{\rm SFR})=0.4$ yields $A=0.75$ and $B=0.08$ for unresolved sources and $A=0.85$ and $B=0.10$ for resolved sources. Our best-fit values for this relationship are given in Table~\ref{tab:r31vssfr} and are shown in Figures~\ref{fig:j0901sfrvsr31} and \ref{fig:j0901sfrvsr31delens}. Based on the uncertainties on the best-fit slopes, we do not find evidence for a trend in $r_{3,1}$ with SFR surface density in the northern image, western image, or the source-plane reconstruction. However, our analysis does suggest that the observed $r_{3,1}$ values are correlated with $\Sigma_{\rm SFR}$ for the southern image and tentatively correlated for all images combined. While the slopes of the best-fit models for southern and total correlations are similar to the results of \citet{narayanan2014} (as is the slope for the source-plane reconstruction, if we neglect its significant uncertainty), the overall normalization is lower. This offset indicates that consistency between the observed integrated $r_{3,1}$ and its predicted value is largely an effect of the asymmetric distribution of $r_{3,1}$ pixels, which has a tail to large values (Figure~\ref{fig:j0901ratiohisto}). 

\begin{deluxetable}{lccc}
\tablewidth{0pt}
\tablecaption{$r_{3,1}$-$\Sigma_{\rm SFR}$ fit parameters \label{tab:r31vssfr}}
\tablehead{ \colhead{Image} & \colhead{$A$} & \colhead{$B$} & \colhead{$\sigma$}}
\startdata
North & $0.69\pm0.04$ & $0.08\pm0.08$ & $0.33\pm0.04$ \\
South & $0.61\pm0.03$ & $0.18\pm0.05$ & $0.18\pm0.05$ \\
West & $0.65\pm0.02$ & $-0.09\pm0.07$ & $0.28\pm0.03$ \\
Total & $0.69\pm0.04$ & $0.09\pm0.04$ & $0.32\pm0.03$ \\
De-lensed & $0.67\pm0.15$ & $0.14\pm0.30$ & $0.03\pm0.03$ \\
\enddata
\end{deluxetable}

All of the best-fit correlations derived for the image plane that are consistent with \citet{narayanan2014} require additional scatter beyond the statistical uncertainty of the individual image-plane maps, suggesting that other physical processes besides the SFR density affect the molecular excitation (if the predicted correlation is real). However, the source-plane reconstruction does not require additional scatter; the spatially varying uncertainty associated with the source-plane CO maps appears to be sufficient, and therefore suggests that the scatter contains no astrophysical information. The larger uncertainty associated with the source plane and relatively limited range in SFR surface densities allowed by the $2\sigma$ significance requirement (not shown in Fig.~\ref{fig:j0901sfrvsr31delens}, but it removes most pixels with $\log(\Sigma_{\rm SFR}/({\rm M_\odot\,yr^{-1}\,kpc^{-2}}))<0.1$ and leaves the spur of high $r_{3.1}$ values at $\log(\Sigma_{\rm SFR}/({\rm M_\odot\,yr^{-1}\,kpc^{-2}}))\sim0.4$) likely contributes to the uncertainty in the correlation. 

We emphasize that although these coefficients are the best-fit values for an assumed linear relationship, that does not mean there is actually a statistically significant correlation between $r_{3,1}$ and $\Sigma_{\rm SFR}$ for J0901. The models of \citet{narayanan2014} were derived to reproduce a much wider range in $\Sigma_{\rm SFR}$ than probed by any one galaxy, and were intended to predict the \emph{global} CO excitation for a galaxy-wide average SFR surface density. Therefore, it is perhaps unsurprising that their correlation does not reproduce our observed distribution of $r_{3,1}$ values for a single galaxy. Our results do not clearly indicate that the SFR surface density depends on the gas excitation for the physical scales probed in our images of J0901. This result is consistent with the unchanging Schmidt-Kennicutt index for the different CO lines. Similar, albeit unresolved, comparisons by \citet{yao2003} for a sample of infrared bright galaxies in the local universe and by \citet{sharon2016} for a sample of $z\sim2$ SMGs also do not find a correlation between the luminosity of a total SFR tracer ($L_{\rm FIR}$) and $r_{3,1}$. However, \citet{kamenetzky2016} do find some correlation of $r_{3,1}$ with $L_{\rm FIR}$ for all nearby galaxies observed with the {\it Herschel} SPIRE Fourier Transform Spectrometer (mostly U/LIRGs and IR-bright AGN, but includes a substantial number of galaxies with $L_{\rm FIR}\sim10^{10}\,{\rm L_\sun}$).

\subsection{Possible AGN origin for excess $35\,{\rm GHz}$ continuum emission}
\label{sec:VLAcont}

Synchrotron emission at radio wavelengths can be an alternative probe of galaxies' SFRs. However, J0901 is known to contain an AGN, which may corrupt long-wavelength estimates of its SFR. In addition, our $35\,{\rm GHz}$ (observed frame) continuum detection is in the region of the SED where the Rayleigh-Jeans tail of the dust emission, synchrotron emission, and free-free emission can all contribute to the observed flux density. We therefore estimate the contribution from each of these emission components to the observed $35\,{\rm GHz}$ flux density to determine whether the observed emission is driven by star formation and/or the AGN.

J0901 falls off the radio-FIR correlation presented in \citet{magnelli2015}. Using the (observed) TIR luminosity from \citet{saintonge2013} rescaled for our magnification, the standard spectral index for synchrotron emission ($S_\nu \propto\nu^{-0.8}$), and the (weakly) redshift-dependent form the of radio-IR correlation from \citet{magnelli2015}, we would expect the $35\,{\rm GHz}$ (observed frame; $115\,{\rm GHz}$ rest frame) continuum emission in J0901 to be only $\sim1\,{\rm \mu Jy}$, which is significantly less than our measured flux density of $0.66\pm0.12\,{\rm mJy}$. Alternatively, we can use J0901's SFR (based on $L_{\rm TIR}$ from \citet{saintonge2013}) and invert the relationship for determining SFR from $1.4\,{\rm GHz}$ continuum luminosity \citep{kennicutt2012} to determine the expected contribution to the $35\,{\rm GHz}$ flux density from synchrotron emission (again, assuming $S_{\nu}\propto\nu^{-0.8}$ and our \mbox{CO(3--2)}-determined magnification factor). Using this method, we expect $\sim9\,{\rm \mu Jy}$ of synchrotron emission at $35\,{\rm GHz}$, which is more than we would expect based on the radio-FIR correlation, but still nearly two orders of magnitude smaller than the observed emission. 

Thermalized free-free emission from the H\,{\sc ii} regions of massive ($>5\,M_\sun$) stars can also be used as a tracer of (high mass) star formation. Following \citet{condon1992}, the TIR-determined SFR yields an expected $35\,{\rm GHz}$ flux density of $\sim60\,{\rm \mu Jy}$ (for an assumed magnification factor of $31.3$; we lack adequate S/N to independently determine the magnification for the VLA continuum map). Since this estimate does not account for the formation of lower mass stars, we estimate that $\lesssim20\%$ of the observed $35\,{\rm GHz}$ flux density comes from free-free emission.


The Rayleigh-Jeans tail of the dust continuum peak is unable to account for the difference between the observed $35\,{\rm GHz}$ continuum emission and the expected contributions from synchrotron and free-free emission associated with star formation. Using our observed $858\,{\rm \mu m}$ SMA detection and $\beta=1.5$ \citep{saintonge2013}, we predict a $11.5\pm2.9\,{\rm \mu Jy}$ contribution to the $35\,{\rm GHz}$ flux density. Similar calculations using the Rayleigh-Jeans tail continuum measurements of J0901 in \citet{saintonge2013} produce consistently low extrapolated $35\,{\rm GHz}$ flux densities.

Regardless of the discrepancy between the two methods for determining the synchrotron emission from star formation, which dust continuum estimate we use, and reasonable perturbations for assumed spectral indices, the expected combined contribution from star formation and dust emission is $\lesssim20\%$ of our observed $35\,{\rm GHz}$ continuum detection. Therefore, either the free-free emission has a significantly different magnification factor from the \mbox{CO(1--0)} emission, or the bulk of the VLA continuum emission is due to an AGN. The magnification factor required to bring our observed $35\,{\rm GHz}$ emission into alignment with our total SFR is $\mu\gtrsim140$ (assuming it is dominated by free-free emission), several times larger than our other magnification factors, which seems unlikely if they are all tracing the same star forming material within J0901 (i.\/e.\/, there should not be much differential lensing). Given that J0901 is known to contain an AGN, we conclude that its large $35\,{\rm GHz}$ flux density is most likely due to synchrotron emission from the AGN. The synchrotron emission from the AGN may also be affected by differential lensing, and to the extent that we trust the morphology of the low-S/N $35\,{\rm GHz}$ emission, it does not appear to originate from the brightest emission regions at longer wavelengths, but may correspond to the bright emission seen in ${\rm H\alpha}$ and [N\,{\sc ii}]. High-resolution observations at lower frequencies are necessary to test this hypothesis.

As a final possibility, some fraction of the radio continuum emission may not be associated with J0901 at all, and may instead be due to members of the foreground group of galaxies. This scenario may explain the inconsistencies between star formation rate predictors and the offset of the peak emission in the northern image. As the central galaxies in the lensing cluster definitely do produce $35\,{\rm GHz}$ continuum emission, the southern emission peak may also be due to emission from the foreground interloper that is nearly aligned with the southern arc. Higher SNR or better spatial resolution observations are necessary to determine what fractions of the radio continuum emission are associated with J0901 and the lensing galaxies.

\section{Summary}
\label{sec:concl}

We present $\sim1^{\prime\prime}$ resolution ($\sim2\,{\rm kpc}$ in the source plane) observations of the \mbox{CO(1--0)}, \mbox{CO(3--2)}, ${\rm H\alpha}$, and [N\,{\sc ii}] lines in a strongly-lensed star-forming galaxy at $z=2.26$, SDSS\,J0901+1814 (J0901). We use our highest S/N line detection (the \mbox{CO(3--2)} line) and existing \emph{HST} data to constrain the lensing potential of the foreground group of galaxies, and find a typical magnification factor $\mu\approx30$ (depending on wavelength). Dynamical modeling of the source-plane reconstruction using both the CO and ${\rm H\alpha}$ data indicates that J0901 is a nearly face-on ($i\approx30\degree$) massive disk with $r_{1/2}\gtrsim4\,{\rm kpc}$. Our ${\rm H\alpha}$ observations of J0901 trace only a small fraction of the total star formation rate implied by the galaxy's $L_{\rm TIR}$. Applying our new magnification factors to $L_{\rm TIR}$ from \citet{saintonge2013}, we find the SFR for J0901 is $268^{+63}_{-61}\,{M_\sun}\,{\rm yr^{-1}}$. J0901's magnification-corrected SFR and stellar mass place it only $\sim0.25\,{\rm dex}$ above the star-forming galaxy main sequence, consistent with its being a ``normal" galaxy considering the significant uncertainty in its sSFR.

Our CO observations yield a total molecular gas mass of $M_{\rm gas}=(1.6^{+0.3}_{-0.2})\times10^{11}(\alpha_{\rm CO}/4.6)\,{M_\sun}$. The molecular gas is nearly equal to the magnification-corrected stellar mass, which yields a total baryonic mass of ($2.6^{+0.5}_{-0.3})\times10^{11}\,{M_\odot}$ that is significantly larger than our dynamical mass estimate of $\sim1.3\times10^{11}\,{M_\odot}$. Reducing the assumed CO-to-${\rm H_2}$ conversion factor to the typical ``starburst" values of $\alpha_{\rm CO}\sim0.8$ would bring the baryonic and dynamical masses into alignment (assuming moreover J0901 is baryon dominated). For our two integrated CO lines, we find an average line ratio of $r_{3,1}=0.79\pm0.12$, which is skewed somewhat higher than the peak of the the pixelized $r_{3,1}$ distribution. After correcting for the inclination angle, we find evidence for a significant decrease in $r_{3,1}$ as a function of radius out to $\sim10\,{\rm kpc}$. However, there is no significant correlation between $r_{3,1}$ and the [N\,{\sc ii}]/${\rm H\alpha}$ ratio (used as a metallicity tracer), nor does there appear to be a significant trend in $r_{3,1}$ with velocity channel.

Using our CO and ${\rm H\alpha}$ maps, we analyze where J0901 falls relative to the Schmidt-Kennicutt relation for other galaxies. The relative positions of galaxies strongly depend on the extinction correction used to determine the spatially resolved SFR and assumed CO-to-${\rm H_2}$ conversion factor. Since we do not have a spatially resolved tracer of the obscured star formation, we plot J0901 using the ${\rm H\alpha}$-derived SFR surface density and show how it would shift assuming the obscured star formation traces the unobscured star formation. With the correction to account for the obscured star formation, J0901 appears to be slightly offset to higher SFEs than ``normal" disk galaxies found in the local universe \citep[e.\/g.\/,][]{leroy2013}. Given J0901's dynamics and high metallicity, we assume a typical Galactic conversion factor. Contrary to results claiming that galaxies are offset relative to the local Schmidt-Kennicutt relation solely due to the assumed conversion factor, J0901 would be offset even further for lower values of ${\rm \alpha_{CO}}$. We find the average slope for the Schmidt-Kennicutt relation for J0901 to be $\bar n=1.54\pm0.13$ in the image plane. We do not find significantly different slopes when using the \mbox{CO(1--0)} and \mbox{CO(3--2)} lines to trace the molecular gas for the matched resolution/inner $uv$-radius data (in either the image or source plane), but we do find some difference for the native resolution CO data. The observed slope of the Schmidt-Kennicutt relation does differ between the \mbox{CO(1--0)} maps using the natural resolution data and the \mbox{CO(1--0)} maps that have been smoothed and $uv$-clipped (to match the \mbox{CO(3--2)} data). We also find a slightly flatter slope of $\bar n = 1.24\pm0.02$ when using the source-plane reconstructions of J0901. While the true index for J0901 is somewhat uncertain, all of these indeces are higher than the average for normal disk galaxies in the local universe \citep[e.\/g.\/,][]{leroy2013} but within their observed scatter. Few measurements of the resolved Schmidt-Kennicutt relation exist at high redshift, but J0901 is within the measured range of indices of $n=1$--$2$. However, our analysis assumes a global extinction correction to the ${\rm H\alpha}$ data used to trace the star formation, and \citet{genzel2013} find significant variation in the index depending on the assumed extinction correction.

We also use our resolved observations to assess whether the correlation between SFR surface density and CO excitation identified in the simulations of \citet{narayanan2014} holds \emph{within} individual galaxies. For the limited range in $\Sigma_{\rm SFR}$ in J0901, we do not reproduce the \citet{narayanan2014} correlation, although the galaxy-wide average $r_{3,1}$ is comparable to the value predicted from the measured average SFR surface density. This distinction is likely tied to the limited range of $\Sigma_{\rm SFR}$ and skewed distributions of $r_{3,1}$ values in our data. However, as any correlations are weak, we do not ascribe much meaning to the difference between the observed fits and the correlation predicted in \citet{narayanan2014}.

We find a significant excess of $35\,{\rm GHz}$ (observed frame) continuum emission relative to the expected contributions from the Rayleigh-Jeans tail of the dust emission, synchrotron emission, and free-free emission predicted for the measured SFR. Given that the magnification factor for the free-free emission would need to be about five times larger than what we have found at other wavelengths to account for the excess flux, we conclude that the $35\,{\rm GHz}$ continuum emission is either synchrotron emission from the AGN and/or contamination from the foreground group of lensing galaxies.

Although J0901's SFE and sSFR are slightly elevated, and it contains an AGN, J0901 appears to be a relatively normal massive disk galaxy at $z=2.26$. The nearly face-on orientation and additional physical resolution enabled by gravitational lensing (for the same observed angular resolution) makes J0901 a valuable laboratory for probing galaxy evolution in galaxies at $z\sim2$ \citep[see also][]{johnson2017b}, such as feedback from star formation and AGN. However, larger samples of resolved systems with comparable-quality multi-wavelength observations are necessary to test whether these results  (e.\/g.\/, pertaining to the Schmidt-Kennicutt index as a function of CO line and spatial variations in CO line ratios) generalize to other (populations of) high-$z$ galaxies. 

\acknowledgments{We thank the anonymous referee for helpful comments. This work has been supported by the National Science Foundation through grant AST-0955810. The National Radio Astronomy Observatory is a facility of the National Science Foundation operated under cooperative agreement by Associated Universities, Inc. The Submillimeter Array is a joint project between the Smithsonian Astrophysical Observatory and the Academia Sinica Institute of Astronomy and Astrophysics and is funded by the Smithsonian Institution and the Academia Sinica. This work is based in part on observations made with the ESO Very Large Telescope at the La Silla Paranal Observatory under program ID 087.A-0972.}

\bibliographystyle{apj}

\end{document}